\documentclass[11pt]{article}
\usepackage{gbmacros}
\usepackage{times,mathpazo,charter,graphicx,color,array,bm}
\usepackage[section]{placeins}
\numberwithin{equation}{section}
\usepackage{subcaption}
\usepackage[colorlinks=true,citecolor=blue]{hyperref}
\usepackage{cite}
\usepackage[tight]{shorttoc}

\def\maketitle{\par\noindent{\LARGE\bf\sffamily\thetitle}\\[1.4ex]
{\large\theauthor}\\[0.6ex]
\textit{\thetextinfo}\\[0.2ex]
{\small\today}\par\vglue1.4\bigskipamount}
\def\title#1{\def\thetitle{#1}}
\def\author#1{\def\theauthor{#1}}
\def\textinfo#1{\def\thetextinfo{#1}}
\allowdisplaybreaks
\usepackage{tocloft}
\makeatletter
\let\tru@int=\int
\def\int{\mathop{\textstyle\tru@int}\limits}
\def\overl@ss#1#2{\vcenter{\offinterlineskip
        \ialign{$\m@th#1\hfil##\hfil$\crcr#2\crcr<\crcr } }}
\def\overgr@at#1#2{\vcenter{\offinterlineskip
        \ialign{$\m@th#1\hfil##\hfil$\crcr#2\crcr>\crcr } }}
\def\gl{\mathrel{\mathpalette\overl@ss>}}
\def\lg{\mathrel{\mathpalette\overgr@at<}}
\def\d{\mathrm{d}}
\def\Natural{\mathbb{N}}

\def\Real{\mathbb{R}}
\def\Complex{\mathbb{C}}
\def\Re{\mathop{\rm Re}\nolimits}
\def\Im{\mathop{\rm Im}\nolimits}

\def\pvint{\int\kern-0.94em-\kern0.2em}
\let\^=\widehat
\let\==\bar
\let\@=\mathbf

\def\d{\mathrm{d}}
\def\e{\mathrm{e}}

\def\Wr{\mathrm{Wr}}

\def\~#1{\widetilde{#1}}

\makeatother

\advance\textwidth -2mm
\advance\hoffset 1mm
\advance\textheight 4mm
\advance\voffset -6mm
\advance\parskip -2pt
\def\be{\begin{equation}}
\def\ee{\end{equation}}
\def\bse{\begin{subequations}}
\def\ese{\end{subequations}}
\def\eqref#1{(\ref{#1})}

\def\re{\mathrm{re}}
\def\im{\mathrm{im}}
\def\I{\mathrm{I}}

\def\conj#1{\overline{#1}}

\def\R{\mathbb{R}}
\def\C{\mathbb{C}}

\newcommand{\norm}[1]{\lVert#1\rVert}

\newcommand{\RN}[1]{%
	\textup{\uppercase\expandafter{\romannumeral#1}}%
}

\def\reftitle#1{``#1''}
\let\booktitle=\emph
\definecolor{darkblue}{rgb}{0, 0, 0.77}
\let\bb=\relax
\let\eb=\relax

\begin{document}
\thispagestyle{empty}

{\title{Inverse scattering transform for the focusing nonlinear\\Schr\"odinger equation with counterpropagating flows}
\author{Gino Biondini$^*$, Jonathan Lottes$^*$, Dionyssios Mantzavinos$^\dag$}
\textinfo{$^*$ Department of Mathematics, State University of New York at Buffalo, Buffalo, New York 14260, USA
\\
$^\dag$ Department of Mathematics, University of Kansas, Lawrence, Kansas 66045, USA}
\maketitle}

\kern-2\medskipamount
\begingroup
\small
\paragraph{Abstract.}
The inverse scattering transform for the focusing nonlinear Schr\"odinger equation 
is presented for 
a general class of initial conditions whose asymptotic behavior at infinity consists 
of counterpropagating waves.
The formulation takes into account the branched nature of the two asymptotic eigenvalues of the associated scattering problem. 
The Jost eigenfunctions and scattering coefficients are defined explicitly as single-valued functions on the complex plane with jump discontinuities along certain branch cuts. 
The analyticity properties, symmetries, discrete spectrum, asymptotics and behavior at the branch points are discussed explicitly. 
The inverse problem is formulated as a matrix Riemann-Hilbert problem with poles.
Reductions to all cases previously discussed in the literature are explicitly discussed.
The scattering data associated to a few special cases consisting of physically relevant Riemann problems are explicitly computed.
\endgroup
%
\kern-\medskipamount
\tableofcontents

\section{Introduction and motivation}

The nonlinear Schr\"odinger (NLS) equation,
$	iq_t + q_{xx} \pm 2 |q|^2q=0  $,	
(``$+$'' for focusing; ``$-$'' for defocusing)
 is one of the most important systems in nonlinear science, 
since it arises as a model in deep water waves, plasmas, acoustics, optics and Bose-Einstein condensation \cite{A2001,IR2000,PS2004,AS1981,SS1999}. 
Indeed, the NLS equation is a universal model for the evolution of a complex envelope of weakly nonlinear dispersive wave trains \cite{BN1967}.
The NLS equation is also one of the most well-known examples of an integrable nonlinear evolution equation.
Infinite-dimensional integrable systems have been studied extensively due to the combination of physical relevance and rich mathematical structure
\cite{AS1981,NMPZ1984,FT1987,BBEIM1994,APT2004}. 
In particular, for the NLS equation, the inverse scattering transform (IST) was developed by Zakharov and Shabat in 1972 
to solve the initial value problem (IVP) in the case of zero boundary conditions  (BCs) at infinity and of initial conditions (ICs) with sufficient smoothness \cite{ZS1972}. Shortly after, the same authors extended the formulation of the IST to solve the IVP with symmetric nonzero boundary conditions (NZBCs) in the defocusing case \cite{ZS1973}. 
The behavior of solutions in these cases has since been extensively studied and unraveled in several works, e.g., see  
\cite{AKNS1974,SY1974,GK1978,KMM2003,TV2000,TVZ2004,BT2013,DPMV2013,JM2013,CJ2016,EKT2016,K1977,KI1978,M1979,BP1982,GK2012}
and the references therein. 
In particular, the case of symmetric NZBCs in the focusing NLS equation has received renewed attention recently \cite{BK2014,BM2016,BM2017,BLM2018}, 
and the case of fully asymmetric NZBCs in both focusing and defocusing NLS equations was also studied \cite{BP1982,DPMV2014,PV2015,BFP2016}.

Importantly, however, all of the above works considered the case of either zero or constant BCs at infinity.    
\bb
In the case of the Korteweg-de Vries equation, solutions with more general kind of behavior were recently studied in \cite{ZDZ2016,GR2020}.
For the NLS equation, however,
\eb
only two works in the more general case of plane-wave BCs are available in the literature, one in the focusing case \cite{BV2007} and one in the defocusing case \cite{J2015}.
Nevertheless, in both of those works  only a specific choice of ICs was considered, corresponding to a Riemann problem,
namely a plane wave in each of the half-lines $x>0$ and $x<0$  with a discontinuity at the origin.
The aim of this work is to develop the IST for solving the IVP for the focusing NLS equation
\be
\label{e:fNLS}
	iq_t + q_{xx} + 2|q|^2q=0\,, \quad (x, t) \in \R \times \R\,,
\ee
with a more general class of ICs $q(x,0)$ which reduce to plane waves only as $x\to\pm\infty$, namely, 
\[
	q(x,0) = A_\pm \e^{-iV|x|\pm i\delta}(1 + o(1))\,, \quad x \rightarrow \pm \infty\,,
\label{e:ICs}
\]
where $A_\pm>0$ and $V,\delta \in \R$. 
Throughout this work, $q:\Real\times\Real^+\to\Complex$, and subscripts $x$ and $t$ denote partial differentiation.
Detailed statements about the precise function spaces required for the various steps in the development of the IST 
will be given later.
Note that one could equally well consider the seemingly more general class of ICs
$q(x,0) = A_\pm \e^{iV_\pm x + i\delta_\pm}(1 + o(1))$ as $x\to\pm\infty$.
However, there is no actual need to do so,
since without loss of generality one can always reduce this latter class to the ICs ~\eqref{e:ICs}, namely $V_\pm = \pm V$ and $\delta_\pm = \pm\delta$,
using the Galilean and phase invariances of the NLS equation. 
Thus, the present work encompasses the most general family of solutions of the focusing NLS equation
which tends asymptotically to genus-0 (i.e., constant or plane wave) behavior at infinity.

The family of ICs~\eqref{e:ICs} includes those studied in all of the aforementioned works on the focusing NLS equation as special cases. 
In particular, the long-time asymptotics of solutions in various subcases when either $A_-$ and/or $A_+$ are nonzero have been studied by various authors in recent years \cite{BM2017,BV2007,BKS2011}. 
Here, we address the general case and show how the various subcases can be obtained as appropriate reductions, thus providing a unified framework for the study of these problems. We also consider various Riemann problems, i.e. pure step ICs.
As usual, the development of the IST
proceeds under the assumption of existence and uniqueness.
Once a representation for the solution of the IVP has been obtained, however, one can use it as the starting point
to rigorously prove the well-posedness of the problem in  appropriate function spaces e.g., see \cite{bdt1988,Z1989,TO2015}.

This work is organized as follows.
Section~\ref{s:Direct problem: Jost solutions and analyticity properties} introduces the Jost solutions and their properties. Section~\ref{Direct problem: Scattering matrix, symmetries and discrete eigenvalues} introduces the scattering matrix and symmetries of the Jost solutions. Section~\ref{Inverse problem: RHP formulation} formulates the inverse problem as a matrix Riemann-Hilbert problem. 
Section~\ref{Reductions} discusses various reductions as special cases, 
such as that of equal amplitudes, zero velocities, or one-sided boundary conditions.
Section~\ref{Riemann problems} is devoted to various explicit initial conditions. 
Proofs of theorems, lemmas and corollaries are provided in section~\ref{s:proofs},
and section~\ref{s:conclusions} ends this work with some concluding remarks.

\section{Direct problem: Jost solutions and analyticity properties}
\label{s:Direct problem: Jost solutions and analyticity properties}

The focusing NLS equation~\eqref{e:fNLS} is the compatibility condition $\phi_{xt} = \phi_{tx}$ or, equivalently, $X_t - T_x + [X,T]=0$, of the following overdetermined linear system of ODEs known as a Lax pair: %
\bse
\label{e:LP}
\begin{gather}
		\label{e:LPX}
		\phi_x(x,t,k) = X(x,t,k)\phi(x,t,k)\,,
		\\
		\label{e:LPT}
		\phi_t(x,t,k) = T(x,t,k)\phi(x,t,k)\,,
\end{gather}
\ese
where
\bse
\begin{gather}
		\begin{align}
		X(x,t,k) &= ik\sigma_3 + Q(x,t)\,,
		\\
		T(x,t,k) &= -2ik^2\sigma_3 + i\sigma_3 \left(Q_x(x,t)-Q^2(x,t)\right) - 2kQ(x,t)\,,
		\end{align}
\end{gather}
\ese
and
\[
	Q(x,t) = \begin{pmatrix}
						 0 				& q(x,t)\\
						 -\conj{q(x,t)} & 0
					 \end{pmatrix},
	\quad
	\sigma_3 = \begin{pmatrix}
							 1 & 0\\
							 0 & -1
						 \end{pmatrix},
	\quad
	\sigma_1 = \begin{pmatrix}
							 0 & 1\\
							 1 & 0
						 \end{pmatrix},
\]
with the bar denoting complex conjugation.
Equation~\eqref{e:LPX} is referred to as the scattering problem,
the complex-valued matrix function $\phi(x,t,k)$ is referred to as the eigenfunction,
$k$ is referred to as the scattering parameter,
and $q(x,t)$ as the scattering potential.
The matrix $\sigma_1$ is defined now for later use.

The IST method can be outlined as follows: first, using appropriate solutions of the Lax pair~\eqref{e:LP} known as Jost solutions, one constructs a map that associates the solution $q(x,t)$ of the NLS equation to a suitable set of 
``scattering data'', which are independent of $x$ and $t$ and depend only on $k$. Then,  inverting this map one recovers the potential in terms of said scattering data.
In this section, we introduce the Jost solutions and we determine their properties. Proofs for all the results in this section are given in Section~\ref{a:Direct Problem}.

\subsection{Jost solutions: Formal definition}

It is useful to first consider the eigenfunctions corresponding to the following two exact plane-wave solutions of the 
NLS equation~\eqref{e:fNLS}:
\begin{align}
\label{e:qpmdef}
		&q_\pm(x,t) = A_\pm\e^{-2if_\pm(x,t) \pm i\delta}\,,
		\\
\noalign{\noindent with}
		&f_\pm(x,t) = \frac{1}{2} \left[\pm Vx + (V^2 - 2A_\pm^2)t\right]\,.
\end{align}
Here and throughout we use the subscripts $\pm$ to relate to behavior as $x \rightarrow \pm \infty$. 
(Note that the labels $q_\pm$ have been used in previous works to denote constant values, independent of $x$ and $t$.  This is not the case here.) 

Observe that the asymptotic behavior~\eqref{e:ICs} for the ICs can be written as 
$q(x,0) = q_\pm(x,0)\,(1+o(1))$, $x\to\pm\infty$.
Thus,  as long  as the IVP is well-posed, the condition~\eqref{e:ICs} implies \[\label{e:nzbc}
q(x,t) = q_\pm(x,t)(1+o(1))\,,\quad x \rightarrow \pm\infty\,,
\]
for all $t\in \R$, so that
\bse
\begin{alignat}{2}
		  Q(x,t) &= Q_\pm(x,t)(1+o(1)), & &x\rightarrow \pm\infty\,,  
		\\
		 X(x,t,k) &= X_\pm(x,t,k)(1+o(1)), \quad & &x \rightarrow \pm\infty\,,  
		\\
		 T(x,t,k) &= T_\pm(x,t,k)(1+o(1)), & &x \rightarrow \pm\infty\,,  
\end{alignat}
\ese
where
\be
\begin{aligned}
		Q_\pm(x,t) &= \e^{-if_\pm(x,t)\sigma_3}\big(A_\pm \sigma_3 \e^{\pm i\delta\sigma_3}\sigma_1\big)\e^{if_\pm(x,t)\sigma_3}\,,
		\\
		X_\pm(x,t,k) &= ik\sigma_3 + Q_\pm(x,t)\,,
		\\
		T_\pm(x,t,k) &= -2ik^2\sigma_3 +i\sigma_3\big((Q_\pm)_x(x,t) - Q_\pm^2(x,t)\big) - 2kQ_\pm(x,t)\,.
\end{aligned}
\ee
In Section~\ref{Jost solutions for the exact potentials q_pm(x,t)}, we derive the following simultaneous solutions $\~\phi_\pm(x,t,k)$ to the Lax pair~\eqref{e:LP} for the exact potentials $q_\pm(x,t)$:%
\[\label{e:phipmtilde}
	\~\phi_\pm(x,t,k) = \e^{-if_\pm(x,t)\sigma_3} E_\pm(k) \e^{i\theta_\pm(x,t,k)\sigma_3}\,,
\]
with
\bse
\begin{align}
		\label{e:lambda_pm(k)}
		\lambda_\pm(k) &= \left((k \pm V/2)^2 + A_\pm^2\right)^{1/2}\,,
		\\
		\label{e:E_pm(k)}
		E_\pm(k) &= \I + \frac{iA_\pm}{\lambda_\pm(k) + (k \pm V/2)}\e^{\pm i\delta\sigma_3}\sigma_1\,,
		\\
		\label{e:theta_pm(k)}
		\theta_\pm(x,t,k) &= \lambda_\pm(k) \left(x-2(k\mp V/2)t\right)\,.
\end{align}
\ese
Note that $\lambda_\pm(k)$ has branch points at $k=p_{\pm}$ and $k=\conj{p_{\pm}}$, where
\be
	\label{branch points}
	p_{\pm}=\mp V/2+ iA_\pm\,.
\ee
\bb
We find that
\be
	\label{e:det E_pm}
  D_\pm(k):=\det E_\pm(k) = \frac{2\lambda_\pm(k)}{\lambda_\pm(k) + (k \pm V/2)}.
\ee
In the special case $A_\pm=0$,~\eqref{e:det E_pm} reduces to $D_\pm(k) \equiv 1$, 
and the whole formalism
reduces to the IST with zero boundary conditions.
When $A_\pm\ne0$, $D_\pm(k)$ vanishes only at the branch points of $\lambda_\pm(k)$. 
Moreover, since
\[
	A_\pm^2 = \left(\lambda_\pm(k) + \left(k\pm V/2\right)\right) \left(\lambda_\pm(k) - \left(k\pm V/2\right)\right),
\]
neither factor on the right-hand side is ever zero, and therefore $D_\pm(k)$ has no poles.

Motivated by~\eqref{e:phipmtilde}, we define the Jost solutions $\phi_\pm(x,t,k)$ for the potential $q(x,t)$ satisfying~\eqref{e:ICs} to be the simultaneous solutions of the Lax pair~\eqref{e:LP} such that
\[
\label{e:phi x asymptotics}
	\phi_\pm(x,t,k) = \frac{1}{d_\pm(k)}\e^{-if_\pm(x,t)\sigma_3} E_\pm(k) \e^{i\theta_\pm(x,t,k)\sigma_3} (1+o(1))\,, \quad x \rightarrow \pm\infty\,,
\]
where the factor
\[
	d_\pm(k):= (D_\pm(k))^{1/2}
\]
is introduced to simplify the resulting symmetries and jump matrices that will be computed (see Sections~\ref{s:symmetries} and~\ref{Jump matrix and residue conditions}).
Moreover, by Abel's theorem, since $X$ and $T$ are traceless, the determinants of $\phi_\pm(x,t,k)$ are independent of $x$ and $t$ and %
\be
	\det \phi_\pm(x,t,k) = \lim_{x \rightarrow \pm \infty} \det \phi_\pm(x,t,k) = \frac{\det E_\pm(k)}{d_\pm(k)^2} = 1.
\ee
On the other hand, the factor $d_\pm(k)$ introduces poles at the branch points which will need to be considered (see Section~\ref{s:branchpoints}). 
We will make an explicit choice of branch cut for $\lambda_\pm(k)$ and $d_\pm(k)$ in Section~\ref{ss:Branch cut for lambda_pm(k)}. A rigorous definition of the Jost solutions and their domains of existence and analyticity will be given in Section~\ref{Rigorous definition of the Jost solutions, analyticity and continuous spectrum}.
\eb

\subsection{Branch cuts for the asymptotic eigenvalues}
\label{ss:Branch cut for lambda_pm(k)}

\begin{figure}[b!]
	\centering
	\includegraphics[width=0.375\textwidth]{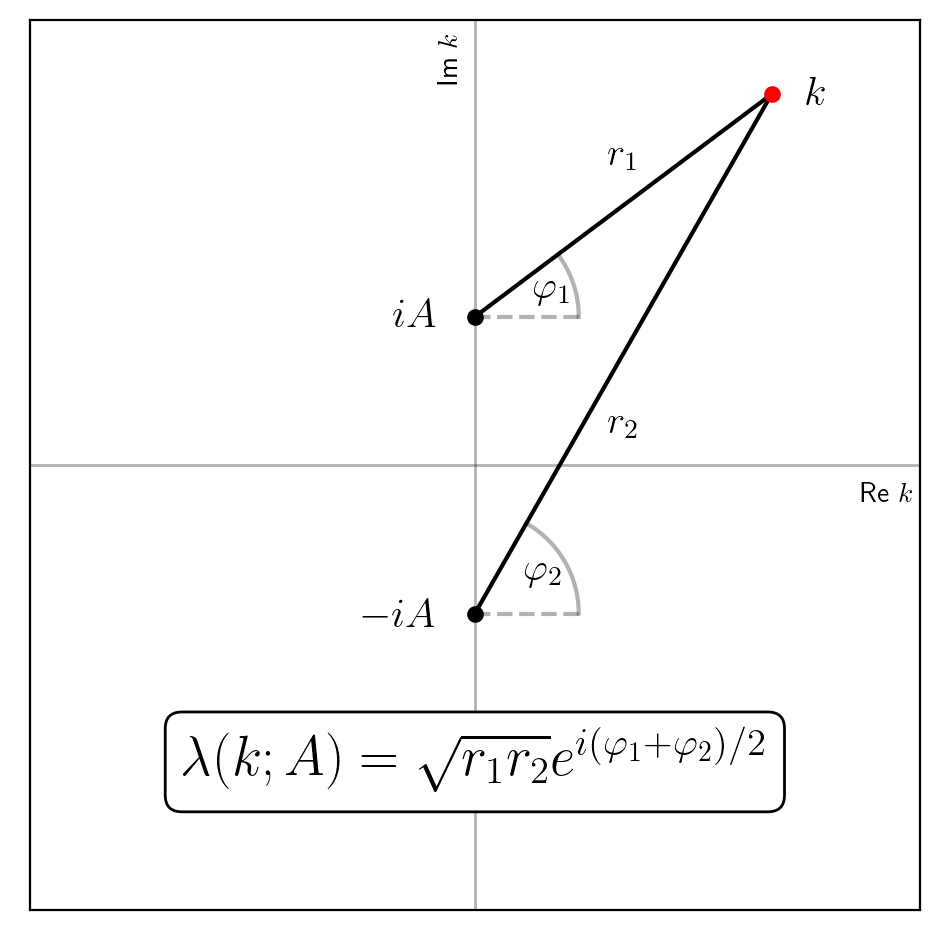}
	\caption{The definition~\eqref{e:explicit lambda(k;A) definition} of the generic square root $\lambda(k;A)$ with $\phi_j \in [-\pi/2, 3\pi/2)$.}
\label{f:lambda}
\end{figure}
In order to discuss the analyticity properties of the Jost solutions defined above, it is necessary to make an explicit choice of branch cut to define $\lambda_\pm(k)$ for all $k \in \C$. To simplify the argument, we first define
\[
\label{e:lambda(k;A)}
	\lambda(k;A) = (k^2 + A^2)^{1/2} = (k-iA)^{1/2} (k+iA)^{1/2}\,.
\]
Note that $\lambda(k;A) \in \R$ exactly when $k \in \R \cup i[-A,A]$. We take the branch cut of $\lambda(k;A)$ to lie along $i[-A,A]$ oriented upward, and define $\lambda(k;A)$ to be continuous from the right. Explicitly, letting $k - iA = r_1\e^{i\varphi_1}$ and $k + iA = r_2\e^{i\varphi_2}$ with $-\pi/2 \leq \varphi_1, \varphi_2 < 3\pi/2$, we define
\be
	\label{e:explicit k-iA, k+iA definition}
	(k-iA)^{1/2} = \sqrt{r_1}\e^{i\varphi_1/2}\,, \quad (k+iA)^{1/2} = \sqrt{r_2}\e^{i\varphi_2/2}\,,
\ee
and
\be
\label{e:explicit lambda(k;A) definition}
	\lambda(k;A) = \sqrt{r_1 r_2}\e^{i(\varphi_1 + \varphi_2)/2}\,,
\ee
so that $\lambda(k;A) = k+O(1/k)$ as $k \rightarrow \infty$ in any direction (cf.\ Fig.~\ref{f:lambda}).
\begin{lemma}
\label{l:lambda symmetries}
The function $\lambda(k;A)$ defined by~\eqref{e:explicit lambda(k;A) definition}  
satisfies the following properties:
	\bse
		\begin{alignat}{2}
		\label{e:Im lambda}
		\Im \lambda(k;A) &\lessgtr 0\,, & &k \in \C^\mp \setminus i[-A,A]\,, 
		\\
		\label{Re lambda}
		\Re \lambda(k;A) &\lessgtr 0\,, & & k \in \R^\mp + i\R\,, 
		\\
		\label{lambda conjugation symmetry}
		 \lambda(\conj{k};A) &= \phantom{-}\overline{\lambda(k;A)}\,, & & k \in \C\,, 
		\\
		\label{lambda negation symmetry}
		 \lambda(-k;A) &= -\lambda(k;A)\,, & & k \in \C \setminus i[-A,A]\,, 
		\\
		\label{lambda jump symmetry}
		 \lambda^\mp(k;A) &= \pm\lambda(k;A)\,, \quad & & k \in i[-A,A]\,. 
\end{alignat}
\ese
\end{lemma}
Here and elsewhere,
$\R^\mp = \{k \in \R: \Re k \lessgtr 0\}$, $\C^\mp = \{k \in \C: \Im k \lessgtr 0\}$ and the superscripts $\mp$ on functions of $k$  denote the limit being taken from the right/left of the negative/positive side of the oriented contour respectively. In particular, for the upward oriented contour $i[-A, A]$, the superscripts $\mp$ denote the limits from the right/left, i.e.
\be
		\lambda^\mp(k;A) := \lim_{\epsilon \downarrow 0} \lambda(k \pm \epsilon;A), \quad k \in i[-A,A]\,.
\ee
With the above definitions,~\eqref{e:lambda_pm(k)} can be expressed as %
\be
	\lambda_\pm(k):=\lambda(k\pm V/2;A_\pm)\,,
\ee
Correspondingly, $\lambda_\pm(k) \in \mathbb{R}$ exactly for $k \in \R \cup \Sigma_\pm$, where 
\be 
	\Sigma_+ = [\conj{p_{+}}, p_{+}] = \Sigma_{+1} \cup \Sigma_{+2}\,,
	\quad  
	\Sigma_- = [\conj{p_{-}}, p_{-}] = \Sigma_{-1} \cup \Sigma_{-2}\,,
\ee
are the upward oriented branch cuts for $\lambda_+(k)$ and $\lambda_-(k)$ respectively, and
\bse
	\label{spectrum}
	\begin{align}
		\Sigma_{\pm 1} &= \Sigma_{\pm}\cap(\C^+ \cup \{\mp V/2\}) = \mp V/2 + i[0,A_\pm]\,,
		\\
		\Sigma_{\pm 2} &= \Sigma_{\pm}\cap(\C^- \cup \{\mp V/2\}) = \mp V/2 + i[-A_\pm,0]\,,
	\end{align}
\ese
with $p_{\pm}$ as defined in~\eqref{branch points} (cf.\ Fig.~\ref{f:spectrum}).
From Lemma~\ref{l:lambda symmetries}, we have
\bse
\label{left/right lambda}
\begin{align}
		\lambda_\pm^-(k) = \phantom{-}\lambda_\pm(k), \quad k \in \Sigma_\pm\,,
		\\
		\lambda_\pm^+(k) = -\lambda_\pm(k), \quad k \in \Sigma_\pm\,.
\end{align}
\ese
Hereafter we will suppress the $k$-dependence of $\lambda_\pm$ when doing so does not create ambiguity. 
For later convenience, we also define the set
\[
\label{e:Sigmadef}
\Sigma = \Real\cup\Sigma_+\cup\Sigma_-\,,
\]
which will comprise the continuous spectrum of the scattering problem (see Section~\ref{Direct problem: Scattering matrix, symmetries and discrete eigenvalues}).

\bb
Recall $D_\pm(k)$ as given in~\eqref{e:det E_pm}.
\eb
With the chosen branch cuts for $\lambda_\pm$, $D_\pm(k)$ are analytic for $k \in \C \setminus \Sigma_\pm$. 
When deriving the jump conditions in the Riemann-Hilbert problem, it will also be necessary to understand the discontinuities of $D_\pm(k)$ across the branch cuts $\Sigma_\pm$.
Explicitly, it is easy to show that
\[
\label{D symmetry}
	D_\pm^+(k) = \frac{4\lambda_\pm^2}{A_\pm^2} \frac{1}{D_\pm(k)}, \quad k \in \Sigma_\pm\,,
\]
where, due to our choice~\eqref{left/right lambda} for $\lambda_\pm$, the values of $D_\pm$ on the branch cuts coincide with their limits from the right, i.e. $D_\pm^-(k) = D_\pm(k)$ for all $k\in\Sigma_\pm$.
\bb
Another choice of branch cut is needed to uniquely define $d_\pm(k) = (D_\pm(k))^{1/2}$, whose discontinuities also must be understood.
Explicitly, we choose
\[\label{e:dpm-def}
d_\pm(k) := \sqrt{D_\pm(k)}\,, \quad k \in \C\,,
\]
where $\sqrt{\cdot}$ denotes the principal square root with branch cut along $\R^-\cup \{0\}$.
\eb
\bb
\begin{lemma}
\label{l:d_pm properties}
	The function $d_\pm(k)$ is analytic in $\C\setminus\Sigma_\pm$ and continuous from the right on $\Sigma_\pm$ with
	\bse
		\begin{alignat}{2}
			d_\pm(k) &= 1 + O(1/k)\,, \quad && k \rightarrow \infty\,,
			\label{e:d_pm asymptotics}
			\\
			\label{d symmetry}
			 d_\pm^+(k) &= \frac{2\lambda_\pm}{A_\pm}\frac{1}{d_\pm(k)}\,, && k \in \Sigma_\pm\,.
		\end{alignat}
	\ese
\end{lemma}
\eb
The limiting values of the Jost solutions on the branch cuts will be discussed later in Section~\ref{s:symmetries}.
%
\subsection{Jost solutions: Rigorous definition, analyticity and continuous spectrum}
\label{Rigorous definition of the Jost solutions, analyticity and continuous spectrum}

\begin{figure}[t!]
	\centering
\begin{subfigure}{.5\textwidth}
		\centering
		\includegraphics[width=0.8\textwidth]{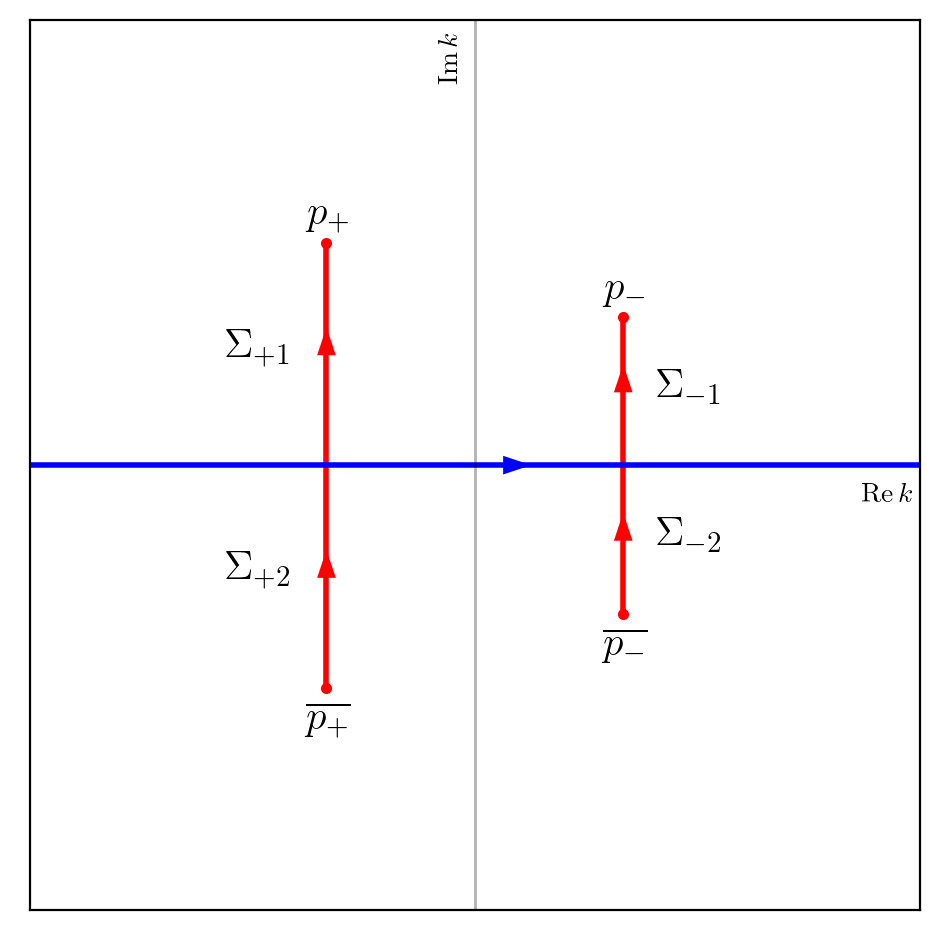}
\end{subfigure}%
\begin{subfigure}{.5\textwidth}
		\centering
		\includegraphics[width=0.8\textwidth]{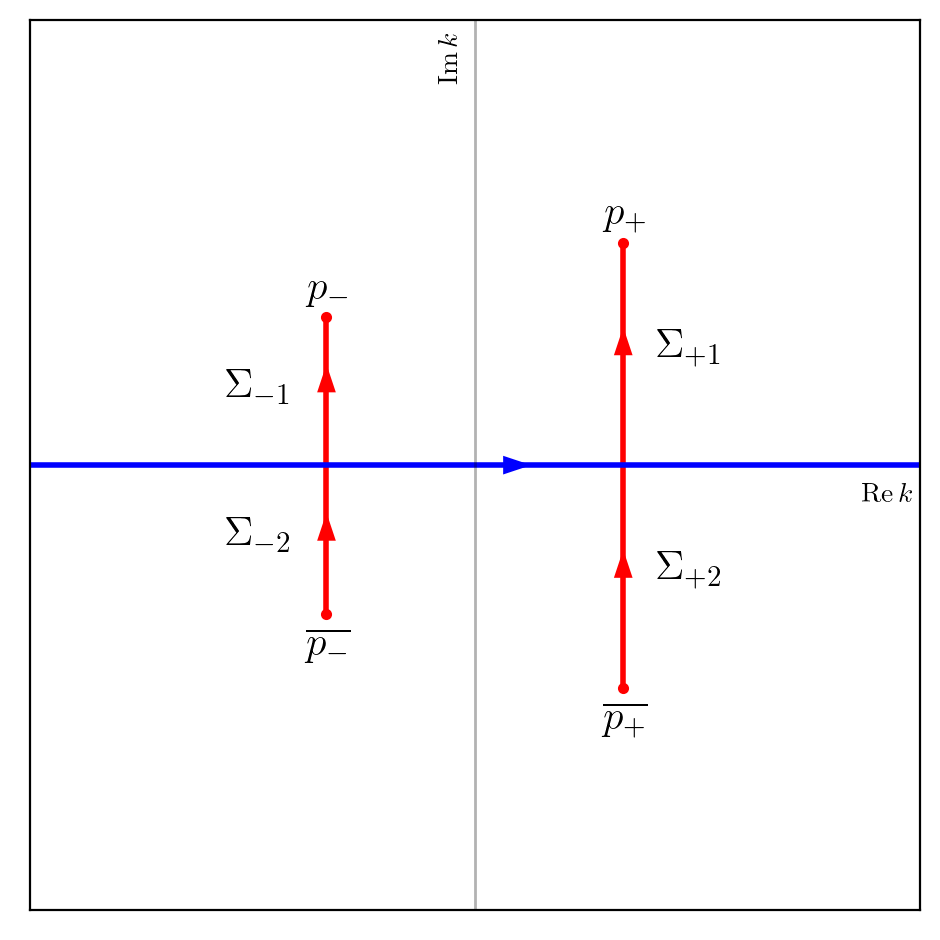}
\end{subfigure}%
	\caption{The contours $\Sigma_{\pm 1}$, $\Sigma_{\pm 2}$ for $V>0$ (left) and $V<0$ (right). Recall that $A_- < A_+$.  Along the real axis (blue line), all four Jost solutions are defined.  On the other hand, on $\Sigma_{\pm 1}$ and $\Sigma_{\pm 2}$ (red segments), only three of the Jost solutions are generically defined.}
\label{f:spectrum}
\end{figure}

We now introduce integral equations that can be used to rigorously define the Jost solutions and establish their regions of existence, continuity and analyticity.
\bb
We first remove the asymptotic oscillations that are present in~\eqref{e:phi x asymptotics} as well as the poles from the factor $d_\pm(k)$ by introducing the modified eigenfunctions
\be
	\mu_\pm(x,t,k) = d_\pm(k)\e^{if_\pm(x,t)\sigma_3}\phi_\pm(x,t,k)\e^{-i\theta_\pm(x,t,k)\sigma_3}\,.
\ee
\eb
The Lax pair~\eqref{e:LP} yields corresponding ODEs for the functions $\mu_\pm$. Noting that %
\begin{align}
	&X(x,t,k) = X_\pm(x,t,k) + \Delta Q_\pm(x,t)\,,
	\\
\noalign{\noindent with} 
	&\Delta Q_\pm(x,t) = Q(x,t) - Q_\pm(x,t)\,,
\end{align}
these ODEs can be formally integrated (see Section~\ref{Integral equations for mu_pm appendix}) to obtain the integral equations
\[
\label{mu integral equations}
	\mu_\pm(x,t,k) = E_\pm(k) + \int_{\pm\infty}^x E_\pm(k) \e^{i\lambda_\pm(x-y)\sigma_3}E_\pm^{-1}(k) \e^{2if_\pm(y,t)\sigma_3} \Delta Q_\pm(y,t)\mu_\pm(y,t,k)\e^{-i\lambda_\pm(x-y)\sigma_3}dy\,.
\]
We now let $\phi_{\pm 1}$ and $\phi_{\pm 2}$ denote the first and second columns of $\phi_\pm$ respectively. 
Using the left- and right-background solutions $q_\pm(x,t)$ defined in~\eqref{e:qpmdef} and the notation for $\Sigma_{\pm,1,2}$ introduced in~\eqref{spectrum} 
(cf.\ Fig.~\ref{f:spectrum}),
we then have the following:
\begin{theorem}
\label{Jost analyticity}
	If $(q - q_\pm) \in L_x^1(\R^\pm)$ for all $t\in \R$,  then
\begin{itemize}
		\item 
			$\phi_{+1}(x,t,k)$ is 
			analytic for $k \in \C^+\setminus\Sigma_{+1}$\,,
			continuous from above on $k\in\R$
			and from the right on $k\in\Sigma_{+1}^o$\,,
			and also defined for $k \in \Sigma_{+2}^o$\,. 
		\item 
			$\phi_{+2}(x,t,k)$ is 
			analytic for $k \in \C^-\setminus\Sigma_{+2}$\,,
			continuous from below on $k \in \R$
			and from the right on $k\in\Sigma_{+2}^o$\,,
			and also defined for $k \in \Sigma_{+1}^o$\,.
		\item 
			$\phi_{-1}(x,t,k)$ is 
			analytic for $k \in \C^-\setminus\Sigma_{-2}$\,, 
			continuous from below on $k\in\R$ 
			and from the right on $k \in \Sigma_{-2}^o$\,, 
			and also defined for $k \in \Sigma_{-1}^o$.
		\item 
			$\phi_{-2}(x,t,k)$ is 
			analytic for $k \in \C^+\setminus\Sigma_{-1}$\,,
			continuous from above on $k\in\R$
			and from the right on $k \in \Sigma_{-1}^o$\,,
			and also defined for $k \in \Sigma_{-2}^o$\,.
\end{itemize}
\end{theorem}
Above and throughout the rest of this work, 
\begin{equation}\label{e:Sigmaopmdef}
\Sigma_{+1}^o = \Sigma_{+1}\setminus\{p_+\,,-V/2\}, 
\quad
\Sigma_{+2}^o = \Sigma_{+2}\setminus\{\conj{p_{+}}\,,-V/2\}
\end{equation}
 and similarly for $\Sigma_{-1}^o$ and $\Sigma_{-2}^o$.
The hypothesis of Theorem~\ref{Jost analyticity} does not allow us to draw any conclusions about the Jost eigenfunctions at the branch points.
The \bb behavior \eb of the eigenfunctions at the branch points $p_{\pm}$ and $\conj{p_{\pm}}$
will be discussed in Section~\ref{s:branchpoints}.

The proof of Theorem~\ref{Jost analyticity} proceeds nearly identically as in \cite{BK2014} 
\bb
by analyzing the Neumann series associated with the Volterra integral equations~\eqref{mu integral equations},
\eb
and is included in Section~\ref{Jost analyticity appendix}. 
Moreover, the proof also implies 
the following:
\begin{corollary}
	\label{phi one-sided boundedness}
	Under the hypothesis of Theorem~\ref{Jost analyticity}, for any $a\in\R$,
	\bse
		\begin{alignat}{2}
			&\phi_{+1}(x,t,k)\in L_x^\infty(a,\infty)\,,
			&& k \in \R\cup\C^+\cup\Sigma_{+2}^o\setminus\{p_{+}\}\,,
			\\
			&\phi_{+2}(x,t,k)\in L_x^\infty(a,\infty)\,,
			&& k \in \R\cup\C^-\cup\Sigma_{+1}^o\setminus\{\conj{p_{+}}\}\,,
			\\
			&\phi_{-1}(x,t,k)\in L_x^\infty(-\infty,a)\,,
			&& k \in \R\cup\C^-\cup\Sigma_{-1}^o\setminus\{\conj{p_{-}}\}\,,  
			\\
			&\phi_{-2}(x,t,k)\in L_x^\infty(-\infty,a)\,,
			\quad && k \in \R\cup\C^+\cup\Sigma_{-2}^o\setminus\{p_{-}\}\,.
		\end{alignat}
	\ese
\end{corollary}

\bb
Theorem~\ref{Jost analyticity} shows that $\phi_\pm(x,t,k)$ are continuous for $k\in\R$. 
Moreover, formally differentiating the Volterra integral equation~\eqref{mu integral equations} with respect to $k$ and performing
a similar Neumann series analysis, one can show the following:
\begin{corollary}
	\label{c:phi differentiable}
	Under the hypothesis of Theorem~\ref{Jost analyticity}, $\phi_+(x,t,k)$ and $\phi_-(x,t,k)$ are $C^1(\R)$, i.e. continuously real-differentiable functions of $k$.
\end{corollary}
\eb

\subsection{Jost solutions: Asymptotic behavior as $k \rightarrow \infty$}

Understanding the asymptotic behavior of $\phi_\pm(x,t,k)$ as $k\to\infty$ is necessary in order to properly formulate the inverse problem,
and will also allows us to recover the potential $q $ from the scattering data.
\begin{lemma}
\label{mu asymptotics}
	If $(q-q_\pm) \in L_x^1(\R^\pm)$ and   $q$ is continuously differentiable  with $(q-q_\pm)_x \in L_x^1(\R^\pm)$  for all $t\in \R$, then
	\be
		\mu_\pm(x,t,k) =\I+ O(1/k)\,, \quad k \rightarrow \infty\,,
	\ee
	within the appropriate region of the complex $k$-plane for each column as outlined in Theorem~\ref{Jost analyticity}. Furthermore, 
	\be
		\label{q from mu}
		q(x,t) = -2i\lim_{k\rightarrow \infty} \e^{-2if_-(x,t)}k \big[\mu_-(x,t,k)\big]_{12}\,.
	\ee
\end{lemma}
As a direct consequence of the above lemma, %
\[
	\phi_\pm(x,t,k) = \e^{i(\theta_\pm(x,t,k) - f_\pm(x,t))\sigma_3}(\I + O(1/k))\,, \quad k \rightarrow \infty\,,
\]
within the appropriate regions of the complex $k$-plane for each column.
Observe that %
\[
	\lambda_\pm(k) = (k \pm V/2) + \frac{A_\pm^2}{2(k \pm V/2)} + O(1/k^3)\,, \quad k \rightarrow \infty\,,
\]
and so
\be\label{e:theta-pm-asy}
\begin{aligned}
		\theta_\pm(x,t,k) &= (k\pm V/2)x - \left(2k^2 - V^2/2 + A_\pm^2\right)t + O(1/k)\,, & &k \rightarrow \infty
											\\
											&= \theta_o(x,t,k) + f_\pm(x,t) + O(1/k)\,, & &k \rightarrow \infty\,,
\end{aligned}
\ee
where we have introduced the controlling phase function for the Jost eigenfunctions 
in the problem with zero boundary conditions:
\[
\label{theta_o}
	\theta_o(x,t,k) = k (x-2kt)\,,
\]
which will also be used in Section~\ref{Inverse problem: RHP formulation}.
Lemma~\ref{mu asymptotics} together with~\eqref{e:theta-pm-asy} imply the following:
\begin{lemma}
\label{Jost asymptotics}
	Under the hypotheses of Lemma~\ref{mu asymptotics},
	\[
		\phi_\pm(x,t,k) = \e^{i\theta_o(x,t,k)\sigma_3}(\I + O(1/k))\,, \quad k \rightarrow \infty\,, 
	\]
	within the appropriate region of the complex $k$-plane for each column as specified by Theorem~\ref{Jost analyticity}. 
	Furthermore,
	\be
		q(x,t) = -2i\lim_{k\rightarrow \infty} k \big[\phi_-(x,t,k)\e^{-i\theta_o(x,t,k)\sigma_3}\big]_{12}\,.
	\ee
\end{lemma}
\bb
Note that the presence of $d_\pm(k)$ in the definition of $\phi_\pm(x,t,k)$ does not change the asymptotic $k$ behavior at infinity due to~\eqref{e:d_pm asymptotics}.
\eb

\subsection{Jost solutions: Behavior at the branch points}
\label{s:branchpoints}
%
As mentioned in Section~\ref{Rigorous definition of the Jost solutions, analyticity and continuous spectrum}, the condition $(q-q_\pm)\in L_x^1(\R^\pm)$ for all $t\in \R$ is enough to guarantee the existence and analyticity of the Jost eigenfunctions in suitable open regions of the complex $k$-plane, as well as their continuity along portions of 
the boundary of these regions.
Notably, however, these regions do not include the branch points $p_{\pm}$ and $\conj{p_{\pm}}$.
On the other hand, the behavior of the eigenfunctions near the branch points must be understood in order to specify appropriate growth conditions for the inverse problem. 
\bb
We next show that, under more strict conditions for the potential than those imposed by Theorem~\ref{Jost analyticity}, it is possible to define the modified eigenfunctions $\mu_\pm(x,t,k)$ at the branch points. This in turn determines the behavior of the Jost solutions $\phi_\pm(x,t,k)$ near the branch points.
\eb
To do so, we introduce the weighted $L^1$ spaces
\begin{equation}
L^{1, j}(\mathbb R^\pm) := \left\{f: \mathbb R \to \mathbb C \, \big| \left(1+|x|\right)^j f \in L^1(\mathbb R^\pm)\right\}, \quad j=1,2.
\end{equation}
\bb
\begin{lemma}
\label{l:mu branch point continuity}
		If $(q-q_\pm) \in L_x^{1,1}(\R^\pm)$  for all $t\in \R$, then the modified eigenfunctions $\mu_\pm(x,t,k)$ are continuous at the branch points $p_\pm\,,\conj{p_\pm}$. Specifically,
	\bse
		\begin{align}
			&\mu_{+1}(x,t,k) = \beta_{p_{+}}^{(0)}(x,t) + o(1), \quad k \rightarrow p_{+}\,,
			\\
			&\mu_{+2}(x,t,k) = \beta_{\conj{p_{+}}}^{(0)}(x,t) + o(1), \quad k \rightarrow \conj{p_{+}}\,,
			\\
			&\mu_{-1}(x,t,k) = \beta_{\conj{p_{-}}}^{(0)}(x,t) + o(1), \quad k \rightarrow \conj{p_{-}}\,,
			\\
			&\mu_{-2}(x,t,k) = \beta_{p_{-}}^{(0)}(x,t) + o(1), \quad k \rightarrow p_{-}\,,
		\end{align}
	\ese
	for some vectors $\beta_{p_\pm}^{(0)}(x,t)$, $\beta_{\conj{p_\pm}}^{(0)}(x,t)$. Moreover, $\beta_{p_\pm}^{(0)}(x,t)$ and $\beta_{\conj{p_\pm}}^{(0)}(x,t)$ are never zero.
\end{lemma}

\begin{lemma}
	\label{l:mu branch point expansion}
	If $(q-q_\pm) \in L_x^{1,2}(\R^\pm)$  for all $t\in \R$, then
	\begin{align}
		&\mu_{+1}(x,t,k) = \beta_{p_{+}}^{(0)}(x,t) + \beta_{p_{+}}^{(1)}(x,t)(k-p_{+})^{1/2} + o(k-p_{+})^{1/2}, \quad k \rightarrow p_{+}\,,
		\\
		&\mu_{+2}(x,t,k) = \beta_{\conj{p_{+}}}^{(0)}(x,t) + \beta_{\conj{p_{+}}}^{(1)}(x,t)(k-\conj{p_{+}})^{1/2} + o(k-\conj{p_{+}})^{1/2}, \quad k \rightarrow \conj{p_{+}}\,,
		\\ 
		&\mu_{-1}(x,t,k) = \beta_{\conj{p_{-}}}^{(0)}(x,t) + \beta_{\conj{p_{-}}}^{(1)}(x,t)(k-\conj{p_{-}})^{1/2} + o(k-\conj{p_{-}})^{1/2}, \quad k \rightarrow \conj{p_{-}}\,,
		\\
		&\mu_{-2}(x,t,k) = \beta_{p_{-}}^{(0)}(x,t) + \beta_{p_{-}}^{(1)}(x,t)(k-p_{-})^{1/2} + o(k-p_{-})^{1/2}, \quad k \rightarrow p_{-}\,,
	\end{align}
	for some vectors $\beta_{p_\pm}^{(1)}(x,t)\,, \beta_{\conj{p_\pm}}^{(1)}(x,t)$ with $\beta_{p_\pm}^{(0)}(x,t)\,, \beta_{\conj{p_\pm}}^{(0)}(x,t)$ as in Lemma~\ref{l:mu branch point continuity}.
\end{lemma}
Higher order expansions in half-integer powers can be found similarly by placing further restrictions on the potential. In order to use the above expansions to describe the behavior of the Jost solutions $\phi_\pm(x,t,k)$ around the branch points, we now clarify the behavior of $d_\pm(k)$.

\begin{lemma}
\label{l:d_pm branch point behavior}
	The asymptotic behavior of $d_\pm(k)$ at the branch points is given by
	\bse
		\begin{alignat}{2}
			 d_\pm(k) &= \Big(\frac{8}{iA_\pm}\Big)^{1/4}(k-p_\pm)^{1/4} + o(1)\,, \quad && k \rightarrow p_\pm\,,
			\\
			 d_\pm(k) &= \Big(\frac{-8}{iA_\pm}\Big)^{1/4}(k-\conj{p_\pm})^{1/4} + o(1)\,, \quad && k \rightarrow \conj{p_\pm}\,.
		\end{alignat}
	\ese
\end{lemma}
The specific branch cuts for the fourth-roots appearing in the above are of little interest, since we will mainly be concerned with the rate of growth of the Jost solutions near the branch points (see Section~\ref{Growth conditions}). Nonetheless, we clarify that
\[
	\label{e:fourth roots}
	(k-p_\pm)^{1/4} := \sqrt{(k-p_\pm)^{1/2}}\,, \qquad (k-\conj{p_\pm})^{1/4} := \sqrt{(k-\conj{p_\pm})^{1/2}}\,, 
\]
where the branch cuts for $(k-p_{\pm})^{1/2}$ and $(k-\conj{p_{\pm}})^{1/2}$ are taken analogously with the definition of $\lambda(k;A)$ in~\eqref{e:lambda(k;A)} so that 
\be
	\lambda_\pm(k) = (k-p_{\pm})^{1/2}(k-\conj{p_{\pm}})^{1/2}\,,
\ee
and $\sqrt{\cdot}$ is the same square root in~\eqref{e:dpm-def}.

We then have the following branch point behavior for the Jost solutions:
\begin{corollary}
\label{phi branch point continuity}
	Under the hypothesis of Lemma~\ref{l:mu branch point continuity},
	\bse
		\begin{align}
			&\phi_{+1}(x,t,k) = b_{p_{+}}^{(0)}(x,t)(k-p_+)^{-1/4} + o(k-p_+)^{-1/4}, \quad k \rightarrow p_{+}\,,
			\\
			&\phi_{+2}(x,t,k) = b_{\conj{p_{+}}}^{(0)}(x,t)(k-\conj{p_+})^{-1/4} + o(k-\conj{p_+})^{-1/4}, \quad k \rightarrow \conj{p_{+}}\,,
			\\
			&\phi_{-1}(x,t,k) = b_{\conj{p_{-}}}^{(0)}(x,t)(k-\conj{p_-})^{-1/4} + o(k-\conj{p_-})^{-1/4}, \quad k \rightarrow \conj{p_{-}}\,,
			\\
			&\phi_{-2}(x,t,k) = b_{p_{-}}^{(0)}(x,t)(k-p_-)^{-1/4} + o(k-p_-)^{-1/4}, \quad k \rightarrow p_{-}\,,
		\end{align}
	\ese
	for some vectors $b_{p_\pm}^{(0)}(x,t)$, $b_{\conj{p_\pm}}^{(0)}(x,t)$. Moreover, $b_{p_\pm}^{(0)}(x,t)$ and $b_{\conj{p_\pm}}^{(0)}(x,t)$ are never zero.
\end{corollary}
\begin{corollary}
\label{phi branch point expansion}
  Under the hypothesis of Lemma~\ref{l:mu branch point expansion},
	\bse
		\begin{align}
			\phi_{+1}(x,t,k) &= b_{p_{+}}^{(0)}(x,t)(k-p_+)^{-1/4} + b_{p_{+}}^{(1)}(x,t)(k-p_{+})^{1/4} + o(k-p_{+})^{1/4}\,, \quad k \rightarrow p_{+}\,,
			\\
			\phi_{+2}(x,t,k) &= b_{\conj{p_{+}}}^{(0)}(x,t)(k-\conj{p_+})^{-1/4} + b_{\conj{p_{+}}}^{(1)}(x,t)(k-\conj{p_{+}})^{1/4} + o(k-\conj{p_{+}})^{1/4}\,, \quad k \rightarrow \conj{p_{+}}\,,
			\\
			\phi_{-1}(x,t,k) &= b_{\conj{p_{-}}}^{(0)}(x,t)(k-\conj{p_-})^{-1/4} + b_{\conj{p_{-}}}^{(1)}(x,t)(k-\conj{p_{-}})^{1/4} + o(k-\conj{p_{-}})^{1/4}\,, \quad k \rightarrow \conj{p_{-}}\,,
			\\
			\phi_{-2}(x,t,k) &= b_{p_{-}}^{(0)}(x,t)(k-p_-)^{-1/4} + b_{p_{-}}^{(1)}(x,t)(k-p_{-})^{1/4} + o(k-p_{-})^{1/4}\,, \quad k \rightarrow p_{-}\,,
		\end{align}
	\ese
for some vectors $b_{p_\pm}^{(1)}(x,t)$, $b_{\conj{p_\pm}}^{(1)}(x,t)$ with $b_{p_\pm}^{(0)}(x,t)$, $b_{\conj{p_\pm}}^{(0)}(x,t)$  as in Corollary~\ref{phi branch point continuity}.
\end{corollary}
\eb

\section{Direct problem: Scattering matrix, symmetries and discrete eigenvalues}
\label{Direct problem: Scattering matrix, symmetries and discrete eigenvalues}

The scattering data are constructed by studying the relations between the two sets of  Jost solutions $\phi_+$ and $\phi_-$.
Proofs for all the results in this section are given in Section~\ref{a:Direct Problem}.

\subsection{Scattering matrix}
\label{ss:Scattering matrix}

For $k \in \R$, both $\phi_+(x,t,k)$ and $\phi_-(x,t,k)$ are  fundamental matrix solutions of both parts of the Lax pair~\eqref{e:LP}. 
Thus, there exists a matrix
\be
\label{scattering matrix}
	S(k) = \begin{pmatrix} 
					 s_{11}(k) & s_{12}(k)\\
					 s_{21}(k) & s_{22}(k)
				 \end{pmatrix},
\ee
independent of $x$ and $t$, such that
\be
\label{scattering matrix relation}
	\phi_-(x,t,k) = \phi_+(x,t,k)S(k)\,,\quad k\in\R\,.
\ee
The matrix $S(k)$ is known as the scattering matrix 
\bb
(from the right)
\eb
 and its entries are known as the scattering coefficients. 
\bb
Note that $\det S(k) = 1$. Writing~\eqref{scattering matrix relation} column-wise, we have
\eb
\bse 
\label{scattering relation}
\begin{align}
		\label{scattering relation 1}
		\phi_{-1}(x,t,k) = s_{11}(k)\phi_{+1}(x,t,k) + s_{21}(k)\phi_{+2}(x,t,k), \quad k \in \R\,,
		\\
		\label{scattering relation 2}
		\phi_{-2}(x,t,k) = s_{12}(k)\phi_{+1}(x,t,k) + s_{22}(k)\phi_{+2}(x,t,k), \quad k \in \R\,.
\end{align}
\ese
Theorem~\ref{Jost analyticity} together with relations~\eqref{scattering relation} gives the following Wronskian representations for the scattering coefficients:
\begin{corollary}
\label{Wronskians}
	Under the hypothesis of Theorem~\ref{Jost analyticity}, the scattering coefficients admit the following Wronskian representations:
	\bse
\label{e:Wronskians}
\bb
	\begin{align}
			\label{s_11}
			s_{11}(k) &= \Wr[\phi_{-1}(x,t,k),\phi_{+2}(x,t,k)]\,, \quad k \in \R\cup\C^-\setminus\{\conj{p_{+}}, \conj{p_{-}}\}\,,
			\\
			\label{s_12}
			s_{12}(k) &= \Wr[\phi_{-2}(x,t,k),\phi_{+2}(x,t,k)]\,, \quad k \in \R\cup\Sigma_{+1}^o\cup\Sigma_{-2}^o\,,
			\\
			\label{s_21}
			s_{21}(k) &= \Wr[\phi_{+1}(x,t,k),\phi_{-1}(x,t,k)]\,, \quad k \in \R\cup\Sigma_{-1}^o\cup\Sigma_{+2}^o\,,
			\\
			\label{s_22}
			s_{22}(k) &= \Wr[\phi_{+1}(x,t,k),\phi_{-2}(x,t,k)]\,, \quad k \in \R\cup\C^+\setminus\{p_{+}, p_{-}\}\,,
		\end{align}
	\eb
	\ese
 with $\Sigma_{\pm 1}^o$ and $\Sigma_{\pm 2}^o$  defined by ~\eqref{e:Sigmaopmdef}. Moreover, $s_{22}(k)$ and $s_{11}(k)$ are analytic in $\C^+ \setminus \Sigma$ and $\C^- \setminus \Sigma$ respectively.
\end{corollary}

Note that the Wronskian representations~\eqref{e:Wronskians} are first defined for $k\in\Real$, where both of the relations~\eqref{scattering relation} hold.
Each of them can then be extended off the real $k$-axis to define the corresponding scattering coefficient wherever the right-hand side of each of representations~\eqref{e:Wronskians} is defined.

In the special case of no counterflows, i.e. $V=0$, the scattering relations~\eqref{scattering relation} and Wronskian representations~\eqref{e:Wronskians} can be further extended (see Section~\ref{Reductions}).

\bb
The Wronskian representations together with Corollary~\ref{c:phi differentiable} and Lemma~\ref{Jost asymptotics} give the following:
\begin{corollary}
Under the hypotheses of Theorem~\ref{Jost analyticity},
the scattering matrix $S(k)$ is $C^1(\Real)$. 
\label{c:Sdifferentiable}
\end{corollary}
\eb

\begin{corollary}
\label{scattering coefficient asymptotics}
	Under the hypotheses of Lemma~\ref{mu asymptotics},
	\bse
		\label{e:scattering coefficient asymptotics}
		\begin{alignat}{2}
			s_{11}(k) &= 1 + O(1/k)\,, \quad &&k \rightarrow \infty\,,
			\\
			s_{12}(k) &= \phantom{--}O(1/k)\,,  && k \rightarrow \infty\,,
			\\
			s_{21}(k) &= \phantom{--}O(1/k)\,,  && k \rightarrow \infty\,,
			\\
			\label{s_22 asymptotics}
			s_{22}(k) &= 1 + O(1/k)\,, && k \rightarrow \infty\,, 
		\end{alignat}
	\ese
	within the appropriate regions of the complex $k$-plane for each column as stated in Corollary~\ref{Wronskians}.
\end{corollary}
%

%
Before the introduction of the scattering matrix, all calculations were symmetric upon exchanging limits as $x\to-\infty$ and as $x\to\infty$, thanks to the symmetry of the NLS equation under space reflections.  
The relation~\eqref{scattering matrix relation}, however, breaks this symmetry.
On the other hand, we can similarly write
\[
\label{R scattering matrix relation}
	\phi_+(x,t,k) = \phi_-(x,t,k)R(k)\,, \quad k \in \R\,,
\]
or, in column form,%
\bse 
\label{R scattering relation}
\begin{gather}
		\label{R scattering relation 1}
		\phi_{+1}(x,t,k) = r_{11}(k)\phi_{-1}(x,t,k) + r_{21}(k)\phi_{-2}(x,t,k)\,, \quad k \in \R\,,
		\\
		\label{R scattering relation 2}
		\phi_{+2}(x,t,k) = r_{12}(k)\phi_{-1}(x,t,k) + r_{22}(k)\phi_{-2}(x,t,k)\,, \quad k \in \R\,,
\end{gather}
\ese
for some scattering matrix 
\bb
(from the left)
\eb
\be
\label{R scattering matrix}
	R(k) = \begin{pmatrix} 
					 r_{11}(k) & r_{12}(k)\\
					 r_{21}(k) & r_{22}(k)
				 \end{pmatrix}.
\ee
The two scattering matrices $S$ and $R$ are related simply by
\[
	R(k) = S^{-1}(k)\,, \quad k \in \R\,.
\]
Moreover, Wronskian representations exist similar to those in Corollary~\ref{Wronskians}:
\begin{corollary}
\label{R Wronskians}
	Under the hypothesis of Theorem~\ref{Jost analyticity}, the left scattering coefficients can be extended through the Wrosnkian representations,
	\bse
		\label{e:R Wronskians}
		\bb
		\begin{align}
			\label{r_11}
			r_{11}(k) &= \Wr[\phi_{+1}(x,t,k),\phi_{-2}(x,t,k)]\,, \quad k \in \R\cup\C^+\setminus\{p_{+}, p_{-}\}\,,
			\\
			\label{r_12}
			r_{12}(k) &= \Wr[\phi_{+2}(x,t,k),\phi_{-2}(x,t,k)]\,, \quad k \in \R\cup\Sigma_{+1}^o\cup\Sigma_{-2}^o\,,
			\\
			\label{r_21}
			r_{21}(k) &= \Wr[\phi_{-1}(x,t,k),\phi_{+1}(x,t,k)]\,, \quad k \in \R\cup\Sigma_{-1}^o\cup\Sigma_{+2}^o\,,
			\\
			\label{r_22}
			r_{22}(k) &= \Wr[\phi_{-1}(x,t,k),\phi_{+2}(x,t,k)]\,, \quad k \in \R\cup\C^-\setminus\{\conj{p_{+}}, \conj{p_{-}}\}\,.
		\end{align}
		\eb
	\ese
Moreover, $r_{11}(k)$ and $r_{22}(k)$ are analytic in $\C^+\setminus\Sigma$ and $\C^-\setminus\Sigma$ respectively.
\end{corollary}
From Corollaries~\ref{Wronskians} and~\ref{R Wronskians}, we see that
\be
\label{R S relations}
\bb
\begin{aligned}
	r_{11}(k) &= \phantom{-}s_{22}(k)\,, \quad &
	r_{12}(k) &= -s_{12}(k)\,,\\
	r_{21}(k) &= -s_{21}(k)\,,&
	r_{22}(k) &= \phantom{-}s_{11}(k)\,,
\end{aligned}
\eb
\ee
wherever the expressions are defined.

\noindent
For later use, we also define the reflection coefficients as
\bse
\label{reflection coefficients}
\begin{alignat}{2}
	\label{rho definition}
	\rho(k) &= s_{12}(k)/s_{22}(k)\,,  &&k \in \R \cup \Sigma_{+1}^o\,, 
	\\
	\label{r definition}
	r(k) &= 1/s_{22}(k)r_{21}(k)\,, \quad &&k \in \R \cup \Sigma_{-1}^o\,. 
\end{alignat}
\ese
More precisely, $\rho(k)$ and $r(k)$ will appear in the jump matrices that define the Riemann-Hilbert problem in Section~\ref{Inverse problem: RHP formulation}.
From Corollary~\ref{scattering coefficient asymptotics}, we see that
\[
\label{rho asymptotics}
	\rho(k) = O(1/k)\,, \quad k \rightarrow \pm\infty\,.\]
One can also show that, generically, $r(k) = O(k)$ as $k\to\pm\infty$. 
This does not pose a problem, however, since $r(k)$ only appears in the jumps across the finite segment $\Sigma_-^o$.

As with the scattering matrix from the right, in the special case of no counterflows, i.e. $V=0$, the scattering relations~\eqref{R scattering relation}, Wronskian representations~\eqref{e:R Wronskians} and domains for $\rho(k)$ and $r(k)$ can be further extended (see Section~\ref{Reductions}).

\bb
Corollary~\ref{c:Sdifferentiable} immediately gives the following:
\begin{corollary}
\label{c:rho differentiable}
Under the hypothesis of Theorem~\ref{Jost analyticity}, the reflection coefficient $\rho(k)$ is $C^1(\R \setminus K)$, where $K$ is the set of zeros of $s_{22}(k)$.
\end{corollary}

In preparation for the formulation of the inverse problem, it is convenient to introduce the following matrix:
\[
	\label{e:Phi}
	\Phi(x,t,k)
  = \begin{cases}
		\Big(\phi_{+1}(x,t,k)\,, \dfrac{\phi_{-2}(x,t,k)}{s_{22}(k)}\Big)\,, \quad k \in \C^+ \setminus \Sigma\,,
		\\[15pt]
		\Big(\dfrac{\phi_{-1}(x,t,k)}{s_{11}(k)}\,, \phi_{+2}(x,t,k)\Big)\,, \quad k \in \C^- \setminus \Sigma\,.
\end{cases}
\]
Note that $\Phi(x,t,k)$ is a simultaneous fundamental matrix solution to the Lax pair~\eqref{e:LP} that is meromorphic for $k\in\C\setminus\Sigma$, with $\det \Phi(x,t,k) = 1$.  
\eb

\subsection{Continuous spectrum}

The spectrum of the scattering problem is defined as the set of all $k\in\C$ for which there exist solutions to the Lax pair~\eqref{e:LP} bounded for all $x\in\Real$. 
As usual, the spectrum consists of a continuum of eigenvalues, \bb$\Sigma^o$\eb, which we refer to as the continuous spectrum, 
together with a discrete set $K\cup\conj{K}$ of eigenvalues (where $\conj{K}$ is the image of $K$ under complex conjugation)  which we refer to as the discrete spectrum (discussed later).
In the case of zero boundary conditions for the potential or of symmetric boundary conditions with zero velocity (i.e., $\lim_{x\to\pm\infty} q(x,t) = 0$ or $\lim_{x\to\pm\infty} q(x,t) = \~q_\pm$ with $|\~q_+|= |\~q_-|$, respectively),
the set where both columns of $\phi_-(x,t,k)$ are defined coincides with that where both columns of $\phi_+(x,t,k)$ are, and this set comprises the continuous spectrum of the scattering problem.
For example, in the case of symmetric boundary conditions with zero velocity the continuous spectrum is the set $\Real\cup(-iA,iA)$, with $A=|\~q_\pm|$.
As discussed in Section~\ref{Rigorous definition of the Jost solutions, analyticity and continuous spectrum}, 
however, this is not the case here.
Specifically, $\phi_\pm(x,t,k)$ can be defined simultaneously only for $k\in\Real$, and indeed this is the only set where the full scattering relations~\eqref{scattering matrix relation} and 
\eqref{R scattering matrix relation}
hold.

Nonetheless, it is still possible to partially extend half of~\eqref{scattering matrix relation} and 
\eqref{R scattering matrix relation} along appropriate segments of \bb$\Sigma_\pm^o$\eb. 
Specifically, taking into account the regions of definition and analyticity of the Jost solutions, 
for all $k\in\Sigma_+^o$ (where $\lambda_+(k)\in\Real$) one can express the analytic column of $\phi_-(x,t,k)$ as a linear combination of the columns of $\phi_+(x,t,k)$,
and viceversa on $\Sigma_{-}^o$.
Specifically:
\begin{corollary}
\label{c:extendedscattering}
Under the hypothesis of Theorem~\ref{Jost analyticity}, the scattering relations~\eqref{scattering relation 1},~\eqref{scattering relation 2},~\eqref{R scattering relation 1} and~\eqref{R scattering relation 2} can be extended to $k \in \Sigma_{+2}^o$, $\Sigma_{+1}^o$, $\Sigma_{-1}^o$ and $\Sigma_{-2}^o$ respectively.  That is:
	\bse
    \label{e:extended scattering relations}
	\begin{align}
			\phi_{-1}(x,t,k) &= s_{11}(k)\phi_{+1}(x,t,k) + s_{21}(k)\phi_{+2}(x,t,k)\,, \quad k \in \R \cup \Sigma_{+2}^o\,,
			\\
			\phi_{-2}(x,t,k) &= s_{12}(k)\phi_{+1}(x,t,k) + s_{22}(k)\phi_{+2}(x,t,k)\,, \quad k \in \R \cup \Sigma_{+1}^o\,,
\\
			\label{linear combination 2}
			\phi_{+1}(x,t,k) &= r_{11}(k)\phi_{-1}(x,t,k) + r_{21}(k)\phi_{-2}(x,t,k)\,, \quad k \in \R \cup \Sigma_{-1}^o\,,
			\\
			\label{linear combination 4}
			\phi_{+2}(x,t,k) &= r_{12}(k)\phi_{-1}(x,t,k) + r_{22}(k)\phi_{-2}(x,t,k)\,, \quad k \in \R \cup \Sigma_{-2}^o\,.
        \end{align}
	\ese
\end{corollary}
Note that the coefficients in the right-hand side of~\eqref{e:extended scattering relations} were labeled consistently with the Wronskian representations in~\eqref{e:Wronskians}.
Together with Corollary~\ref{phi one-sided boundedness}, 
the expressions in Corollary~\ref{c:extendedscattering} allow us to conclude that the corresponding eigenfunctions are bounded 
over all $x\in\Real$:
\begin{corollary}
\label{phi boundedness}
Under the hypothesis of Theorem~\ref{Jost analyticity}, for all $t\in \R$ we have 	\bse
		\begin{align}
				&\phi_{+1}(x,t,k)\in L_x^\infty(\R)\,, \quad k \in \R\cup\Sigma_{-1}^o\,,
				\\
				&\phi_{+2}(x,t,k)\in L_x^\infty(\R)\,, \quad k \in \R\cup\Sigma_{-2}^o\,,
				\\
				&\phi_{-1}(x,t,k)\in L_x^\infty(\R)\,, \quad k \in \R\cup\Sigma_{+2}^o\,,
				\\
				&\phi_{-2}(x,t,k)\in L_x^\infty(\R)\,, \quad k \in \R\cup\Sigma_{+1}^o\,.
		\end{align}
	\ese
\end{corollary}
In turn, defining 
\[
\label{e:Sigmaodef}
\Sigma^o := \Real\cup\Sigma_+^o\cup\Sigma_-^o\,,
\] %
Corollary~\ref{phi boundedness} implies:
\begin{corollary}
	\label{c:continuous spectrum}
	Under the hypothesis of Theorem~\ref{Jost analyticity} and for $V\not=0$, the continuous spectrum is given by $\Sigma^o$.
\end{corollary}
In the special case of no counterflows, i.e. $V=0$, the relations~\eqref{e:extended scattering relations} can be further extended so that the regions of boundedness over $x\in\R$ in Corollary~\ref{phi boundedness} are correspondingly extended. The continuous spectrum also requires further consideration in this case (see Section~\ref{Reductions}).

\subsection{Symmetries}
\label{s:symmetries}
%
The symmetries
\be
	\label{X symmetry}
	\conj{X(x,t,\conj{k})} = -\sigma_\ast X(x,t,k) \sigma_\ast\,, \quad
	\conj{T(x,t,\conj{k})} = -\sigma_\ast T(x,t,k) \sigma_\ast\,, \quad k\in\C\,,\ee
where
\be
	\label{sigma_ast}
	\sigma_\ast = \begin{pmatrix}
									0  & 1\\
									-1 & 0
								\end{pmatrix},
\ee
lead to the following symmetry relations:
\begin{lemma}[First symmetry, Jost solutions]
\label{first symmetry Jost} 
	Under the hypothesis of Theorem~\ref{Jost analyticity}, we have the symmetries
	\bse
		\label{phi conjugates}
		\begin{align}
			\label{phi_+1 conjugate}
			\conj{\phi_{+1}(x,t,\conj{k})} &= \phantom{-}\sigma_* \phi_{+2}(x,t,k)\,, \quad k \in \R\cup\C^-\cup\Sigma_{+1}^o\setminus\{\conj{p_{+}}\}\,,
			\\
			\label{phi_+2 conjugate}
			\conj{\phi_{+2}(x,t,\conj{k})} &= -\sigma_* \phi_{+1}(x,t,k)\,, \quad k \in \R\cup\C^+\cup\Sigma_{+2}^o\setminus\{p_{+}\}\,,
			\\
			\label{phi_-1 conjugate}
			\conj{\phi_{-1}(x,t,\conj{k})} &= \phantom{-}\sigma_* \phi_{-2}(x,t,k)\,, \quad k \in \R\cup\C^+\cup\Sigma_{-2}^o\setminus\{p_{-}\}\,,
			\\
			\label{phi_-2 conjugate}
			\conj{\phi_{-2}(x,t,\conj{k})} &= -\sigma_* \phi_{-1}(x,t,k)\,, \quad k \in \R\cup\C^-\cup\Sigma_{-1}^o\setminus\{\conj{p_{-}}\}\,.
		\end{align}
	\ese
\end{lemma}
\begin{lemma}[First symmetry, scattering coefficients]
\label{first symmetry extended} 
	Under the hypothesis of Theorem~\ref{Jost analyticity}, we have the symmetries
	\bse
		\label{e:first symmetry}
		\begin{align}
			\label{first symmetry a extended}
			\conj{s_{22}(\conj{k})} &= \phantom{-}s_{11}(k)\,, \quad k \in \R\cup \C^-\setminus\{\conj{p_{+}}\,, \conj{p_{-}}\}\,,
			\\
			\label{first symmetry b extended}
			\conj{s_{12}(\conj{k})} &= -s_{21}(k)\,, \quad k \in \R\cup\Sigma_{+2}^o\cup\Sigma_{-1}^o\,,
			\\
			\label{first symmetry Sigma_{-2}}
			\conj{r_{11}(\conj{k})} &= \phantom{-}r_{22}(k)\,, \quad k \in \R\cup\C^-\setminus\{\conj{p_{+}}\,, \conj{p_{-}}\}\,, 
			\\
			\label{first symmetry Sigma_{-1}}
			\conj{r_{21}(\conj{k})} &= -r_{12}(k)\,, \quad k \in  \R \cup \Sigma_{+2}^o \cup \Sigma_{-1}^o\,.
		\end{align}
	\ese
\end{lemma}
\bb
Lemmas~\ref{first symmetry Jost} and~\ref{first symmetry extended} then give the following symmetry for $\Phi(x,t,k)$ as given in~\eqref{e:Phi}:
\begin{corollary}
\label{c:first symmetry Phi}
Under the hypothesis of Theorem~\ref{Jost analyticity}, we have the symmetry
\be
\Phi^\dagger(x,t,k) = \Phi(x,t,k)^{-1}\,, \quad k \in \C\setminus\Sigma\,,
\ee
where $\dagger$ denotes the Schwarz conjugate-transpose with respect to $k$.
\end{corollary}
\eb
Lemma~\ref{first symmetry extended} also gives the following symmetries for the reflection coefficients:
\begin{corollary}[First symmetry, reflection coefficients]
	Under the hypothesis of Theorem~\ref{Jost analyticity}, we have the symmetries
	\begin{alignat}{2}
		\label{rho symmetry}
		\conj{\rho(\conj{k})} &=s_{21}(k)/s_{11}(k)\,,  &&k \in \R \cup \Sigma_{+2}^o\,,
		\\
		\label{r symmetry}
		\conj{r(\conj{k})} &= -1/s_{11}(k)r_{12}(k)\,, \quad && k \in \R \cup \Sigma_{-2}^o\,.
	\end{alignat}
\end{corollary}
Let us again use the superscript $\pm$ to denote the left/right limits along an oriented contour in the complex $k$-plane, as in~\eqref{left/right lambda}.
The symmetry $\lambda_\pm^+(k) = -\lambda_\pm(k)$ for $k \in \Sigma_\pm$ leads to the following:
\begin{lemma}[Second symmetry, Jost solutions]
\label{second symmetry}
	Under the hypothesis of Theorem~\ref{Jost analyticity},
	\bse
		\label{e:second symmetry}
		\bb
		\begin{align}
			\label{second symmetry Sigma_{+1}}
			\phi_{+1}^+(x,t,k) &= -i\e^{-i\delta} \phi_{+2}(x,t,k)\,, \quad k \in \Sigma_{+1}^o\,,
			\\
			\label{second symmetry Sigma_{+2}}
			\phi_{+2}^+(x,t,k) &= -i\e^{+i\delta} \phi_{+1}(x,t,k)\,, \quad k \in \Sigma_{+2}^o\,,
			\\
			\label{second symmetry Sigma_{-2}}
			\phi_{-1}^+(x,t,k) &= -i\e^{+i\delta} \phi_{-2}(x,t,k)\,, \quad k \in \Sigma_{-2}^o\,,
			\\
			\label{second symmetry Sigma_{-1}}
			\phi_{-2}^+(x,t,k) &= -i\e^{-i\delta} \phi_{-1}(x,t,k)\,, \quad k \in \Sigma_{-1}^o\,.
		\end{align}
		\eb
	\ese
\end{lemma}
%
%
\begin{lemma}[Second symmetry, scattering coefficients]
	\label{l:second symmetry scattering coefficients}
	Under the hypothesis of Theorem~\ref{Jost analyticity} and for $V\not=0$,
	\bse
		\label{e:second symmetry scattering coefficients}
		\bb
		\begin{alignat}{3}
		s_{22}^+(k) &= \phantom{-}i\e^{-i\delta}s_{12}(k)\,, \quad
		&
		r_{11}^+(k) &= -i \e^{-i\delta}r_{12}(k)\,, 
		\quad 
		&k \in \Sigma_{+1}^o\,,
		\\
		s_{22}^+(k) &= -i\e^{-i\delta}s_{21}(k)\,, 
		&
		r_{11}^+(k) &= \phantom{-}i \e^{-i\delta}r_{21}(k)\,,
		\quad
		&k \in \Sigma_{-1}^o\,,
		\\
		s_{11}^+(k) &= \phantom{-}i\e^{+i\delta}s_{21}(k)\,, 
		&
		r_{22}^+(k) &= -i \e^{+i\delta}r_{21}(k)\,,
		\quad
		&k \in \Sigma_{+2}^o\,,
		\\
		s_{11}^+(k) &= -i\e^{+i\delta}s_{12}(k)\,, 
		&
		r_{22}^+(k) &= \phantom{-}i \e^{+i\delta}r_{12}(k)\,,
		\quad
		&k \in \Sigma_{-2}^o\,.
		\end{alignat}
		\eb
	\ese
\end{lemma}
\bb
Note that, without the scaling factor $d_\pm(k)$ in~\eqref{e:phi x asymptotics} to define $\phi_\pm(x,t,k)$, the symmetries~\eqref{e:second symmetry} and~\eqref{e:second symmetry scattering coefficients} would change to include factors of $iA_\pm/(\lambda_\pm + (k\pm V/2))$ (cf. \cite{BM2017}).
\eb

In the special case of no counterflows, i.e. $V=0$, the symmetries~\eqref{phi conjugates},~\eqref{e:first symmetry},~\eqref{rho symmetry} and~\eqref{r symmetry} can be extended. The symmetries~\eqref{e:second symmetry} are unchanged; however, the symmetries~\eqref{e:second symmetry scattering coefficients} must be adjusted (see Section~\ref{Reductions}).

\subsection{Discrete eigenvalues}
%
The discrete eigenvalues of the scattering problem are those values of $k\in\C\setminus\Sigma$ (i.e., away from the continuous spectrum and the branch points) for which there exist solutions of the Lax pair~\eqref{e:LP}
bounded for all $x\in\R$. 
As usual, the discrete eigenvalues are in one-to-one correspondence with the zeros of the analytic scattering coefficients:
\begin{lemma}
\label{discrete eigenvalues}
	Under the hypothesis of Theorem~\ref{Jost analyticity}, there exists an eigenfunction bounded for all $x\in\R$ satisfying 
	the Lax pair~\eqref{e:LP} at $k=k_o \in \C^+\setminus\Sigma$ [respectively $\conj{k_o}\in\C^-\setminus\Sigma$] if and only if $s_{22}(k_o)=0$ [respectively $s_{11}(\conj{k_o})=0$]. Moreover, whenever such conditions are satisfied, the corresponding eigenfunctions decay exponentially at both spatial infinities.
\end{lemma}
%
%
\begin{corollary}
\label{c:discrete_eigenvalues}
The set $K\cup\conj{K}$ of discrete eigenvalues (with $K \subset \C^+ \setminus \Sigma$) is comprised of a possibly infinite set of isolated points in $\Complex\setminus\Sigma$. 
\end{corollary}
Note that the set of discrete eigenvalues could possibly have one or more accumulation points in $\Sigma= \Real\cup\Sigma_+\cup\Sigma_-$ 
since, generically, $s_{11}(k)$ and $s_{22}(k)$ are not analytic there.
Indeed, it is well known that such situations occur for the focusing NLS equation with zero boundary conditions \cite{DZ1991}.
It is also possible that 
$s_{11}(k)$ could possess zeros along real $k$-axis
even if the set of discrete eigenvalues $K\cup\conj{K}$ is finite.
Indeed, while non-generic, such situations are fairly common in the case of zero boundary conditions \cite{AS1981}.
Zeros of $s_{11}(k)$ and $s_{22}(k)$ along the continuous spectrum are referred to as spectral singularities \cite{Z1989-2}.
In contrast, the scattering coefficients do not vanish on $\Sigma_+^o$ or $\Sigma_-^o$:

\begin{lemma}
\label{nonzero scattering entries on branch cuts}
	Under the hypothesis of Theorem~\ref{Jost analyticity} and for $V\not=0$,
	\bse
		\begin{align}
			s_{11}(k) \not= 0\,, \quad r_{22}(k) \not=0\,, \quad k \in \Sigma_{+2}^o\cup\Sigma_{-2}^o\,,
			\\
			s_{12}(k) \not= 0\,, \quad r_{12}(k) \not=0\,, \quad k \in \Sigma_{+1}^o\cup\Sigma_{-2}^o\,,
			\\
			s_{21}(k) \not= 0\,, \quad r_{21}(k) \not=0\,, \quad k \in \Sigma_{+2}^o\cup\Sigma_{-1}^o\,,
			\\
			s_{22}(k) \not= 0\,, \quad r_{11}(k) \not=0\,, \quad k \in \Sigma_{+1}^o\cup\Sigma_{-1}^o\,.
		\end{align}
	\ese
\end{lemma}
Note that the above statements do not hold in the case $V=0$ (see Section~\ref{Reductions}).
Indeed, when $V=0$, the limiting case of a discrete eigenvalue on the branch cut gives rise to Akhmediev breathers \cite{BK2014}.
Also, the coefficients can vanish on the boundary of $\Sigma_+$ or $\Sigma_-$, i.e., at $k  =  p_\pm$ and $k  = \conj{p_\pm}$.
Indeed, in the case $V=0$, such zeros lead to rational solutions such as the Peregrine breather and its generalizations
\cite{BK2014,BilmanMiller2019}.

\bb
Recall that reflectionless potentials,  i.e. those for which $\rho(k)\equiv 0$, correspond to pure soliton solutions. As a consequence of Lemma~\ref{nonzero scattering entries on branch cuts}, we see that there are no reflectionless potentials when $V\not=0$, and all solutions must have a radiative component.
\eb

\bb
Recalling $\Phi(x,t,k)$ as given by~\eqref{e:Phi}, we see that the discrete eigenvalues correspond to singularities of $\Phi(x,t,k)$. It will be important to understand the residues of $\Phi(x,t,k)$ at the discrete eigenvalues for the inverse problem. Letting $\Phi_1(x,t,k)$ and $\Phi_2(x,t,k)$ denote the first and second columns of $\Phi(x,t,k)$ respectively, we have the following:
\begin{lemma}
	\label{l:residues}
	Under the hypothesis of Theorem~\ref{Jost analyticity}, if $s_{22}(k)$ has a finite set of simple zeros, $K=\{k_1, \dots, k_N\} \subset \C^+\setminus\Sigma$\,, there are norming constants $c_1, \dots, c_n \in \C$ such that
	%
	\bse
		\label{residues}
		\begin{gather}
			\Res_{k=k_n} \Phi(x,t,k) = \big(\hspace{1.5mm} 0\,, \phantom{-}c_n \Phi_1(x,t,k_n)\, \big)\,,
			\quad n = 1,\dots, N\,.
			\\
			\Res_{k=\conj{k_n}} \Phi(x,t,k) = \big(-\conj{c_n} \Phi_2(x,t,\conj{k_n})\,, 0\,\big)\,,
			\quad n = 1,\dots, N\,.
		\end{gather}
	\ese
\end{lemma}
\eb

Higher order zeros \bb of the analytic scattering coefficients \eb can be dealt with similarly, but we omit such cases for \bb
brevity.
\eb
%
%

\bb
\subsection{Scattering coefficients: Behavior at the branch points}
\label{ss:S(k) branch point behavior}

Understanding the  behavior of the scattering coefficients at the branch points will be necessary to properly formulate the inverse problem. Corollary~\ref{phi branch point continuity} together with the Wronksian definitions~\eqref{e:Wronskians} gives the behavior of the scattering coefficients at the branch points:
\begin{corollary}
\label{wr branch point continuity} 
	Under the hypothesis of Lemma~\ref{l:mu branch point continuity} and for $V\not=0$,
	\bse
	\begin{align}
	  s_{11}(k) = b_{\pm 11}^{(0)}(k-\conj{p_\pm})^{-1/4} + o(k-\conj{p_{\pm}})^{-1/4}\,, \quad k \rightarrow \conj{p_{\pm}}\,,
		\\
		s_{12}(k) = b_{+12}^{(0)}(k-p_+)^{-1/4} + o(k-p_+)^{-1/4}\,, \quad k \rightarrow p_+\,,
		\\
		s_{12}(k) = b_{-12}^{(0)}(k-\conj{p_-})^{-1/4} + o(k-\conj{p_-})^{-1/4}\,, \quad k \rightarrow \conj{p_-}\,,
		\\
		s_{21}(k) = b_{+21}^{(0)}(k-\conj{p_+})^{-1/4} + o(k-\conj{p_+})^{-1/4}\,, \quad k \rightarrow \conj{p_+}\,,
		\\
		s_{21}(k) = b_{-21}^{(0)}(k-p_-)^{-1/4} + o(k-p_-)^{-1/4}\,, \quad k \rightarrow p_-\,,
		\\
		s_{22}(k) = b_{\pm 22}^{(0)}(k-p_\pm)^{-1/4} + o(k-p_\pm)^{-1/4}\,, \quad k \rightarrow p_\pm\,,
	\end{align}
	\ese
	for some constants $b_{\pm 11}^{(0)}$, $b_{\pm 12}^{(0)}$, $b_{\pm 21}^{(0)}$, $b_{\pm 22}^{(0)}$ where the limits must be taken from within the regions described by Corollary~\ref{Wronskians}.
\end{corollary}
Analogous expansions can easily be given for the entries of $R(k)$, but we omit them for brevity.

Note that $b_{\pm 11}^{(0)}=0$ exactly when $\mu_{+2}(x,t,k)$ and $\mu_{-1}(x,t,k)$ are linearly dependent at the branch points $\conj{p_\pm}$. Similarly, $b_{\pm 22}^{(0)} = 0$ exactly when the modified eigenfunctions $\mu_{+1}(x,t,k)$ and $\mu_{-2}(x,t,k)$ are linearly dependent at the branch points $p_\pm$. The symmetry \eqref{first symmetry a extended} shows that $s_{\pm 11}=0$ exactly when $s_{\pm 22}=0$. Moreover, note that $\mu_{+1}(x,t,k)$ and $\mu_{+2}(x,t,k)$ are always linearly dependent at the branch points $p_{+}$ and $\conj{p_+}$ (whenever they can be defined there) while $\mu_{-1}(x,t,k)$ and $\mu_{-2}(x,t,k)$ are always linearly dependent at the branch points $p_-$ and $\conj{p_-}$. As a result, $b_{\pm 11}^{(0)}\,$, $b_{\pm 12}^{(0)}\,$, $b_{\pm 21}^{(0)}$ and $b_{\pm 22}^{(0)}$ are either all zero or all nonzero depending on the linear dependence of $\mu_{+1}(x,t,k)$ and $\mu_{-2}(x,t,k)$ at $p_\pm$. We say that the \textbf{generic case} holds when $\mu_{+1}(x,t,k)$ and $\mu_{-2}(x,t,k)$ are linearly independent at both branch points so that $b_{\pm 11}^{(0)}\,$, $b_{\pm 12}^{(0)}\,$, $b_{\pm 21}^{(0)}$ and $b_{\pm 22}^{(0)}$ are all nonzero.

\begin{corollary}
\label{c:reflection coefficients branch point behavior generic case}
Under the hypothesis of Lemma~\ref{l:mu branch point continuity} and for $V\not=0$, in the generic case that $\mu_{+1}(x,t,k)$ and $\mu_{-2}(x,t,k)$ are linearly independent at the branch points $p_\pm$, we have
\bse
\begin{align}
	&\qquad & \rho(k) &= \rho_o + o(1)\,, & & k \rightarrow p_+\,, & &\qquad\\
	&\qquad & r(k) &= r_o(k-p_-)^{1/2} + o(k-p_-)^{1/2}\,, & & k \rightarrow p_-\,, & &\qquad
\end{align}
\ese
for some nonzero constants $\rho_o\,, r_o$ where the limits must be taken from within the regions described by~\eqref{reflection coefficients}.
\end{corollary}

There are numerous exceptional cases beyond the generic case discussed above. When necessary, higher order expansions for the scattering coefficients about the branch points can be found using Corollary~\ref{phi branch point expansion} together with the Wronskian representations~\eqref{e:Wronskians}:
\begin{corollary}
\label{wr branch point expansion} 
	Under the hypothesis of Lemma~\ref{l:mu branch point expansion} and for $V\not=0$,
		\bse
	\begin{align}
	  s_{11}(k) = b_{\pm 11}^{(0)}(k-\conj{p_\pm})^{-1/4} + b_{\pm 11}^{(1)}(k-\conj{p_\pm})^{1/4} + o(k-\conj{p_{\pm}})^{1/4}\,, \quad k \rightarrow \conj{p_{\pm}}\,,
		\\
		s_{12}(k) = b_{+12}^{(0)}(k-p_+)^{-1/4} + b_{+12}^{(1)}(k-p_+)^{1/4} + o(k-p_+)^{1/4}\,, \quad k \rightarrow p_+\,,
		\\
		s_{12}(k) = b_{-12}^{(0)}(k-\conj{p_-})^{-1/4} + b_{-12}^{(1)}(k-\conj{p_-})^{1/4} + o(k-\conj{p_-})^{1/4}\,, \quad k \rightarrow \conj{p_-}\,,
		\\
		s_{21}(k) = b_{+21}^{(0)}(k-\conj{p_+})^{-1/4} + b_{+21}^{(1)}(k-\conj{p_+})^{1/4} + o(k-\conj{p_+})^{1/4}\,, \quad k \rightarrow \conj{p_+}\,,
		\\
		s_{21}(k) = b_{-21}^{(0)}(k-p_-)^{-1/4} + b_{-21}^{(1)}(k-p_-)^{1/4} + o(k-p_-)^{1/4}\,, \quad k \rightarrow p_-\,,
		\\
		s_{22}(k) = b_{\pm 22}^{(0)}(k-p_\pm)^{-1/4} + b_{\pm 22}^{(1)}(k-p_\pm)^{1/4} + o(k-p_\pm)^{1/4}\,, \quad k \rightarrow p_\pm\,,
	\end{align}
	\ese
	for some constants $b_{\pm 11}^{(0)}$, $b_{\pm 12}^{(0)}$, $b_{\pm 21}^{(0)}$, $b_{\pm 22}^{(0)}$, $b_{\pm 11}^{(1)}$, $b_{\pm 12}^{(1)}$, $b_{\pm 21}^{(1)}$, $b_{\pm 22}^{(1)}$ where the limits must be taken from within the regions described by Corollary~\ref{Wronskians}.
\end{corollary}

We consider here only the exceptional case in which $b_{\pm 11}^{(0)}$, $b_{\pm 12}^{(0)}$, $b_{\pm 21}^{(0)}$, $b_{\pm 22}^{(0)}$ all zero and $b_{\pm 11}^{(1)}$, $b_{\pm 12}^{(1)}$, $b_{\pm 21}^{(1)}$, $b_{\pm 22}^{(1)}$ all nonzero. Other cases can be treated similarly.
\begin{corollary}
\label{c:reflection coefficients branch point behavior exceptional case}
Under the hypothesis of Lemma~\ref{l:mu branch point expansion} and for $V\not=0$, in the exceptional case that $\mu{+1}(x,t,k)$ and $\mu_{-2}(x,t,k)$ are linearly dependent at the branch points $p_\pm$ with $b_{\pm 11}^{(1)}$, $b_{\pm 12}^{(1)}$, $b_{\pm 21}^{(1)}$, $b_{\pm 22}^{(1)}$ all nonzero, we have
\bse
\begin{align}
	&\qquad & \rho(k) &= \rho_o + o(1)\,, & & k \rightarrow p_+\,, & &\qquad\\
	&\qquad & r(k) &= r_o(k-p_-)^{-1/2} + o(k-p_-)^{-1/2}\,, & & k \rightarrow p_-\,, & &\qquad
\end{align}
\ese
for some nonzero constants $\rho_o\,, r_o$ where the limits must be taken from within the regions described by~\eqref{reflection coefficients}.
\end{corollary}
\eb

\bb
\subsection{Alternative solutions of the Lax pair}
\label{ss:U_in}

The Jost solutions $\phi_\pm(x,t,k)$ were defined to satisfy the asymptotic boundary conditions~\eqref{e:phi x asymptotics}. The preceding sections have explored the resulting analyticity properties. One could also seek solutions to an initial value problem for the Lax pair. We follow the recent work by Bilman and Miller (see Ref.~\cite{BilmanMiller2019} for details). 

\begin{lemma}
\label{l:LP IVP}
	Suppose $q(x,t)$ is a bounded classical solution of~\eqref{e:fNLS} defined for $(x,t) \in \R \times [0,\infty)$. Then for each $k\in\C$, there exists a unique simultaneous fundamental solution 
	$
	\Psi(x,t,k)
	$
	of both parts of the Lax pair~\eqref{e:LP} together with the initial condition 
	\[
	\Psi(0,0,k) = \I\,.
	\]
	Moreover, $\Psi(x,t,k)$ is an entire function of $k$ and $\det \Psi(x,t,k) \equiv 1$.
\end{lemma}

The proof proceeds by identifying $\Psi(x,t,k)$ as the value at $u=1$ of the unique solution of the integral equation
\[
\Psi(\xi(u),\tau(u),k) = \I + \int_0^u \big[ \, X(\xi(s),\tau(s),k)x'(s) + T(\xi(s),\tau(s),k) t'(s)\big]\, \Psi(\xi(s),\tau(s),k) \,ds\,,
\]
where $(\xi(\cdot),\tau(\cdot)):[0,1] \to \R\times[0,\infty)$ is a smooth path from
$(\xi(0),\tau(0)) = (0,0)$ to $(\xi(1),\tau(1)) = (x,t)$. 
One could choose any other base point $(x_o,t_o)$ at which to normalize $\Psi(x,t,k)$ instead of $(0,0)$. 
While the choice $x_o=0$ is inconsequential, it is important to take $t_o=0$, so that the relevant scattering data 
(see below) can be computed using the initial condition $q(x,0)$.
Incidentally, the same fundamental matrix solution $\Psi(x,t,k)$ also proves to be useful 
when studying boundary value problems on the half line \cite{FIS2005,F2008}.

In contrast to Theorem~\ref{Jost analyticity}, Lemma~\ref{l:LP IVP} makes no requirement that $(q-q_\pm) \in L^1_x(\R^\pm)$. Moreover, whereas the regions of analyticity for $\phi_\pm(x,t,k)$ can only generically be shown to include the appropriate half-planes (minus the relevant branch cuts), $\Psi(x,t,k)$ is entire. These differences can be intuitively understood by recognizing that, for any fixed $(x,t)$, the integral equations defining $\phi_\pm(x,t,k)$ integrate along infinitely long paths from the base points $(\pm\infty,t)$, whereas the integral equation for $\Psi(x,t,k)$ integrates along a finite path from the base point $(0,0)$. 

On the other hand, given any fundamental matrix solution to the Lax pair~\eqref{e:LP}, one can readily obtain $\Psi(x,t,k)$. 
Indeed, we have the following:

\begin{corollary}
\label{c:phi_in}
Under the hypotheses of Theorem~\ref{Jost analyticity} and Lemma~\ref{l:LP IVP}, we have
\[
	\label{e:phi_in}
	\Psi(x,t,k) = \Phi(x,t,k)C^{-1}(k)\,, \quad k\in\C\setminus(\Sigma\cup K\cup\conj{K})\,,
\]
where $C(k) = \Phi(0,0,k)$.
\end{corollary}

The above corollary follows immediately from the uniqueness of the initial-value problem in Lemma~\ref{l:LP IVP}. Moreover, one could use~\eqref{e:phi_in} as a definition for $\Psi(x,t,k)$, with the exception of the removable singularities for $k\in\Sigma\cup K\cup\conj{K}$. 
Note that $C(k)$ is not an entire function of $k$.  
However, the discontinuities of $\Phi(x,t,k)$ and $C^{-1}(k)$ across $\Sigma$ exactly cancel so that $\Psi(x,t,k)$ is entire. The symmetries~\eqref{phi conjugates} are then passed to $\Psi(x,t,k)$, though it follows directly from ~\eqref{X symmetry} and Lemma~\ref{l:LP IVP} that $\Psi(x,t,k)$ satisfies the following:

\begin{corollary}
	\label{c:first symmetry phi_in}
	Under the hypotheses of Lemma~\ref{l:LP IVP}, we have
	\[
		\conj{\Psi(x,t,\conj{k})} = -\sigma_\ast \Psi(x,t,k)\sigma_\ast\,, \quad k \in \C\,.
	\]
\end{corollary}

\eb

\section{Inverse problem: Riemann-Hilbert problem formulation}
\label{Inverse problem: RHP formulation}
%
We now turn our attention to the inverse problem (namely, recovering the solution of the NLS equation from its scattering data), which we will formulate as a matrix Riemann-Hilbert problem (RHP). Proofs for all the results in this section are given in Section~\ref{a:Inverse problem}.
We begin by introducing the sectionally meromorphic matrix function
\bb
\be
\label{M definition}
	M(x,t,k) = 
	\Phi(x,t,k)\e^{-i\theta_o(x,t,k)\sigma_3}\,,\quad k\in\C\setminus\Sigma\,,
\ee
\eb
where $\Phi(x,t,k)$ is given by~\eqref{e:Phi} and $\theta_o(x,t,k) = k(x-2kt)$, as in~\eqref{theta_o}.
In light of Lemma~\ref{Jost asymptotics} and Corollary~\ref{scattering coefficient asymptotics}, we see that $M(x,t,k) = \I+O(1/k)$ as $k \rightarrow \infty$. Note that $\det M(x,t,k) = 1$. 
\bb
Since $\Phi(x,t,k)$ satisfies
 the Lax pair~\eqref{e:LP}, we have the following:
\eb
\begin{lemma}[Modified Lax pair]
The matrix $M(x,t,k)$ defined by~\eqref{M definition} satisfies the modified Lax pair%
\bse
		\label{e:modified Lax pair}
		\begin{align}
		M_x(x,t,k) - ik\big[\sigma_3\,, M(x,t,k)\big] &= Q(x,t)M(x,t,k)\,,
		\\
		M_t(x,t,k) + 2ik^2\big[\sigma_3\,, M(x,t,k)\big] &= \big(i\sigma_3\big(Q_x(x,t)-Q^2(x,t)\big) + 2kQ(x,t)\big)M(x,t,k)\,.
		\end{align}
	\ese
\end{lemma}

\subsection{Jump matrix and residue conditions}
\label{Jump matrix and residue conditions}

As before, we use the superscripts $\pm$ to denote the non-tangential left/right limits toward the oriented contours, with $\Sigma$ oriented as in Fig~\ref{f:spectrum}. 
We first express the discontinuity of $M(x,t,k)$ across $\Sigma^o$. 
We should note that in many previous works, $\theta_\pm(x,t,k)$ were used instead 
of $\theta_o(x,t,k)$ in the definition~\eqref{M definition}.
The use of $\theta_o(x,t,k)$, however, 
results in a considerable simplification of the jumps across the branch cuts
compared to $\theta_\pm(x,t,k)$,
since, unlike $\theta_\pm(x,t,k)$, $\theta_o(x,t,k)$ is entire.
\begin{lemma}
	\label{l:jump matrix}
	Under the hypothesis of Theorem~\ref{Jost analyticity} and for $V\not=0$, the matrix $M(x,t,k)$ defined in~\eqref{M definition} satisfies the jump condition
	\be
\label{e:M jump condition generic}
	M^+(x,t,k) = M^-(x,t,k)J(x,t,k)\,,\quad k\in\Sigma^o,
	\ee
where the jump matrix $J(x,t,k)$ is given by 
\begin{equation}
\label{e:JfromJo}
J(x,t,k) = \e^{i\theta_o(x,t,k)\sigma_3} J_o(k) \,\e^{-i\theta_o(x,t,k)\sigma_3}\,, 
\end{equation}
with
\begin{align}
		\label{nonzero velocity jump matrix}
J_o(k) = 
		\begin{cases} 
			\begin{pmatrix}
				1 & \rho(k)\\ 
				\conj{\rho(k)} & 1+\rho(k)\conj{\rho(k)}
			\end{pmatrix}, 
			&k \in \R\,,
			\\[15pt]
			\begin{pmatrix}
				i\rho(k)\e^{-i\delta} & 0\\
				-i\e^{-i\delta} & -i\e^{i\delta}/\rho(k)
			\end{pmatrix}, 
			&k \in \Sigma_{+1}^o\,,
			\\[15pt]
			\begin{pmatrix}
				i\e^{-i\delta}/\conj{\rho(\conj{k})} & -i\e^{i\delta}\\
				0 & -i\conj{\rho(\conj{k})}\e^{i\delta}
			\end{pmatrix}, 
			&k \in \Sigma_{+2}^o\,,
			\\[15pt]
			\begin{pmatrix}
				1 & -r(k)\\
				0 & 1
			\end{pmatrix}, 
			&k \in \Sigma_{-1}^o\,,
			\\[15pt]
			\begin{pmatrix}
				1 & 0\\
				\conj{r(\conj{k})} & 1
			\end{pmatrix}, 
			&k \in \Sigma_{-2}^o\,,
		\end{cases}
\end{align}
where $\rho(k)$ and $r(k)$ are given in~\eqref{reflection coefficients}.
\end{lemma}
Special cases, as well as the 
reduction to $V=0$, 
are discussed in Sections~\ref{Reductions} and~\ref{Riemann problems}.

As a consequence of Lemma~\ref{nonzero scattering entries on branch cuts} we see that the entries of the jump matrix are non-singular 
\bb
on $\Sigma_+^o \cup \Sigma_-^o$, but potentially have unbounded growth toward the branch points. Corollaries~\ref{c:reflection coefficients branch point behavior generic case} and~\ref{c:reflection coefficients branch point behavior exceptional case} describe some of the possible behaviors of the jump matrix at these potential singularities. In particular, in the generic case we see that $J_o(k)$ is continuous at the branch points.
\eb

Note that we have broken the $x\mapsto -x$ symmetry in the definition of $M(x,t,k)$, which rescales the columns of $\phi_-(x,t,k)$ using the analytic scattering coefficients from the right, $s_{11}(k)$ and $s_{22}(k)$. One could instead define $M(x,t,k)$ using the analytic scattering coefficients from the left, $r_{11}(k)$ and $r_{22}(k)$ to rescale $\phi_+(x,t,k)$. Indeed, defining
\bb
\be
	\label{hat M(x,t,k)}
	\^M(x,t,k) = 
\begin{cases}
\displaystyle
\bigg(\phi_{-2}(x,t,k), \frac{\phi_{+1}(x,t,k)}{r_{11}(k)}\bigg)\e^{i\theta_o(x,t,k)\sigma_3}\,, \quad k \in \C^+ \setminus \Sigma\,,
		\\[12pt]
\displaystyle
		\bigg(\frac{\phi_{+2}(x,t,k)}{r_{22}(k)}, \phi_{-1}(x,t,k)\bigg)\e^{i\theta_o(x,t,k)\sigma_3}\,, \quad k \in \C^- \setminus \Sigma\,,
\end{cases}
\ee
\eb
we find that $\^M^+(x,t,k) = \^M^-(x,t,k)\^J(x,t,k)$, where 
\be
	\^J(x,t,k) = J(x,t,k)\big[(\rho,r,\theta_o,\Sigma_+,\Sigma_-) \mapsto (\^\rho, \^r, -\theta_o\,, \Sigma_-, \Sigma_+)\big],
\ee
with $\^\rho(k) = r_{21}(k)/r_{11}(k)$ and $\^r(k) = 1/r_{11}(k)s_{12}(k)$.

As usual, when a non-empty discrete spectrum is present, the matrix $M(x,t,k)$ acquires pole singularities at the eigenvalues forming the discrete spectrum, and these singularities must be taken into account to complete the formulation of the RHP.
Specifically, 
\bb
letting $M_1(x,t,k)$ and $M_2(x,t,k)$ denote the first and second columns of $M(x,t,k)$ respectively,
\eb
from
\eqref{residues} we have the following residue conditions for $M(x,t,k)$:
\begin{lemma}
\bb
	Under the hypothesis of Theorem~\ref{Jost analyticity}, 
\eb
	if $s_{22}(k)$ has a finite set of simple zeros, $K=\{k_1, \dots, k_N\} \subset \C^+\setminus\Sigma$\,, then $M(x,t,k)$ is analytic in $\C\setminus(\Sigma\cup K \cup \conj{K})$. Moreover, $M(x,t,k)$ has simple poles at each $k_n\in K$ and $\conj{k_n}\in\conj{K}$\,, and there are norming constants $c_1, \dots, c_n \in \C$ such that
	%
	\bb
	\bse
		\label{residue conditions}
		\begin{gather}
			\Res_{k=k_n} M(x,t,k) = \big(\hspace{2.6mm}0\,,\, \phantom{-}c_n\e^{2i\theta_o(x,t,k_n)}M_1(x,t,k_n)\,\big)\,,
			\quad n = 1,\dots, N\,,
			\\
			\Res_{k=\conj{k_n}} M(x,t,k) = \big(-\conj{c_n}\e^{-2i\theta_o(x,t,\conj{k_n}}M_2(x,t,\conj{k_n})\,,\,0\,\big)\,,
			\quad n = 1,\dots, N\,.
		\end{gather}
	\ese
	\eb
\end{lemma}

\subsection{Growth conditions}
\label{Growth conditions}

In addition to the normalization, jump condition and residue conditions for $M(x,t,k)$, one must specify appropriate growth conditions near the branch points~\cite{BilmanMiller2019}.

\bb
Corollary~\ref{phi branch point continuity} describes the behavior of the Jost solutions near the branch points. Specifically, if $(q-q_\pm) \in L_x^{1,1}(\R^\pm)$ then the eigenfunctions $\phi_{+1}(x,t,k)$ and $\phi_{-2}(x,t,k)$ have $-1/4$ power growth toward their respective branch points $p_\pm$.
The behavior of $s_{11}(k)$ and $s_{22}(k)$ near the branch points then determines the growth conditions for the inverse problem. Recall that in the generic case (see Section~\ref{ss:S(k) branch point behavior}), $s_{11}(k)$ and $s_{22}(k)$ have $-1/4$ power growth toward the branch points $\conj{p_\pm}$ and $p_\pm$ respectively. 
%
In such cases, Corollaries~\ref{phi branch point continuity} and~\ref{wr branch point continuity} give the following result: %
\begin{lemma}
\label{nonzero velocity growth conditions generic case}
Let $V\ne0$ and the hypothesis of Lemma~\ref{l:mu branch point continuity} be satisfied.
In the generic case that $\mu_{+1}(x,t,k)$ and $\mu_{-2}(x,t,k)$ are linearly independent at the branch points $p_\pm$\,, we have
\[
	M(x,t,k) = \begin{cases}
		\big( B_{p_+}^{(0)}(x,t) + o(1)\big)(k-p_+)^{-\sigma_3/4} \,, &k\rightarrow p_{+}\,,
		\\
		\big( B_{\conj{p_+}}^{(0)}(x,t) + o(1)\big)(k-\conj{p_+})^{+\sigma_3/4} \,, &k\rightarrow \conj{p_{+}}\,,
		\\
		B_{p_-}^{(0)}(x,t) + o(1) \,, &k\rightarrow p_{-}\,,
		\\
		B_{\conj{p_-}}^{(0)}(x,t) + o(1) \,, &k\rightarrow \conj{p_{-}}\,,
	\end{cases}
\]	
for some invertible matrices $B^{(0)}_{p_\pm}(x,t)$, $B_{\conj{p_\pm}}^{(0)}(x,t)$.
\end{lemma}
Note that the invertibility of the matrices $B_{p_\pm}^{(0)}(x,t)\,$, $B_{\conj{p_\pm}}^{(0)}(x,t)$ is an immediate consequence of the linear independence of $\mu_{+1}(x,t,k)$ and $\mu_{-2}(x,t,k)$ at the branch points. In particular, Lemma~\ref{nonzero velocity growth conditions generic case} shows that the following limits exist:
\be
	\begin{aligned}
	&\lim_{k\to p_+} M(x,t,k)(k-p_+)^{+\sigma_3/4}\,,\qquad & &\lim_{k\to p_-} M(x,t,k)\,,\\
	&\lim_{k\to \conj{p_+}} M(x,t,k)(k-\conj{p_+})^{-\sigma_3/4}\,,\qquad & &\lim_{k\to\conj{p_-}}M(x,t,k)\,.
	\end{aligned}
	\label{e:growth condition generic case}
\ee
The requirement that these limits exist will serve as the growth conditions for the RHP in the generic case in order to guarantee uniqueness of solutions.

The asymptotic behavior changes in the exceptional case in which $\mu_{+1}(x,t,k)$ and $\mu_{-2}(x,t,k)$ are linearly dependent at one or both of the branch points $p_{\pm}$. 
Specifically, suppose we are in the case where $b_{\pm 11}^{(0)}$, $b_{\pm 22}^{(0)}$ are zero and $b_{\pm 11}^{(1)}$, $b_{\pm 22}^{(1)}$ are nonzero, where $b_{\pm 11}^{(0)}$, $b_{\pm 22}^{(0)}$, $b_{\pm 11}^{(1)}$ and $b_{\pm 22}^{(1)}$ are as in Corollary~\ref{wr branch point expansion}. In such cases, Corollaries~\ref{phi branch point expansion} and~\ref{wr branch point expansion} give the following result:
%

%
\begin{lemma}
\label{nonzero velocity growth conditions exceptional case}
	Let $V\ne0$ and the hypothesis of Lemma~\ref{l:mu branch point expansion} be satisfied.  
	In the exceptional case that $\mu_{+1}(x,t,k)$ and $\mu_{-2}(x,t,k)$ are linearly dependent at the branch points $p_{\pm}$ with $b_{\pm 11}^{(1)}$ and $b_{\pm 22}^{(1)}$ (as given by Corollary~\ref{wr branch point expansion}) nonzero, we have
\[
	M(x,t,k) = \begin{cases}
	 \big(B_{p_{+}}^{(0)}(x,t) + B_{p_{+}}^{(1)}(x,t)(k-p_+)^{1/2} + o(k-p_+)^{1/2}\big)(k-p_+)^{-1/4}\,, &k\rightarrow p_{+}\,,
		\\
		\big(B_{\conj{p_+}}^{(0)}(x,t) + B_{\conj{p_+}}^{(1)}(x,t)(k-\conj{p_+})^{1/2} + o(k-p_+)^{1/2}\big)(k-\conj{p_+})^{-1/4}\,, &k\rightarrow \conj{p_{+}}\,,
		\\
		\big(B_{p_-}^{(0)}(x,t) + B_{p_-}^{(1)}(x,t)(k-p_-)^{1/2} + o(k-p_+)^{1/2}\big)(k-p_-)^{-1/4+\sigma_3/4}\,, &k\rightarrow p_{-}\,,
		\\
		\big(B_{\conj{p_-}}^{(0)}(x,t) + B_{\conj{p_-}}^{(1)}(x,t)(k-\conj{p_-})^{1/2} + o(k-p_+)^{1/2}\big)(k-\conj{p_-})^{-1/4-\sigma_3/4}\,, &k\rightarrow \conj{p_{-}}\,,
	\end{cases}
	\label{e:growth condition exceptional case}
\]
for some matrices $B_{p_\pm}^{(0)}(x,t)\,$, $B_{\conj{p_\pm}}^{(0)}(x,t)\,$, $B_{p_\pm}^{(1)}(x,t)\,$, $B_{\conj{p_\pm}}^{(1)}(x,t)\,$, with $\det B^{(0)}_{p_{\pm}}(x,t) = \det B_{\conj{p_\pm}}^{(0)}(x,t) = 0$.
\end{lemma}
Other exceptional cases can be treated similarly. Higher order expansions for $M(x,t,k)$ about the branch points can be found by placing further restrictions on the potential.

Note that the asymmetry between the growth conditions at $p_-$ and $\conj{p_-}$ on one hand and those at $p_+$ and $\conj{p_+}$ on the other is a result of the choice to rescale $\phi_-(x,t,k)$ using the analytic scattering coefficients from the right to define $M(x,t,k)$. If one took $\^M(x,t,k)$ as defined in \eqref{hat M(x,t,k)} instead, the growth conditions would match exactly those for $M(x,t,k)$ but with $p_+$ and $p_-$ interchanged.
\eb

\subsection{Riemann-Hilbert problem, linear algebraic-integral equations and reconstruction formula}
\label{reconstruction}

Together, the results of Sections~\ref{Jump matrix and residue conditions} and~\ref{Growth conditions}
describe the properties of the matrix $M(x,t,k)$ through its definition~\eqref{M definition}
in terms of the Jost eigenfunctions.  
Specifically, in the generic case in which the analytic Jost solutions are linearly independent at the branch points, we have:
\begin{definition}[Riemann-Hilbert problem]
\label{nonzero velocity RHP}
Determine a matrix $M(x,t,k)$ satisfying the following conditions:
\vspace*{-1ex}
\begin{itemize}
\advance\itemsep-4pt
\item[(i)] 
$M(x,t,k)$  is analytic for $k\in\C\setminus(\Sigma\cup K\cup\conj{K})$, with $K \subset \C^+\setminus\Sigma$ finite,
\item[(ii)]
$M(x,t,k)$ satisfies the jump condition~\eqref{e:M jump condition generic},
with $J(x,t,k)$ given by~\eqref{nonzero velocity jump matrix},
\item[(iii)]
$M(x,t,k) =\I+ O(1/k)\,,\ k\rightarrow\infty$,
\item[(iv)]
$M(x,t,k)$ has simple poles at each $k_n\in K$ and $\conj{k_n}\in\conj{K}$ satisfying the residue conditions~\eqref{residue conditions},
\item[(v)]
$M(x,t,k)$ satisfies the growth conditions~\eqref{e:growth condition generic case} at the branch points $p_\pm$ and $\conj{p_\pm}$ 
\bb
(i.e. the limits exist).
\eb
\end{itemize}
\label{d:RHP}
\end{definition}

\bb
\begin{theorem}
\label{t:direct RHP solution}
For $V\not=0$, if $(q-q_\pm) \in L^{1,1}_x(\R^\pm)$ and $(q-q_\pm)_x \in L^1_x(\R^\pm)$ for all $t\in\R$ with $\mu_{+1}(x,t,k)$ and $\mu_{-2}(x,t,k)$ linearly independent at both branch points $p_\pm$, then the matrix $M(x,t,k)$ defined by~\eqref{M definition} satisfies Riemann-Hilbert problem~\ref{nonzero velocity RHP}.
\end{theorem}
\eb

We now invert the perspective and seek to recover $M(x,t,k)$ just from the five properties in
Definition~\ref{d:RHP}.
That is, we show how the solution of the above RHP can be converted into that of a suitable set of
linear algebraic-integral equations. 
\bb
We first have:
\begin{lemma}
\label{l:det RHP}
If $M(x,t,k)$ is any solution of the RHP~\ref{nonzero velocity RHP}, then 
$\det M(x,t,k) = 1$ for all $k \in \C\setminus(\Sigma \cup K \cup \conj{K})$.
Moreover, $M^\dagger(x,t,k)^{-1}$ also solves the RHP.
\end{lemma}
%
The first part of Lemma~\ref{l:det RHP} is easily proved by applying the determinant to the conditions in RHP~\ref{nonzero velocity RHP} to arrive at a scalar RHP seeking an entire function which is $1 + O(1/k)$ as $k\to\infty$. 
The second part of Lemma~\ref{l:det RHP} is verified through straightforward calculations.
\eb

For brevity, in what follows 
we suppress the dependence of $M$, $J$ and $\theta_o$ on $x$ and $t$ wherever this does not cause ambiguity. Again letting $M_1(k)$ and $M_2(k)$ denote the first and second columns of $M(k)$ respectively, we have the following:

\begin{theorem}
	\label{t:integral equation for M}
	\bb
	If the RHP~\ref{nonzero velocity RHP} admits a solution $M(k)$, it
	is given as a solution
	\eb
	to the following system of linear algebraic-integral equations:
	\bse
		\label{e:algebraic linear system}
		\begin{align}
			\begin{split}
			\label{e:integral equation for M}
			M(k) &=\I+ \sum_{n=1}^N \Big(\frac{-\conj{c_n}\e^{-2i\theta_o(\conj{k_n})}M_2(\conj{k_n})}{k-\conj{k_n}}, \frac{c_n\e^{2i\theta_o(k_n)}M_1(k_n)}{k-k_n}\Big)
\\
                &\kern1.85em 
			+ \frac{1}{2\pi i} \int_\Sigma \frac{M^-(\xi)(J(\xi)-I)}{\xi-k}d\xi\,,
                \quad k \in \C\setminus(\Sigma\cup K\cup\overline{K})\,,			\end{split}
			\\
\label{e:M algebraic 1}
			M_1(k_n) &= 
				\begin{pmatrix}
					1\\
					0
				\end{pmatrix}
				- \sum_{m=1}^N \frac{\conj{c_m}\e^{-2i\theta_o(\conj{k_m})}M_2(\conj{k_m})}{k_n-\conj{k_m}} + \frac{1}{2\pi i} \int_\Sigma \frac{\big[ M^-(\xi)(J(\xi)-\I) \big]_1}{\xi-k_n}d\xi\,,
			\\
\label{e:M algebraic 2}
			M_2(\conj{k_n}) &= 
				\begin{pmatrix}
					0\\
					1
				\end{pmatrix}
				+ \sum_{m=1}^N \frac{c_m\e^{2i\theta_o(k_m)}M_1(k_m)}{\conj{k_n}-k_m} + \frac{1}{2\pi i} \int_\Sigma \frac{\big[ M^-(\xi)(J(\xi)-\I) \big]_2}{\xi-\conj{k_n}}d\xi\,.
		\end{align}
	\ese	
\end{theorem}
The final \bb step \eb in the inverse problem is to reconstruct the solution of the NLS equation from that of the RHP. This is done without any appeal to the direct problem, but instead by using only those conditions on $M(x,t,k)$ imposed by the RHP~\ref{nonzero velocity RHP}.

\begin{lemma}
	\label{l:M Lax pair}
	Let $M(x,t,k)$ solve the RHP~\ref{nonzero velocity RHP}. Then $M(x,t,k)$ satisfies the modified Lax pair~\eqref{e:modified Lax pair} with
	\[
		\label{e:Q(x,t) inverse problem definition}
		Q(x,t) := -i\lim_{k\rightarrow \infty} k\big[\sigma_3\,, M(x,t,k)\big]\,.
	\]
\end{lemma}

\begin{corollary}
	[Reconstruction formula]
	\label{c:reconstruction formula}
	Let $M(x,t,k)$ solve the RHP~\ref{nonzero velocity RHP}.  
	The corresponding solution of the NLS equation~\eqref{e:fNLS} is given by
	\be
	\label{e:reconstruction formula}
		q(x,t) = -2i\sum_{n=1}^n c_n\e^{2i\theta_o(x,t,k_n)}M_{11}(x,t,k_n) - \frac{1}{\pi} \int_\Sigma \big[ M^-(x,t,\xi)(J(x,t,\xi)-I) \big]_{12} d\xi \,.
	\ee
\end{corollary}

\bb
\subsection{Existence and uniqueness of solutions of the Riemann-Hilbert problem}
\label{ss:Alternative RHP}

The issue of the existence and uniqueness of a solution to the RHP~\ref{nonzero velocity RHP} is nontrivial because of the singular behavior of $M(x,t,k)$ at the branch points (see Section~\ref{Growth conditions}). On the other hand, following recent work~\cite{BilmanMiller2019}, one can define an alternative matrix and RHP that is also regular at the branch points, as we show next.

\begin{figure}[t!]
	\centering
\begin{subfigure}{.5\textwidth}
		\centering
		\includegraphics[width=0.8\textwidth]{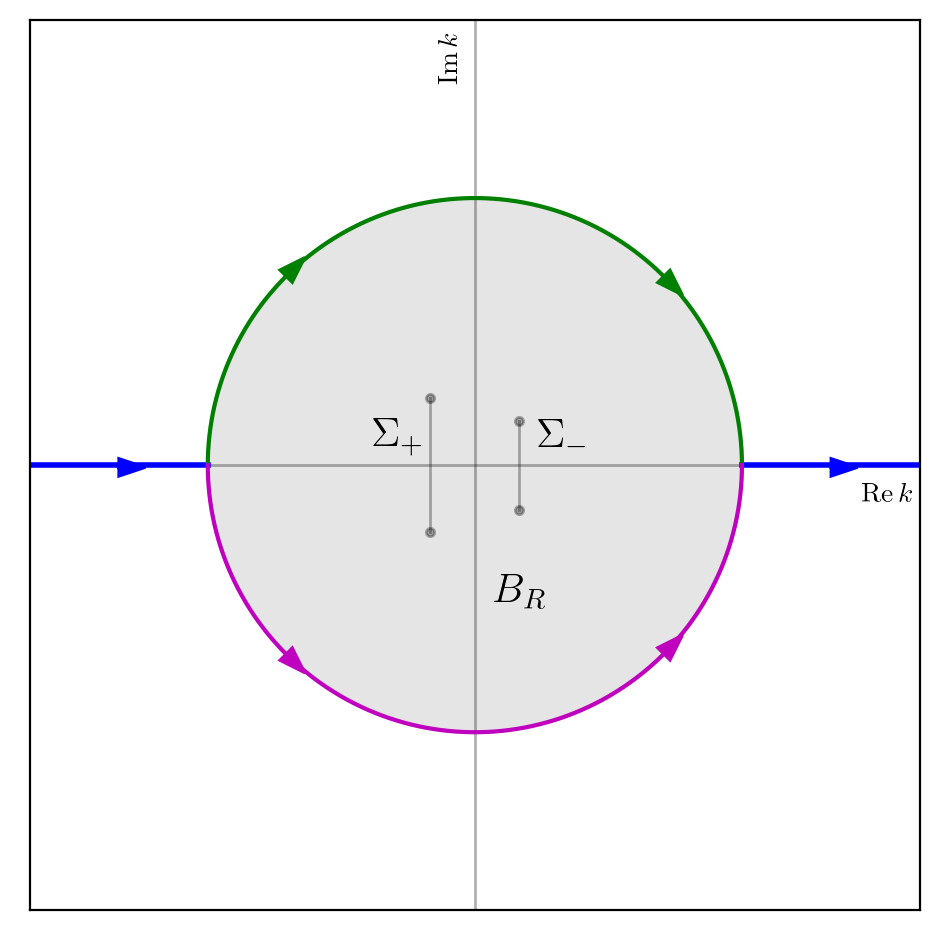}
\end{subfigure}%
\begin{subfigure}{.5\textwidth}
		\centering
		\includegraphics[width=0.8\textwidth]{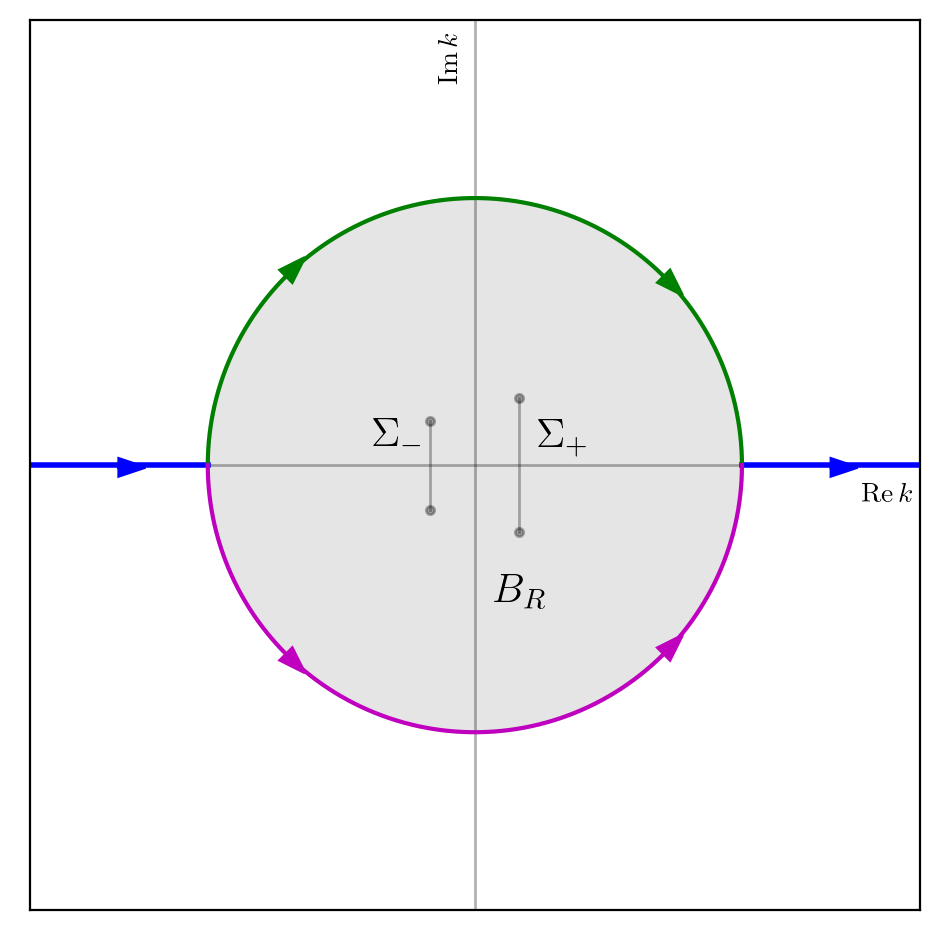}
\end{subfigure}
	\caption{Orientation of the contour $\~\Sigma=(-\infty,-R)\cup\partial B_R\cup(R,\infty)$. The branch cuts $\Sigma_\pm$ are included for $V>0$ (left) and $V<0$ (right) with $A_- < A_+$. Note that $\~M(x,t,k)$ is analytic on the branch cuts and along $(-R,R)$.}
\label{f:alternative contours}
\end{figure}

To begin, we choose $R>0$ large enough so that the ball $B_R$ of radius $R$ centered at the origin of the complex $k$-plane contains both branch cuts $\Sigma_+$, $\Sigma_-$ and all zeros of the analytic scattering coefficients $s_{11}(k)$ and $s_{22}(k)$. Note that the large $k$  behavior~\eqref{e:scattering coefficient asymptotics} of the scattering coefficients guarantees that such an $R$ always exists.
We then introduce a modified, sectionally analytic matrix, 
\[
	\label{e:alternative M}
	\~M(x,t,k) = 
	\begin{cases}
		\Phi(x,t,k)\,\e^{-i\theta_o(x,t,k)\sigma_3}\,, &k \in \C\setminus((-\infty,-R] \cup B_R \cup [R,\infty))\,,
		\\
		\Psi(x,t,k)\,\e^{-i\theta_o(x,t,k)\sigma_3}\,,&k \in B_R\,,
	\end{cases}
\]
with $\Phi(x,t,k)$ and $\Psi(x,t,k)$ given by~\eqref{e:Phi} and~\eqref{e:phi_in} respectively. Corollaries~\ref{c:phi_in} and Lemma~\ref{l:jump matrix} immediately give the following:

\begin{corollary}
Under the hypotheses of Theorem~\ref{Jost analyticity} and Lemma~\ref{l:LP IVP}, the matrix $\~M(x,t,k)$ defined in \eqref{e:alternative M} satisfies the jump condition
\[
	\label{e:alternative jump condition}
	\~M^+(x,t,k) = \~M^-(x,t,k)\~J(x,t,k)\,, \quad k \in \~\Sigma\,,
\]
where the contour 
\[
\~\Sigma =(-\infty,-R) \cup \partial B_R \cup (R,\infty)
\]
is oriented as in Fig~\ref{f:alternative contours} and the jump matrix $\~J(x,t,k)$ is given by
\[
	\~J(x,t,k) = \e^{i\theta_o(x,t,k)\sigma_3} \~J_o(k)\e^{-i\theta_o(x,t,k)\sigma_3}\,,
\]
with
\[
\label{e:alternative jump matrix}
\~J_o(k) = 
	\begin{cases}
	J_o(k)\,, & k \in (-\infty,R)\cup(R,\infty)\,,
	\\
	C(k)\,, &k \in \partial B_R \cap \C^+\,,
	\\
	C(k)^{-1}\,, &k \in \partial B_R \cap \C^-\,,
	\end{cases}
\]
where $J_o(k)$ is given in~\eqref{nonzero velocity jump matrix} and $C(k)$ is given by~\eqref{e:phi_in}.
\label{c:Mtilde properties}
\end{corollary}

Note that $\~M(x,t,k)$ matches exactly $M(x,t,k)$ for large $k$, and as such has the same normalization at infinity. 
Importantly, not also that, since all the zeros of $s_{11}(k)$ and $s_{22}(k)$ are inside $B_R$ where $\Psi(x,t,k)$ is analytic instead, $\~M(x,t,k)$ is sectionally analytic (not just sectionally meromorphic like $M(x,t,k)$). 
This means that no residues conditions will be needed in the modified RHP.
Moreover, $\~M(x,t,k)$ is analytic on both branch cuts $\Sigma_\pm$, including at the branch points. 
The absence of jumps across the branch cuts is a departure from the formalism of~\cite{BilmanMiller2019}, and results from the use of $\theta_o(x,t,k)$ instead of $\theta_\pm(x,t,k)$ to define $\~M(x,t,k)$.

The definition~\eqref{e:alternative M} and Corollary~\ref{c:Mtilde properties} give rise to the following modified RHP:

\begin{definition}[Modified Riemann-Hilbert problem]
\label{d:alternative RHP}
Determine a matrix $\~M(x,t,k)$ satisfying the following conditions:
\vspace*{-1ex}
\begin{itemize}
\advance\itemsep-4pt
\item[(i)]
$\~M(x,t,k)$ is analytic for $k\in\C\setminus\~\Sigma$\,,
\item[(ii)]
$\~M(x,t,k)$ satisfies the jump condition~\eqref{e:alternative jump condition}\,,
with $\~J(x,t,k)$ given by~\eqref{e:alternative jump matrix}\,,
\item[(iii)]
$\~M(x,t,k) =\I+ O(1/k)\,,\ k\rightarrow\infty$\,.
\end{itemize}
\end{definition}

\begin{theorem}
\label{t:alternative RHP solution}
	If $(q-q_\pm) \in L^{1}_x(\R^\pm)$ and $(q-q_\pm)_x \in L^{1}_x(\R^\pm)$ for all $t\in\R$, then the matrix $\~M(x,t,k)$ defined by~\eqref{e:alternative M} satisfies the modified Riemann-Hilbert problem~\ref{d:alternative RHP}.
\end{theorem}

The important difference between the RHP~\ref{nonzero velocity RHP} and the RHP~\ref{d:alternative RHP} is that the latter has no singular behavior at the branch points.
All the information about the behavior of $M(x,t,k)$ near the branch points, as well as all possible remnants of any discrete spectrum, 
are encoded into the jump matrix $\~J(x,t,k)$ along $\partial B_R$. 
This new RHP then falls under the framework developed by Zhou \cite{Z1989}.
For convenience, we state Zhou's vanishing lemma explicitly:

\begin{lemma}(Zhou's vanishing lemma, theorem~9.3 in Ref.~\cite{Z1989})
\label{l:Zhou's lemma}
Let $\~\Sigma$ be an oriented contour which is a finite union of simple smooth closed curves (possibly extending to infinity) with a finite number of self intersections.
Consider a generic Riemann-Hilbert problem that consists of finding a matrix $\~M(x,t,k)$ satisfying the following conditions:
\vspace*{-1ex}
\begin{itemize}
\advance\itemsep-4pt
\item[(i)]
$\~M(x,t,k)$ is analytic for $k\in\C\setminus \~\Sigma$\,,
\item[(ii)] $\~M^+(x,t,k) = \~M^-(x,t,k)\~J(x,t,k)\,, \ k \in \~\Sigma$\,,
\item[(iii)]
$\~M(x,t,k) =\I+ O(1/k)\,,\ k\rightarrow\infty$\,.
\end{itemize}
Suppose the contour $\~\Sigma$ is Schwarz symmetric and if the jump matrix $\~J(x,t,k)$ satisfies the following:
\vspace*{-1ex}
\begin{itemize}
\advance\itemsep-4pt
\item[(a)]$\~J(x,t,k)$ is $C^1(\~\Sigma)$\,, 
\item[(b)] $\~J(x,t,k) = \~J^\dagger(x,t,k)\,,\ k \in \~\Sigma\setminus\R$\,, where $\dagger$ denotes the Schwarz conjugate-transpose, 
\item[(c)] 
$\Re \~J(x,t,k)$ is positive definite for $k \in \~\Sigma\cap\R$\,.
\end{itemize}
Then the Riemann-Hilbert problem admits a unique solution.
\end{lemma}
%
Lemma~\ref{l:Zhou's lemma} does not apply to the original RHP~\ref{d:RHP}
(since, for example, the contour is not closed), but it does apply to the modified RHP~\ref{d:alternative RHP}: 
%
\begin{theorem}
	\label{c:alternative RHP existence and uniqueness}
	Suppose $\rho(k) \in C^1(\R\setminus(-R,R))$ and $C(k) \in C^1(\partial B_R\cap\C^\pm)$ with $C(k)C^\dagger(k)=\I$.
	The modified Riemann-Hilbert problem~\ref{d:alternative RHP} admits a unique solution.
\end{theorem}
Importantly, 
note that the above conditions on $\rho(k)$ and the matrix $C(k)$ are automatically satisfied when they are generated through the direct problem (see section~\ref{Direct problem: Scattering matrix, symmetries and discrete eigenvalues} for details).

Finally, we show how solutions of the modified RHP are related to those of the original RHP in order to establish the uniqueness of solutions of the original RHP~\ref{nonzero velocity RHP}.
We do so by constructing an appropriate map between solutions of the two RHPs.
Specifically, if $M_o(x,t,k)$ is any solution of the original RHP~\ref{nonzero velocity RHP}, 
let $C_o(k) = M_o(0,0,k)$.  We know $C_o(k)$ has unit determinant by Lemma~\ref{l:det RHP}. 
With $C_o(k)$ fixed, we now define the following map $F_o$ operating on matrix-valued functions $m(x,t,k)$:
\be
	\label{e:M to tilde M}
F_o(m)(x,t,k) = \begin{cases}
  m(x,t,k)\,,\qquad &k \in \C\setminus(\Sigma \cup B_R)\,,\\
  m(x,t,k)\,\e^{i\theta_o(x,t,k)\sigma_3}\,C_o^{-1}(k)\,\e^{-i\theta_o(x,t,k)\sigma_3}\,, &k \in B_R\setminus\Sigma\,,\end{cases}
\ee
where the ball $B_R$ of radius $R$ is taken large enough to contain all singularities of the original RHP. 
We then show that, for \textit{any} solution $M(x,t,k)$ of the original RHP, 
$\~M(x,t,k) := F_o(M)(x,t,k)$ is indeed a solution of the modified RHP, with $C(k)$ replaced by $C_o(k)$. 
The map therefore allows one to establish the uniqueness of solutions to the original RHP~\ref{nonzero velocity RHP}:
\begin{theorem}
	\label{t:RHP uniqueness}
	If $\rho(k) \in C^1(\R\setminus(-R,R))$ and if there exists a solution $C_o(k)$ of the original Riemann-Hilbert problem~\ref{nonzero velocity RHP} at $(x,t)=(0,0)$ satisfying $C_o(k)C_o^\dagger(k) = \I$, then any solution of the original Riemann-Hilbert problem is unique for all $(x,t)\in\Real\times\Real^+$.
\end{theorem}
\eb

\section{Reductions: symmetric amplitudes; one-sided boundary conditions; zero velocity}
\label{Reductions}

The general framework of the previous sections admits several distinguished reductions, as we discuss next.

\textbf{Symmetric amplitudes.}
The case of symmetric amplitudes is obtained when $A_- = A_+ = A$.  
This is a straightforward reduction of the general formalism of the previous sections, 
as all of the individual results go through without adjustment in this case. 
The only difference is simply that now the  two branch cuts $\Sigma_+$ and $\Sigma_-$ in the complex $k$-plane have the same height.

\textbf{One-sided boundary conditions.}
The general formalism also goes through in the case of one-sided BCs, namely the case $A_-=0$
(or, equivalently, $A_+=0$, due to the invariance of NLS under space reflection).
Actually, due to the Galilean and phase invariance of NLS, in this case one can also take $V=0$ and $\delta=0$ without loss of generality. 
This particular reduction was studied in \cite{PV2015,BKS2011}.
The only differences from the general formalism of the previous sections is that now $\lambda_-\equiv k$ for all $k\in\Complex$ and hence $\Sigma_- = \{0\}$.

\begin{figure}[b!]
	\centering
	\includegraphics[width=0.375\textwidth]{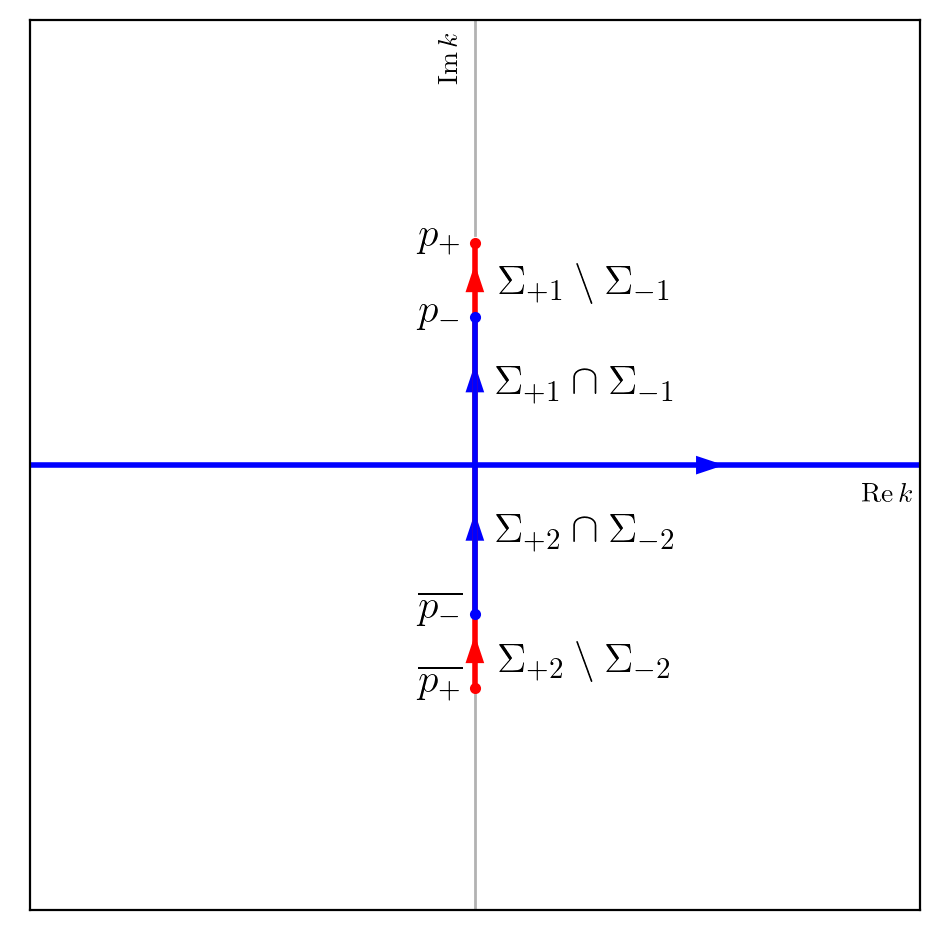}
	\caption{Continuous spectrum for $V=0$ assuming $A_- < A_+$. The blue segments indicate where all four Jost solutions are defined generically, while the red segments indicate where only three of the solutions are defined generically.}
\label{fig:spectrumzerovelocity}
\end{figure}

\textbf{No counterpropagating flows.}
 This reduction corresponds to the case of zero asymptotic velocity, i.e. $V=0$, and was studied in \cite{BK2014,BM2017} for equal amplitudes and in \cite{DPMV2014} for unequal amplitudes.
The general approach presented in the previous sections can be successfully  implemented in this case as well. However, special consideration is required because when $V=0$ the segments $\Sigma_+$ and $\Sigma_-$ are partially overlapping (cf.\ Fig.~\ref{fig:spectrumzerovelocity}) and, therefore, the domains of applicability of certain results change. In what follows, we take $A_-\leq A_+$  so that the overlapping portion of the branch cuts is given by $\Sigma_-$ and the non-overlapping portion is given by $\Sigma_+\setminus\Sigma_-$. This is done without loss of generality thanks to the reflection symmetry of NLS, i.e., the symmetry under the transformation $x\mapsto -x$.
Specifically, Corollaries~\ref{Wronskians},~\ref{R Wronskians},~\ref{c:extendedscattering} and~\ref{phi boundedness} are modified as follows:
\begin{corollary}[Analogue of Corollary~\ref{Wronskians}]
\label{Wronskians zero velocity}
	Under the hypothesis of Theorem~\ref{Jost analyticity} and for $V=0$, the scattering coefficients admit the following Wronskian representations:
	\bb
	\bse
		\label{e:Wronskians zero velocity}
		\begin{align}
			\label{s_11 zero velocity}
			s_{11}(k) &= \Wr[\phi_{-1}(x,t,k),\phi_{+2}(x,t,k)]\,, \quad k \in \R\cup\C^-\cup\Sigma_{-1}^o\setminus\{\conj{p_{+}},\conj{p_{-}}\}\,,
			\\
			\label{s_12 zero velocity}
			s_{12}(k) &= \Wr[\phi_{-2}(x,t,k),\phi_{+2}(x,t,k)]\,, \quad k \in \R\cup\Sigma_{+1}^o\cup\Sigma_{-2}^o\,,
			\\
			\label{s_21 zero velocity}
			s_{21}(k) &= \Wr[\phi_{+1}(x,t,k),\phi_{-1}(x,t,k)]\,, \quad k \in \R\cup\Sigma_{-1}^o\cup\Sigma_{+2}^o\,,
			\\
			\label{s_22 zero velocity}
			s_{22}(k) &= \Wr[\phi_{+1}(x,t,k),\phi_{-2}(x,t,k)]\,, \quad k \in \R\cup\C^+\cup\Sigma_{-2}^o\setminus\{p_{+}, p_{-}\}\,.
		\end{align}
	\ese
	\eb
Moreover, $s_{22}(k)$ and $s_{11}(k)$ are analytic in $\C^\pm \setminus \Sigma$ respectively.
\end{corollary}
\begin{corollary}[Analogue of Corollary~\ref{R Wronskians}]
\label{R Wronskians zero velocity}
	Under the hypothesis of Theorem~\ref{Jost analyticity} and for $V=0$, the left scattering coefficients can be extended through the Wrosnkian representations,
	\bb
	\bse
		\label{e:R Wronskians zero velocity}
		\begin{align}
			\label{r_11 zero velocity}
			r_{11}(k) &= \Wr[\phi_{+1}(x,t,k),\phi_{-2}(x,t,k)]\,, \quad k \in \R\cup\C^+\cup\Sigma_{-2}^o\setminus\{p_{+}, p_{-}\}\,,
			\\
			\label{r_12 zero velocity}
			r_{12}(k) &= \Wr[\phi_{+2}(x,t,k),\phi_{-2}(x,t,k)]\,, \quad k \in \R\cup\Sigma_{+1}^o\cup\Sigma_{-2}^o\,,
			\\
			\label{r_21 zero velocity}
			r_{21}(k) &= \Wr[\phi_{-1}(x,t,k),\phi_{+1}(x,t,k)]\,, \quad k \in \R\cup\Sigma_{-1}^o\cup\Sigma_{+2}^o\,,
			\\
			\label{r_22 zero velocity}
			r_{22}(k) &= \Wr[\phi_{-1}(x,t,k),\phi_{+2}(x,t,k)]\,, \quad k \in \R\cup\C^-\cup\Sigma_{-1}^o\setminus\{\conj{p_{+}}, \conj{p_{-}}\}\,.
		\end{align}
	\ese
	\eb
	Moreover, $r_{11}(k)$ and $r_{22}(k)$ are analytic in $\C^{\pm}\setminus\Sigma$ respectively.
\end{corollary}
\begin{corollary}[Analogue of Corollary~\ref{c:extendedscattering}]
\label{c:extendedscattering zero velocity}
Under the hypothesis of Theorem~\ref{Jost analyticity} and for $V=0$,
\eqref{scattering relation 1},~\eqref{scattering relation 2} and~\eqref{R scattering relation} can be extended to $k \in \Sigma_{-1}^o\cup\Sigma_{+2}^o$, $\Sigma_{+1}^o\cup\Sigma_{-2}^o$ and $\Sigma_{-}^o$ respectively.  That is:
	\bse
    \label{e:extended scattering relations zero velocity}
	\begin{align}
			\label{linear combination 1 zero velocity}
			\phi_{-1}(x,t,k) &= s_{11}(k)\phi_{+1}(x,t,k) + s_{21}(k)\phi_{+2}(x,t,k)\,, \quad k \in \R \cup \Sigma_{-1}^o\cup\Sigma_{+2}^o\,,
			\\
			\label{linear combination 3 zero velocity}
			\phi_{-2}(x,t,k) &= s_{12}(k)\phi_{+1}(x,t,k) + s_{22}(k)\phi_{+2}(x,t,k)\,, \quad k \in \R \cup \Sigma_{+1}^o\cup\Sigma_{-2}^o\,,
\\
			\label{linear combination 2 zero velocity}
			\phi_{+1}(x,t,k) &= r_{11}(k)\phi_{-1}(x,t,k) + r_{21}(k)\phi_{-2}(x,t,k)\,, \quad k \in \R \cup \Sigma_{-}^o\,,
			\\
			\label{linear combination 4 zero velocity}
			\phi_{+2}(x,t,k) &= r_{12}(k)\phi_{-1}(x,t,k) + r_{22}(k)\phi_{-2}(x,t,k)\,, \quad k \in \R \cup \Sigma_{-}^o\,.
        \end{align}
	\ese
\end{corollary}
\begin{corollary}[Analogue of Corollary~\ref{phi boundedness}]
	\label{phi boundedness zero velocity}
	Under the hypothesis of Theorem~\ref{Jost analyticity} and for $V=0$,
	\bse
		\begin{align}
				&\phi_{+1}(x,t,k)\in L_x^\infty(\R)\,, \quad k \in \R\cup\Sigma_{-}^o\,,
				\\
				&\phi_{+2}(x,t,k)\in L_x^\infty(\R)\,, \quad k \in \R\cup\Sigma_{-}^o\,,
				\\
				\label{phi_{-1} boundedness zero velocity}
				&\phi_{-1}(x,t,k)\in L_x^\infty(\R)\,, \quad k \in \R\cup\Sigma_{-1}^o\cup\Sigma_{+2}^o\,,
				\\
				\label{phi_{-2} boundedness zero velocity}
				&\phi_{-2}(x,t,k)\in L_x^\infty(\R)\,, \quad k \in \R\cup\Sigma_{+1}^o\cup\Sigma_{-2}^o\,.
		\end{align}
	\ese
\end{corollary}
\bb
Recall that the condition $(q-q_\pm)\in L_x^1(\R)$ assumed in Theorem~\ref{Jost analyticity} does not allow one to generically obtain solutions to the Lax pair at $k=\pm iA_-$ bounded as $x\to-\infty$. Consequently, these points cannot generically be included as part of the continuous spectrum. On the other hand, Lemma~\ref{l:mu branch point continuity} allows one to define $\mu_{-}(x,t,k)$ at the branch points under more strict conditions on the potential. In such cases, the columns of $\e^{-if_-(x,t)\sigma_3}\mu_{-}(x,t,k)\e^{i\theta_-(x,t,k)\sigma_3}$ (although linearly independent) are solutions to the Lax pair at $k=\pm iA_-$ bounded as $x \to -\infty$. Since $\phi_+(x,t,\pm iA_-)$ remains a fundamental matrix solution bounded as $x \to \infty$ when $A_- < A_+$, relations analogous to~\eqref{e:extended scattering relations zero velocity} show that $\e^{-if_-(x,t)\sigma_3}\mu_{-}(x,t,k)\e^{i\theta_-(x,t,k)\sigma_3}$ is a solution to the Lax pair bounded for all $x\in\R$, and the branch points $\pm iA_-$ can be included in the continuous spectrum.
\eb
Next,  Corollary~\ref{c:continuous spectrum}, Lemmas~\ref{first symmetry extended},~\ref{l:second symmetry scattering coefficients} and~\ref{nonzero scattering entries on branch cuts} are modified as follows:
\begin{corollary}[Analogue of Corollary~\ref{c:continuous spectrum}]
	Under the hypothesis of Theorem~\ref{Jost analyticity} and for $V=0$, the continuous spectrum is given by $\Sigma^o\setminus\{\pm iA_-\}$.
If, in addition,  $(q-q_-)\in L_x^{1,1}(-\infty,a)$ for some $a\in\R$ and $A_-<A_+$, then the continuous spectrum is given by $\Sigma^o$. 
\end{corollary}
\begin{lemma}[Analogue of Lemma~\ref{first symmetry extended}]
	\label{first symmetry extended zero velocity} 
	Under the hypothesis of Theorem~\ref{Jost analyticity} and for $V=0$, we have the symmetries
	\bse
		\label{e:first symmetry zero velocity}
		\begin{align}
			\label{first symmetry a extended zero velocity}
			\conj{s_{22}(\conj{k})} &= \phantom{-}s_{11}(k)\,, \quad k \in \R\cup \C^-\cup\Sigma_{-1}^o\setminus\{\conj{p_{+}}\,, \conj{p_{-}}\}\,,
			\\
			\label{first symmetry b extended zero velocity}
			\conj{s_{12}(\conj{k})} &= -s_{21}(k)\,, \quad k \in \R\cup\Sigma_{+2}^o\cup\Sigma_{-1}^o\,,
			\\
			\label{first symmetry Sigma_{-2} zero velocity}
			\conj{r_{11}(\conj{k})} &= \phantom{-}r_{22}(k)\,, \quad k \in \R\cup\C^-\cup\Sigma_{-1}^o\setminus\{\conj{p_{+}}\,, \conj{p_{-}}\}\,, 
			\\
			\label{first symmetry Sigma_{-1} zero velocity}
			\conj{r_{21}(\conj{k})} &= -r_{12}(k)\,, \quad k \in  \R \cup \Sigma_{+2}^o \cup \Sigma_{-1}^o\,.
		\end{align}
	\ese
\end{lemma}
\begin{lemma}[Analogue of Lemma~\ref{l:second symmetry scattering coefficients}]
	\label{l:second symmetry scattering coefficients zero velocity}
	Under the hypothesis of Theorem~\ref{Jost analyticity} and for $V=0$,
	\bb
	\bse
		\label{e:second symmetry scattering coefficients zero velocity}
		\begin{align}
		s_{22}^+(k) &= i\e^{-i\delta}s_{12}(k)\,, 
		&
		r_{11}^+(k) &= -i\e^{-i\delta}r_{12}(k)\,, 
		&
		&k \in \Sigma_{+1}^o\setminus\Sigma_{-1}\,,
		\\
		s_{22}^+(k) &= e^{-2i\delta}s_{11}(k)\,, 
		&
		r_{11}^+(k) &= \phantom{-}\e^{-2i\delta}r_{22}(k)\,,
		&
		&k \in \Sigma_{+1}^o\cap\Sigma_{-1}^o\,,
		\\
		s_{11}^+(k) &= i\e^{+i\delta}s_{21}(k)\,, 
		&
		r_{22}^+(k) &= -i\e^{+i\delta}r_{21}(k)\,,
		&
		&k \in \Sigma_{+2}^o\setminus\Sigma_{-2}\,,
		\\
		s_{11}^+(k) &= e^{+2i\delta}s_{22}(k)\,, 
		&
		r_{22}^+(k) &= \phantom{-}\e^{+2i\delta}r_{11}(k)\,,
		&
		&k \in \Sigma_{+2}^o\cap\Sigma_{-2}^o\,.
		\end{align}
	\ese
	\eb
\end{lemma}
\begin{lemma}[Analogue of Lemma~\ref{nonzero scattering entries on branch cuts}]
	\label{l:nonvanishing scattering coefficients zero velocity}
	Under the hypothesis of Theorem~\ref{Jost analyticity} and for $V=0$,
	\bse
		\begin{align}
			s_{11}(k) \not= 0\,, \quad r_{22}(k) \not= 0\,, \quad k \in \Sigma_{+2}^o\setminus\Sigma_{-2}\,,
			\\
			s_{12}(k) \not= 0\,, \quad r_{12}(k) \not= 0\,, \quad k \in \Sigma_{+1}^o\setminus\Sigma_{-1}\,,
			\\
			s_{21}(k) \not= 0\,, \quad r_{21}(k) \not= 0\,, \quad k \in \Sigma_{+2}^o\setminus\Sigma_{-2}\,,
			\\
			s_{22}(k) \not= 0\,, \quad r_{11}(k) \not= 0\,, \quad k \in \Sigma_{+1}^o\setminus\Sigma_{-1}\,.
		\end{align}
	\ese
\end{lemma}
Note, however, that no statement can be made about the possibility of zeros of the scattering coefficients on $\Sigma_+\cap\Sigma_-$.
The  above  results   can be proved exactly as their counterparts for $V\not=0$.

\bb
If $A_- \not= A_+$, then the behavior of the scattering coefficients follows exactly as in the case of $V\not=0$ treated in Section~\ref{ss:S(k) branch point behavior}. On the other hand, if $A_- = A_+$ then the branch points come together and Corollaries~\ref{wr branch point continuity}, \ref{c:reflection coefficients branch point behavior generic case}, \ref{wr branch point expansion} and~\ref{c:reflection coefficients branch point behavior exceptional case}
 must be adjusted accordingly.

\begin{corollary}[Analogue of Corollary~\ref{wr branch point continuity} when $A_-=A_+$]
\label{wr branch point continuity zero velocity} 
	Under the hypothesis of Lemma~\ref{l:mu branch point continuity} and for $V=0$ and $A_+ = A_- = A$,
	\bse
	\begin{align}
	  s_{11}(k) = b_{\pm 11}^{(0)}(k\mp iA)^{-1/2} + o(k\mp iA)^{-1/2}\,, \quad k \rightarrow \pm iA\,,
		\\
		s_{12}(k) = b_{\pm 12}^{(0)}(k\mp iA)^{-1/2} + o(k\mp iA)^{-1/2}\,, \quad k \rightarrow \pm iA\,,
		\\
		s_{21}(k) = b_{\pm 21}^{(0)}(k\mp iA)^{-1/2} + o(k\mp iA)^{-1/2}\,, \quad k \rightarrow \pm iA\,,
		\\
		s_{22}(k) = b_{\pm 22}^{(0)}(k\mp iA)^{-1/2} + o(k\mp iA)^{-1/2}\,, \quad k \rightarrow \pm iA\,,
	\end{align}
	\ese
	for some constants $b_{\pm 11}^{(0)}$, $b_{\pm 12}^{(0)}$, $b_{\pm 21}^{(0)}$, $b_{\pm 22}^{(0)}$, where each limit must be taken from within the regions described by Corollary~\ref{Wronskians zero velocity}.
\end{corollary}

\begin{corollary}[Analogue of Corollary~\ref{c:reflection coefficients branch point behavior generic case} when $A_-=A_+$]
\label{c:reflection coefficients branch point behavior generic case zero velocity}
Under the hypothesis of Lemma~\ref{l:mu branch point continuity} and for $V=0$ and $A_-=A_+=A$, in the generic case that $\mu_{+1}(x,t,k)$ and $\mu_{-2}(x,t,k)$ are linearly independent at the branch point $iA$, we have
\bse
\begin{align}
	&\qquad & \rho(k) &= \rho_o + o(1)\,, & & k \rightarrow iA\,, & &\qquad\\
	&\qquad & r(k) &= r_o(k-iA) + o(k-iA)\,, & & k \rightarrow iA\,, & &\qquad
\end{align}
\ese
for some nonzero constants $\rho_o\,, r_o$ where the limits must be taken from within the regions described by~\eqref{reflection coefficients}.
\end{corollary}

\begin{corollary}[Analog of Corollary~\ref{wr branch point expansion} when $A_-=A_+$]
\label{wr branch point expansion zero velocity} 
	Under the hypothesis of Lemma~\ref{l:mu branch point expansion} and for $V=0$ and $A_-=A_+=A$,
	\bse
	\begin{align}
	  s_{11}(k) = b_{\pm 11}^{(0)}(k\mp iA)^{-1/2} + b_{\pm 11}^{(1)} + o(1)\,, \quad k \rightarrow \pm iA\,,
		\\
		s_{12}(k) = b_{\pm 12}^{(0)}(k\mp iA)^{-1/2} + b_{\pm 12}^{(1)} + o(1)\,, \quad k \rightarrow \pm iA\,,
		\\
		s_{21}(k) = b_{\pm 21}^{(0)}(k\mp iA)^{-1/2} + b_{\pm 21}^{(1)} + o(1)\,, \quad k \rightarrow \pm iA\,,
		\\
		s_{22}(k) = b_{\pm 22}^{(0)}(k\mp iA)^{-1/2} + b_{\pm 22}^{(1)} + o(1)\,, \quad k \rightarrow \pm iA\,,
	\end{align}
	\ese
	for some constants $b_{\pm 11}^{(0)}$, $b_{\pm 12}^{(0)}$, $b_{\pm 21}^{(0)}$, $b_{\pm 22}^{(0)}$, $b_{\pm 11}^{(1)}$, $b_{\pm 12}^{(1)}$, $b_{\pm 21}^{(1)}$, $b_{\pm 22}^{(1)}$, where each limit must be taken from within the regions described by Corollary~\ref{Wronskians zero velocity}.
\end{corollary}

\begin{corollary}[Analogue of Corollary~\ref{c:reflection coefficients branch point behavior exceptional case} when $A_-=A_+$]
\label{c:reflection coefficients branch point behavior exceptional case zero velocity}
Under the hypothesis of Lemma~\ref{l:mu branch point continuity} and for $V=0$ and $A_-=A_+=A$, in the exceptional case that $\mu_{+1}(x,t,k)$ and $\mu_{-2}(x,t,k)$ are linearly dependent at the branch point $iA$ with $b_{\pm 11}^{(1)}\,$, $b_{\pm 12}^{(1)}\,$, $b_{\pm 21}^{(1)}\,$, $b_{\pm 22}^{(1)}$ all nonzero, we have 
\bse
\begin{align}
	&\qquad & \rho(k) &= \rho_o + o(1)\,, & & k \rightarrow iA\,, & &\qquad\\
	&\qquad & r(k) &= r_o + o(1)\,, & & k \rightarrow iA\,, & &\qquad
\end{align}
\ese
for some nonzero constants $\rho_o\,, r_o$ where the limits must be taken from within the regions described by~\eqref{reflection coefficients}.
\end{corollary}

Lemma~\ref{l:jump matrix} is also adjusted as follows:
\eb
\begin{lemma}[Analogue of Lemma~\ref{l:jump matrix}]
\label{l:jump matrix zero velocity}
Under the hypothesis of Theorem~\ref{Jost analyticity} and for $V=0$, the matrix $M(x,t,k)$ defined by~\eqref{M definition} satisfies the jump condition
\begin{equation}
		M^+(x,t,k) = M^-(x,t,k)\,J(x,t,k)\,,\quad k\in\Sigma^o\,, 
\end{equation}
where

$J(x,t,k)$ is still given by~\eqref{e:JfromJo}  but now with
\begin{align}
		\label{zero velocity jump matrix}
J_o(k) = 
		\begin{cases} 
			\begin{pmatrix}
				1 & \rho\\ 
				\conj{\rho} & 1+\rho(k)\conj{\rho(\conj{k})}
			\end{pmatrix}, 
			&k \in \R\,,
			\\[15pt]
			\begin{pmatrix}
				i\rho(k)\e^{-i\delta} & 0\\
				-i\e^{-i\delta} & -i\e^{i\delta}/\rho(k)
			\end{pmatrix}, 
			&k \in \Sigma_{+1}^o\setminus \Sigma_{-1}\,,
			\\[15pt]
			\begin{pmatrix}
				i\e^{-i\delta}/\conj{\rho(\conj{k})} & -i\e^{i\delta}\\
				0 & -i\conj{\rho(\conj{k})}\e^{i\delta}
			\end{pmatrix}, 
			&k \in \Sigma_{+2}^o\setminus \Sigma_{-2}\,,
			\\[15pt]
			\begin{pmatrix}
				i\rho(k)\e^{-i\delta} & -i(1+\rho(k)\conj{\rho(\conj{k})})\e^{i\delta}\\
				-i\e^{-i\delta} & i\conj{\rho(\conj{k})}\e^{i\delta}
			\end{pmatrix}, 
			&k \in \Sigma_{+1}^o \cap \Sigma_{-1}^o\,,
			\\[15pt]
			\begin{pmatrix}
				-i\rho(k)\e^{-i\delta} & -i\e^{i\delta}\\
				-i(1+\rho(k)\conj{\rho(\conj{k})})\e^{-i\delta} & -i\conj{\rho(\conj{k})}\e^{i\delta}
			\end{pmatrix}, 
			&k \in \Sigma_{+2}^o \cap \Sigma_{-2}^o\,.
		\end{cases}
\end{align}
\end{lemma}
The calculation of the jump matrix $J_o(k)$  in Lemma~\ref{l:jump matrix zero velocity} is carried out in section~\ref{Zero velocity appendix}.
As a consequence of Lemma~\ref{l:nonvanishing scattering coefficients zero velocity}, the entries of the jump matrix are non-singular
\bb
$\Sigma_+^o\setminus\Sigma_-^o$, potentially with unbounded growth toward the branch points. 
Corollaries~\ref{c:reflection coefficients branch point behavior generic case} and~\ref{c:reflection coefficients branch point behavior exceptional case} describe some of the possible behaviors of the jump matrix at these potential singularities when $A_-\not=A_+$, while Corollaries~\ref{c:reflection coefficients branch point behavior generic case zero velocity} and~\ref{c:reflection coefficients branch point behavior exceptional case zero velocity} treat the analogous scenarios for $A_-=A_+$. In particular, in the generic case we see that $J_o(k)$ is continuous at the branch points.
\eb
Also notice that the jumps for $V=0$ on $\R$, $\Sigma_{+1}^o\setminus\Sigma_{-1}$ and $\Sigma_{+2}^o\setminus\Sigma_{-2}$ as given in~\eqref{zero velocity jump matrix} respectively match the jumps for $V\not=0$ on $\R$, $\Sigma_{+1}^o$ and $\Sigma_{+2}^o$ as given in~\eqref{nonzero velocity jump matrix}.

We have broken the $x \mapsto -x$ symmetry twice to arrive at the  jump matrix $J_o(k)$: once in the definition of $M(x,t,k)$ by~\eqref{M definition}, which rescaled the columns of $\phi_-(x,t,k)$ using the analytic scattering coefficients from the right, $s_{11}(k)$ and $s_{22}(k)$, and a second time by taking $A_-\leq A_+$. If instead one takes $A_- \geq A_+$, then the overlapping portion of the branch cuts is given by $\Sigma_{+}$ while the non-overlapping portion is given by $\Sigma_-\setminus\Sigma_+$. In that case, the jumps for $V=0$ on $\R$, $\Sigma_{-1}^o\setminus\Sigma_{+1}$ and $\Sigma_{-2}^o\setminus\Sigma_{+2}$ respectively match the jumps for $V=0$ on $\R$, $\Sigma_{-1}^o$ and $\Sigma_{-2}^o$ in~\eqref{nonzero velocity jump matrix}, while the jumps on the overlap $\Sigma_-\cap\Sigma_+$ match the corresponding jumps in~\eqref{zero velocity jump matrix}. For $V=0$, we make the choice $A_-\leq A_+$ so that all jumps of $M(x,t,k)$ as defined in~\eqref{M definition} can be expressed in terms of $\rho(k)$.

If $A_- \not= A_+$, then the growth conditions follow exactly as in the case of $V\not=0$   treated in Section~\ref{Growth conditions} (cf. Lemmas~\ref{nonzero velocity growth conditions generic case} and~\ref{nonzero velocity growth conditions exceptional case}). 
\bb
On the other hand, if $A_- = A_+$ then, as remarked above, the branch points come together and Lemmas~\ref{nonzero velocity growth conditions generic case} and~\ref{nonzero velocity growth conditions exceptional case} must also be adjusted accordingly:
\eb
\begin{lemma}[Analogue of Lemma~\ref{nonzero velocity growth conditions generic case} when $A_-=A_+$]
\label{zero velocity growth conditions generic case}
	Under the hypothesis of Lemma~\ref{l:mu branch point continuity} and for $V= 0$ and $A_-=A_+=A$, 
	\bb
	in the generic case that $\mu_{+1}(x,t,k)$, $\mu_{-2}(x,t,k)$ are linearly independent at the branch point $iA$,
	\eb
	we have
	\bb
	\[
		M(x,t,k) = \big[ B_{\pm iA}^{(0)}(x,t) + o(1)\big](k\mp iA)^{\mp\sigma_3/4}\,, \quad k \to \pm iA\,,
	\]
	for some invertible matrices $B_{\pm iA}^{(0)}(x,t)$.
	\eb
\end{lemma}
\begin{lemma}[Analogue of Lemma~\ref{nonzero velocity growth conditions exceptional case} when $A_-=A_+$]
\label{zero velocity growth conditions exceptional case}
	Under the hypothesis of Lemma~\ref{l:mu branch point expansion}  and for $V = 0$ and $A_-=A_+=A$, 
	\bb
	in the exceptional case that $\mu_{+1}(x,t,k)$, $\mu_{-2}(x,t,k)$ are linearly dependent at $k = iA$ with $b_{-11}^{(1)}$ and $b_{+22}^{(1)}$ (as given by Corollary~\ref{wr branch point expansion zero velocity}) nonzero,
	we have 
	\[
		\label{e:zero velocity growth conditions exceptional case}
		M(x,t,k) = \big[B_{\pm iA}^{(0)}(x,t) + B_{\pm iA}^{(1)}(x,t)(k\mp iA)^{1/2} + o(k\mp iA)^{1/2}\big](k\mp iA)^{-1/4}\,, \quad k\to \pm iA\,,
	\]
	for some matrices $B_{\pm iA}^{(0)}(x,t)\,$, $B_{\pm iA}^{(1)}(x,t)$ with $\det B_{\pm iA}^{(0)}(x,t) = 0$.
	\eb
\end{lemma}
\bb
As a result of Lemma~\ref{zero velocity growth conditions generic case}, we see that, in the generic case, the following limits exist:
\be
	\lim_{k\to\pm iA} M(x,t,k)(k\mp iA)^{\pm \sigma_3/4}\,.
	\label{e:zero velocity growth conditions generic case}
\ee

Finally, Theorem~\ref{t:direct RHP solution} is adjusted as follows:
\begin{theorem}[Analogue of Theorem~\ref{t:direct RHP solution}]
\label{t:direct RHP solution zero velocity}
For $V=0$ and $A_-<A_+$, if $(q-q_\pm)\in L^{1,1}_x(\R)$ and $(q-q_\pm)_x \in L^1_x(\R)$ for all $t \in \R$ with $\mu_{+1}(x,t,k)$ and $\mu_{-2}(x,t,k)$ linearly independent at both branch points $p_\pm$, then $M(x,t,k)$ as given by~\eqref{M definition} satisfies the modified Riemann-Hilbert problem~\ref{nonzero velocity RHP} with $J(x,t,k)$ given by~\eqref{zero velocity jump matrix}.
\end{theorem}

\begin{theorem}[Analogue of Theorem~\ref{t:direct RHP solution}]
\label{t:direct RHP solution zero velocity equal amplitube}
For $V=0$ and $A_-=A_+=A$, if $(q-q_\pm)\in L^{1,1}_x(\R)$ and $(q-q_\pm)_x \in L^1_x(\R)$ for all $t \in \R$ with $\mu_{+1}(x,t,k)$ and $\mu_{-2}(x,t,k)$ linearly independent at the branch point $iA$, then $M(x,t,k)$ as given by~\eqref{M definition} satisfies the modified Riemann-Hilbert problem~\ref{nonzero velocity RHP} with $J(x,t,k)$ given by~\eqref{zero velocity jump matrix} and growth conditions given by~\eqref{e:zero velocity growth conditions generic case}.
\end{theorem}
\eb

The formal expressions for the system of algebraic-integral equations in Theorem~\ref{t:integral equation for M} and reconstruction formula in Corollary~\ref{c:reconstruction formula} are unchanged for $V=0$. 
\bb
Also, the work in Section~\ref{ss:Alternative RHP} is entirely unchanged.
\eb

\section{Riemann problems}
\label{Riemann problems}

\bb
Recall that Riemann problems are initial-value problems with step-like ICs \cite{L1973,CF1976,B2018}. We now compute explicitly the scattering data and growth conditions for the various Riemann problems described by the general framework of this paper.
\eb
Proofs for all the results in this section are given in section~\ref{a:Riemann problems}.

\subsection{Riemann problem for a pure two-sided step with counterpropagating flows}

Consider the IC 
\[\label{e:ic-rp1}
	q(x,0) = 
\begin{cases}
		A_+\e^{-iVx+i\delta} \,, &x\geq 0\,, \\
		A_-\e^{+iVx-i\delta} \,, 				&x<0\,, 
\end{cases}
\qquad 0 < A_- \leq A_+\,.\]
The special case $A_- = A_+$, $\delta=0$ and $V>0$ for this problem was considered in \cite{BV2007}. 
Note, however, that the normalization for the Jost solutions and the sectionally meromorphic matrices is different in this work.


Explicitly, at $t=0$ we have
\bb
\bse\label{e:jost-rp1}
\label{Pure two-sided step IC, Vnot=0 Jost solutions}
\begin{equation}
	\phi_+(x,0,k) = \frac{1}{d_+(k)}
		\begin{cases}
			\e^{-iVx\sigma_3/2}E_+(k)\e^{i\lambda_+ x\sigma_3}\,, &x\geq 0\,,\\
			\e^{+iVx\sigma_3/2}E_-(k)\e^{i\lambda_- x\sigma_3}E_-^{-1}(k)E_+(k)\,, &x<0\,,
		\end{cases}
\end{equation}
and \begin{equation}
		\phi_-(x,0,k) = \frac{1}{d_-(k)}
		\begin{cases}
			\e^{-iVx\sigma_3/2}E_+(k)\e^{i\lambda_+ x\sigma_3}E_+^{-1}(k)E_-(k)\,, &x\geq 0\,,\\
			\e^{+iVx\sigma_3/2}E_-(k)\e^{i\lambda_- x\sigma_3}\,, &x<0\,.
		\end{cases}
\end{equation}
\ese
%
%
Then, $\phi_-(x,0,k) = \phi_+(x,0,k)S(k)$ with $S(k) = (d_+(k)/d_-(k))E_+^{-1}(k)E_-(k).$
Hence, we find
\begin{gather}
\bse
\begin{aligned}
\label{e:scat-rp1}
\hspace*{-2mm} s_{22}(k) &= \frac{1}{d_+(k)d_-(k)(\lambda_- + (k-V/2))A_+\e^{i\delta}}\big[(\lambda_+ - (k+V/2))A_-\e^{-i\delta} + (\lambda_- + (k-V/2))A_+\e^{i\delta}\big]\,,
		\\
		\rho(k) &= \frac{iA_+\e^{i\delta}}{\lambda_+ + (k+V/2)}\phantom{\cdot}\frac{(\lambda_+ + (k+V/2))A_-\e^{-i\delta} - (\lambda_- + (k-V/2))A_+\e^{i\delta}}{(\lambda_+ - (k+V/2))A_-\e^{-i\delta} + (\lambda_- + (k-V/2))A_+\e^{i\delta}}\,,
		\\
		r(k) &= \frac{4i\lambda_+\lambda_-\e^{i\delta}}{[ (\lambda_+ + (k+V/2))\e^{2i\delta} - (\lambda_+ - (k+V/2))\e^{-2i\delta} ]A_- - 2(k-V/2)A_+}\,.
\end{aligned}
\ese
\end{gather}
\eb
\begin{lemma}\label{Pure two-sided step IC, Vnot=0 zeros}
	For the pure two-sided step initial condition with $V\not=0$ and $0<A_- \leq A_+$, there are no discrete eigenvalues. Furthermore, if $\delta=0$ then
\begin{itemize}
\advance\itemsep -4pt
		\item[(i)] 
			If $A_+ = A_-$ then
			$\rho(k)$ has no zeros
			and $r(k)$ has no poles.
		\item[(ii)]
			If $A_+ \not= A_-$ then
			$\rho(k)$ has a zero and $r(k)$ has a pole only at
			\be
				k = \frac{V}{2}\Big(\frac{A_+ + A_-}{A_+ - A_-}\Big)\,.
			\ee
\end{itemize}
\end{lemma}
%

In section~\ref{Pure two-sided step IC, Vnot=0 appendix}, we show that the 
\bb
modified eigenfunctions $\mu_{+1}(x,t,k)$ and $\mu_{-2}(x,t,k)$
\eb
are linearly independent at the branch points, so that the growth conditions are given by Lemma~\ref{nonzero velocity growth conditions generic case}. The matrix $M(x,t,k)$ as defined in~\eqref{M definition} then satisfies the RHP~\ref{nonzero velocity RHP} 
with $K=\emptyset$
and with $\rho(k)$ and $r(k)$ given by~\eqref{e:scat-rp1}.

\subsection{Riemann problem for a pure two-sided step without counterpropagating flows}

Consider now the IC %
\[\label{e:ic-rp2}
	q(x,0) = 
\begin{cases}
		A_+\e^{+i\delta} \,, &x\geq 0\,, \\
		A_-\e^{-i\delta} \,, 				&x<0\,,
\end{cases}
\qquad 0 < A_- \leq A_+\,,\]
which corresponds to~\eqref{e:ic-rp1} with $V=0$. Much of the work from the preceding section is valid with $V=0$. In particular, the explicit Jost solutions $\phi_\pm(x,t,k)$ at $t=0$ and the scattering coefficients are given again by  expressions~\eqref{e:jost-rp1} and~\eqref{e:scat-rp1}, now with $V=0$. Specifically, the reflection coefficient $\rho(k)$ is given by
\[
	\rho(k) = \frac{iA_+\e^{i\delta}}{\lambda_+ + k}\phantom{\cdot}\frac{(\lambda_+ + k)A_-\e^{-i\delta} - (\lambda_- + k)A_+\e^{i\delta}}{(\lambda_+ - k)A_-\e^{-i\delta} + (\lambda_- + k)A_+\e^{i\delta}}\,.
\] %
\begin{lemma}
\label{Pure two-sided step IC, V=0 zeros}
	For the pure two-sided step initial condition with $0<A_-\leq A_+$\,,  $\delta=0$ and $V=0$, there are no discrete eigenvalues. Furthermore, 
	\begin{itemize}
\advance\itemsep -4pt
		\item[(i)] 
			If $A_+ = A_-$ then $\rho(k) \equiv 0$.
		\item[(ii)]
			If $A_+ \not= A_-$ then $\rho(k)$ has a zero only at $k = 0$.
\end{itemize}
\end{lemma}
%
%
In section~\ref{Pure two-sided step IC, V=0 appendix}, we show that for $A_+ \not= A_-$ and $V=0$ the 
\bb
modified eigenfunctions $\mu_{+1}(x,t,k)$ and $\mu_{-2}(x,t,k)$ 
\eb
are linearly independent at the branch points, so that the growth conditions are given by Lemma~\ref{nonzero velocity growth conditions generic case}. Then, the matrix $M(x,t,k)$ defined by~\eqref{M definition}   satisfies a modified RHP~\ref{nonzero velocity RHP} 
with $K=\emptyset$, $J(x,t,k)$ given by~\eqref{zero velocity jump matrix},
and   $\rho(k)$ and $r(k)$ given by~\eqref{e:scat-rp1}.

On the other hand, if $A_+ = A_- = A$, $V=0$ and $\delta=0$, then the 
\bb
modified eigenfunctions are linearly dependent at the branch points and $s_{11}(k) \equiv s_{22}(k) \equiv 1$, so that the growth conditions are instead given by~\eqref{e:zero velocity growth conditions exceptional case}.
\eb

\subsection{Riemann problem for a pure one-sided step}
%
Finally, consider the IC %
\[
	q(x,0) = 
\begin{cases}
		A\e^{-iVx+i\delta} \,, &x\geq 0\,, \\
		0 \,, 				&x<0\,,
\end{cases}
\qquad A>0\,. \]
To fit the previous framework, we take $A_+ = A$ and $A_- = 0$. Due to the Galilean and phase invariances of NLS, here we can actually assume $V=\delta=0$. Thus, much of the work for the two-sided Riemann problem~\eqref{e:ic-rp2} can be reused after setting $A_+ = A$, $A_-=0$ and $\delta=0$.
%
%
Note that with $A_-=0$ and $V=0$ we have $\lambda_- = k$. 

Explicitly, at $t=0$ we have
\bb
\bse
\label{e:jost rp-3}
\begin{equation}
		\phi_+(x,0,k) = 
		\frac{1}{d(k)}\begin{cases}
			E(k)\e^{i\lambda x\sigma_3}\,, &x\geq 0\,,\\
			\e^{ikx\sigma_3}E(k)\,, &x<0\,,
		\end{cases}
\end{equation}
and \begin{equation}
		\phi_-(x,0,k) = 
		\begin{cases}
			E(k)\e^{i\lambda x\sigma_3}E^{-1}(k)\,, &x\geq 0\,,\\
			\e^{ikx\sigma_3}\,, &x<0\,,
		\end{cases}
\end{equation}
\ese
\eb
where $E(k) = E_+(k)$, 
\bb
$d(k) = d_+(k)$
\eb
with $V=\delta=0$.
%
%
Then $\phi_-(x,0,k) = \phi_+(x,0,k)S(k)$ with 
\bb
$S(k) = d(k)E^{-1}(k)$.
\eb
We then find
\bb
\be
\begin{aligned}\label{Pure one-sided step IC rho}
		s_{22}(k) &= \frac{1}{d(k)}\,,
		&\quad
		\rho(k) &= -\frac{iA}{\lambda + k}\,.
\end{aligned}
\ee
\eb
We see that there are no discrete eigenvalues, $\rho(k)$ has no singularities, and  $\rho(k)\not= 0$ for any $k \in \C$.
%
\bb
Moreover, we easily see that the growth conditions are given by Lemma~\ref{nonzero velocity growth conditions generic case} (ignoring those on $p_-$ and $\conj{p_-}$).
\eb
%

Then, $M(x,t,k)$ satisfies a modified version of RHP~\ref{nonzero velocity RHP} with $K = \emptyset$ and $J(x,t,k)$ given by
\be
J(x,t,k) = 
\begin{cases}
		\begin{pmatrix}
			1 & \rho(k)\e^{2i\theta_o(x,t,k)}\\ 
			-\rho(k)\e^{-2i\theta_o(x,t,k)} & -\frac{2\lambda}{iA}\rho
		\end{pmatrix}, &k \in \R\,,
		\\[15pt]
		\begin{pmatrix}
			i\rho(k) & 0\\
			-i\e^{-2i\theta_o(x,t,k)} & -i/\rho(k)
		\end{pmatrix}, &k \in \Sigma_{+1}^o\,,
		\\[15pt]
		\begin{pmatrix}
			-i/\rho(k) & -i\e^{2i\theta_o(x,t,k)}\\
			0 & i\rho(k)
		\end{pmatrix}, &k \in \Sigma_{+2}^o\,,
  \end{cases}
\ee
with $\rho(k)$ given by~\eqref{Pure one-sided step IC rho}.

\section{Proofs}
\label{s:proofs}

In this Section, we include proofs and calculations for the various theorems, lemmas and corollaries stated in the previous sections.

\subsection{Direct problem}
\label{a:Direct Problem}
%
\paragraph{Jost solutions for the exact potentials $q_\pm(x,t)$.}
\label{Jost solutions for the exact potentials q_pm(x,t)}
We begin by obtaining solutions $\~\psi_\pm(x,t,k)$ to the first part of the Lax pair~\eqref{e:LP}.
Writing
\bse
\be\label{Xhat_pm(k)}
		X_\pm(x,t,k) = \e^{-if_\pm(x,t)\sigma_3}\^X_\pm(k)\e^{if_\pm(x,t)\sigma_3}\,,
\ee
with\be
\^X_\pm(k) = ik\sigma_3 + A_\pm\sigma_3\e^{\pm i\delta}\sigma_1\,,
\ee
\ese
we can write the first part of the Lax pair equivalently as
\be
\begin{aligned}
		( \e^{if_\pm(x,t)\sigma_3}\~\psi_\pm(x,t,k) )_x = \^X_\pm(k \pm V/2) \e^{if_\pm(x,t)\sigma_3}\~\psi_\pm(x,t,k)\,.
\end{aligned}
\ee
Now $\^X_\pm(k\pm V/2)$ has eigenvector and eigenvalue matrices $E_\pm(k)$ and $i\lambda(k;A_\pm)\sigma_3$, with $E_\pm(k)$ and $\lambda(k;A)$ as defined in~\eqref{e:E_pm(k)} and~\eqref{e:lambda(k;A)} respectively. Then
\be
\begin{aligned}
		\~\psi_\pm(x,t,k) = \e^{-if_\pm(x,t)\sigma_3}E_\pm(k)\e^{i\lambda_\pm(k) x\sigma_3}\,,
\end{aligned}
\ee
is a fundamental matrix solution to the first part of the Lax pair~\eqref{e:LP}.
We now seek simultaneous solutions $\~\phi_\pm(x,t,k)$ of both parts of~\eqref{e:LP}.
To this end, note that, since $\~\phi_\pm(x,t,k)$ and $\~\psi_\pm(x,t,k)$ are both solutions to the first part of~\eqref{e:LP}, we have
\[
	\~\phi_\pm(x,t,k) = \~\psi_\pm(x,t,k)B_\pm(t,k)\,,
\]
for some matrix $B_\pm(t,k)$. Differentiating with respect to $t$, we find
\[
	\frac{dB_\pm(t,k)}{dt} = \~\psi_\pm^{-1} ( T_\pm \~\psi_\pm - ( \~\psi_\pm )_t )B_\pm(t,k) = -2i\lambda_\pm(k) (k \mp V/2)\sigma_3 B_\pm(t,k)\,,
\nonumber
\]
(where we suppressed the dependence of $\~\psi$ and $T_\pm$ for brevity), so that
\[
	B_\pm(t,k) = \e^{-2i\lambda_\pm(k) (k\mp V/2)t\sigma_3}B_\pm(0,k)\,.
\]
Taking $B_\pm(0,k) = \I$ gives the simultaneous fundamental matrix solutions
\[
	\~\phi_\pm(x,t,k) = \e^{-if_\pm(x,t)\sigma_3} E_\pm(k) \e^{i\theta_\pm(x,t,k)\sigma_3}\,.
\]
%
\paragraph{Proof of Lemma~\ref{l:lambda symmetries} (Branch cut for $\lambda_\pm(k)$).}
\label{lambda symmetries proof}
The analyticity properties of $\lambda(k;A)$ together with the facts that $\lambda(k;A) = k + O(1/k)$ as $k\rightarrow \infty$ and $\lambda(k;A) \in \R$ exactly when $k \in \R \cup i[-A,A]$ while $\lambda(k;A) \in i\R$ exactly when $k \in i\R \setminus i(-A,A)$ establish~\eqref{e:Im lambda} and~\eqref{Re lambda}.
Let $k-iA = r_1\e^{i\varphi_1}$ and $k+iA = r_2\e^{i\varphi_2}$, with $-\pi/2 \leq \varphi_1, \varphi_2 < 3\pi/2$. 
Then $\conj{k}-iA = \conj{k+iA} = r_2\e^{i\~\varphi_1}$ and $\conj{k}+iA = \conj{k-iA} = r_1\e^{i\~\varphi_2}$, with
\begin{equation}
		\~\varphi_1 = 
		\begin{cases}
			\phantom{2\pi}-\varphi_2\,, &-\pi/2 \leq \varphi_2 \leq \phantom{3}\pi/2\,,\\
			2\pi - \varphi_2\,, &\phantom{-}\pi/2 < \varphi_2 < 3\pi/2\,,
		\end{cases}
\nonumber
\end{equation}
and \begin{equation}
		\~\varphi_2 = 
		\begin{cases}
			\phantom{2\pi}-\varphi_1\,, &-\pi/2 \leq \varphi_1 \leq \phantom{3}\pi/2\,,\\
			2\pi - \varphi_1\,, &\phantom{-}\pi/2 < \varphi_1 < 3\pi/2\,,
		\end{cases}
\nonumber
\end{equation}
so that $-\pi/2 \leq \~\varphi_1, \~\varphi_2 < 3\pi/2$. 
If $-\pi/2\leq \varphi_1,\varphi_2 \leq \pi/2$, then 
\[
\lambda(\conj{k};A) = \sqrt{r_1 r_2} \e^{i(-\varphi_2 - \varphi_1)/2} = \conj{\lambda(k;A)}.
\]
On the other hand, 
if $\pi/2 \leq \varphi_1,\varphi_2 < 3\pi/2$, then 
\[
\lambda(\conj{k};A) = \sqrt{r_1 r_2}\e^{i(-\varphi_2 -\varphi_1 + 4\pi)/2} = \conj{\lambda(k;A)}.
\]
The remaining two cases never occur, so~\eqref{lambda conjugation symmetry} is proved.
Also, $(-k)-iA = -(k+iA) = r_2\e^{i\^\varphi_1}$ and $(-k)+iA = -(k-iA) = r_1\e^{i\^\varphi_2}$, with 
\begin{equation}
		\^\varphi_1 = 
		\begin{cases}
			\varphi_2 + \pi\,, &-\pi/2 \leq \varphi_2 < \phantom{3}\pi/2\,,\\
			\varphi_2 - \pi\,, &\phantom{-}\pi/2 \leq \varphi_2 < 3\pi/2\,,
		\end{cases}
\nonumber
\end{equation}
and \begin{equation}
		\^\varphi_2 = 
		\begin{cases}
			\varphi_1 + \pi\,, &-\pi/2 \leq \varphi_1 < \phantom{3}\pi/2\,,\\
			\varphi_1 - \pi\,, &\phantom{-}\pi/2 \leq \varphi_1 < 3\pi/2\,,
		\end{cases}
\nonumber
\end{equation}
so that $-\pi/2 \leq \^\varphi_1, \^\varphi_2 < 3\pi/2$.
If $-\pi/2 \leq \varphi_1, \varphi_2 < \pi/2$, then 
\[
\lambda(-k;A) = \sqrt{r_1 r_2}\e^{i(\varphi_1 + \varphi_2 + 2\pi)/2} = -\lambda(k;A).
\]
On the other hand, 
if $\pi/2 \leq \varphi_1, \varphi_2 < 3\pi/2$, then 
\[
\lambda(-k;A) = \sqrt{r_1, r_2}\e^{i(\varphi_1 + \varphi_2 - 2\pi)/2} = -\lambda(k;A).
\]
The case of $-\pi/2 \leq \varphi_1 < \pi/2$ and $\pi/2 \leq \varphi_2 < 3\pi/2$ can only happen when $\varphi_1 =-\pi/2$ and $\varphi_2 = \pi/2$ so that $k \in i(-A,A)$. In such case, 
\[
\lambda(-k;A) = \sqrt{r_1 r_2}\e^{i(-\pi/2 + \pi + \pi/2 - \pi)/2} = \sqrt{r_1 r_2}\e^{i(-\pi/2 + \pi/2)/2} = \lambda(k;A).
\]
The remaining case cannot occur, so~\eqref{lambda negation symmetry} is proved. Combining~\eqref{lambda conjugation symmetry} and~\eqref{lambda negation symmetry} and recalling that $\lambda(k;A) \in \R$ for $k \in i[-A,A]$ gives~\eqref{lambda jump symmetry}.
\endproof
%
\paragraph{Proof of Lemma~\ref{l:d_pm properties} (Properties of $d_\pm(k)$).}
\label{d_pm existence proof}
\bb
Here we compute the jump of $d_\pm(k)$ across the branch cut $\Sigma_\pm$.
\eb
To simplify the argument, we first explicity define $( D(k;A) )^{1/2}$, where $D(k;A) = \frac{2\lambda(k;A)}{\lambda(k;A) + k}$ with $\lambda(k;A)$ as defined in~\eqref{e:lambda(k;A)}.

We claim that $\Re D(k;A) >0$ except where $\lambda(k;A) = 0$, in which case $D(k;A) = 0$ (i.e. at the branch points $k = \pm iA$). Indeed,
\[
	D(k;A) = \frac{2\lambda}{\lambda + k} = \frac{2\lambda}{|\lambda + k|^2} ( \conj{\lambda} + \conj{k} )\,,
\nonumber
\]
so that
\[
	\Re D(k;A) = \frac{2}{|\lambda + k|^2} \left(|\lambda|^2 + \lambda_{\re}k_{\re} + \lambda_{\im}k_{\im}\right).
\nonumber
\]
Lemma~\ref{l:lambda symmetries} shows that $\lambda_{\re}$ and $k_{\re}$ have the same sign, as do $\lambda_{\im}$ and $k_{\im}$. This then proves the claim.

Next, note that, similar to~\eqref{D symmetry},
\[
	D^+(k;A) =  \frac{4\lambda^2}{A^2} \frac{1}{D(k;A)}, \quad k \in i(-A,A)\,.
\nonumber
\]
Since $\lambda >0$ on $i(-A,A)$, then using $\sqrt{\cdot}$ defined as the principal square root with branch cut along $\{0\}\cup\R^-$,
\[
	\sqrt{D^+(k;A)} = \frac{2\lambda}{A}\frac{1}{\sqrt{D(k;A)}}\,.
\nonumber
\]
Since $\Re D(k+\epsilon;A) >0$ for $k$ in $i(-A,A)$ and any $\epsilon>0$, then
\[
	\sqrt{D(k;A)}^+ = \sqrt{D^+(k;A)} = \frac{2\lambda}{A}\frac{1}{\sqrt{D(k;A)}}\,, \quad k \in i(-A,A)\,.
\nonumber
\]
Correspondingly, $d_\pm(k) := \sqrt{D(k\pm V/2; A_\pm)}$ satisfies the jump in Lemma~\ref{l:d_pm properties}. 

The asymptotic behavior of $d_\pm(k)$ follows directly from that of $D_\pm(k)$. 
\endproof
%
\paragraph{Integral equations for $\mu_\pm(x,t,k)$.}
\label{Integral equations for mu_pm appendix}
We now establish the integral equations~\eqref{mu integral equations} for $\mu_\pm(x,t,k)$.
We first define
\[
	\psi_\pm(x,t,k) = \e^{-i\theta_\pm(x,t,k)\sigma_3}E_\pm^{-1}(k)\e^{if_\pm(x,t)\sigma_3}\phi_\pm(x,t,k)\,,
\nonumber
\]
so that
$	\psi_\pm(x,t,k) = \I+o(1)$ as $x \rightarrow \pm \infty$.
Recalling that $\phi_\pm(x,t,k)$ 
satisfies the Lax pair, we have
\begin{multline*}
	\e^{-if_\pm(x,t)\sigma_3}\Big( \mp i\frac{V}{2}\sigma_3 E_\pm(k) + E_\pm(k)i\lambda_\pm\sigma_3\Big)\e^{i\theta_\pm(x,t,k)\sigma_3}\psi_\pm(x,t,k) 
+ \e^{-if_\pm(x,t)\sigma_3}E_\pm(k)\e^{i\theta_\pm\sigma_3} ( \psi_\pm )_x
	\\
	= ( X_\pm + \Delta Q_\pm )\e^{-if_\pm(x,t)\sigma_3}E_\pm(k)\e^{i\theta_\pm(x,t,k)\sigma_3}\psi_\pm\,.
\end{multline*}
Now with $\^X_\pm(k)$ as defined in~\eqref{Xhat_pm(k)}, we have
\be
\label{X_pm shift calculation}
\begin{aligned}
		\e^{if_\pm(x,t)\sigma_3}X_\pm(x,t,k)\e^{-if_\pm(x,t)\sigma_3} E_\pm(k)
			= \Big( \^X_\pm(k\pm V/2) \mp i\frac{V}{2}\sigma_3 \Big)E_\pm(k)
			= E_\pm(k)i\lambda_\pm\sigma_3 \mp i\frac{V}{2}\sigma_3 E_\pm(k)\,.
\end{aligned}
\ee
Thus,
\[
	\e^{-if_\pm(x,t)\sigma_3}E_\pm(k)\e^{i\theta_\pm(x,t,k)\sigma_3} ( \psi_\pm )_x = \Delta Q_\pm(x,t)\e^{-if_\pm(x,t)\sigma_3}E_\pm(k)\e^{i\theta_\pm(x,t,k)\sigma_3}\psi_\pm\,.
\nonumber
\]
Formally integrating, we arrive at the integral equations for $\psi_\pm(x,t,k)$,
\[
	\psi_\pm(x,t,k) = \I + \int_{\pm \infty}^x \e^{-i\theta_\pm(y,t,k)\sigma_3}E_\pm^{-1}(k)
		\e^{2if_\pm(y,t)\sigma_3}\Delta Q_\pm(y,t)
		E_\pm(k)\e^{i\theta_\pm(y,t,k)\sigma_3}\psi_\pm(y,t,k) dy\,.
\nonumber
\]
Finally, recognizing that
$\mu_\pm(x,t,k) = E_\pm(k)\e^{i\theta_\pm(x,t,k)\sigma_3}\psi_\pm(x,t,k)\e^{-i\theta_\pm(x,t,k)\sigma_3}
$
gives the corresponding integral equations~\eqref{mu integral equations} for $\mu_\pm(x,t,k)$.
%
\paragraph{Proof of Theorem~\ref{Jost analyticity} (Analyticity of the Jost solutions).}
\label{Jost analyticity appendix}
\bb
Here we use the integral equations for $\mu_\pm(x,t,k)$ to find and prove the regions of analyticity and continuity for the Jost solutions under the assumption that $(q-q_\pm)\in L^1_x(\R)$ for all $t\in\R$.
\eb
The proof follows nearly identically to the proof in \cite{BK2014}. Comparing the integral equations there and here, we have only trivial differences:
(i)~In \cite{BK2014}, the eigenfunctions are expressed in terms of the uniformization variable $z = \lambda + k$.
(ii)~Here, the definition of $\lambda_\pm(k)$ causes a shift of $\mp V/2$ compared to the $\lambda(k)$ that appears in \cite{BK2014}.
(iii)~Here, we have an extra factor $\e^{2if_\pm(y,t)\sigma_3}$.
These differences cause no issue with the analysis of the Neumann iterates. We start by rewriting the integral equations~\eqref{mu integral equations} as
\be
	\mu_\pm(x,t,k) = E_\pm(k)\Big[I + \int_{\pm\infty}^x \e^{i\lambda_\pm(x-y)\sigma_3}E_\pm^{-1} \Delta\^Q_\pm(y,t)\mu_\pm(y,t,k)\e^{-i\lambda_\pm(x-y)\sigma_3}dy\Big],
\ee
where
\bse
\begin{gather}
		\^Q_-(x,t) = \e^{2if_-(x,t)\sigma_3}Q(x,t)\,,
		\\
		\Delta\^Q_\pm(x,t) = \e^{2if_\pm(x,t)\sigma_3}\Delta Q_\pm(x,t)\,.
\end{gather}
\ese
Letting $w(x,t,k)$ be the first column of $W(x,t,k) = E_-^{-1}(k)\mu_-(x,t,k)$, we have
\bse
\be
	w(x,t,k) = \begin{pmatrix}1\\0\end{pmatrix} + \int_{-\infty}^x C(x,y,t,k)w(y,t,k)dy\,,
\ee
where
\be
	C(x,y,t,k) = \diag\big(1, \e^{-2i\lambda_-(x-y)}\big)E_-^{-1}(k)\Delta\^Q_-(y,t)E_-(k)\,.
\ee
\ese
Note that the bounds of integration imply $x-y\geq 0$. Now we introduce a Neumann series for $w$,
\bse
	\[
		w(x,t,k) = \sum_{n=0}^\infty w^{(n)}(x,t,k)\,,
	\]
	with
	\[
		w^{(0)} = \begin{pmatrix}1\\0\end{pmatrix}, \quad w^{(n+1)}(x,t,k) = \int_{-\infty}^x C(x,y,t,k)w^{(n)}(y,t,k)dy\,.
	\]
\ese
Introducing the $L^1$ vector norm 
$\norm{w} := |w_1| + |w_2|$
and the corresponding subordinate matrix norm $\norm{C}$, we then have
\be
	\norm{w^{(n+1)}(x,t,k)} \leq \int_{-\infty}^x \norm{C(x,y,t,k)} \norm{w^{(n)}(y,t,k)}dy\,.
\nonumber
\ee
Note that 
\be
	\norm{E_-(k)} = 1 + \frac{A_-}{|\lambda_- + (k-V/2)|}, \quad \norm{E_-^{-1}(k)} = \frac{1}{|D_-(k)|}\Big(1 + \frac{A_-}{|\lambda_- + (k-V/2)|}\Big)\,.
\nonumber
\ee
Thus,
\begin{multline*}
		\norm{C(x,y,t,k)} \leq \norm{\diag\big(1,\e^{-2i\lambda_-(x-y)}\big)}\norm{E_-^{-1}(k)}\norm{\Delta\^Q_-(y,t)}\norm{E_-(k)}
		\\
		 = c(k)\big(1 + \e^{2\Im \lambda_-(x-y)}\big)|q(y,t) - q_-(y,t)|\,,
\end{multline*}
where 
\be
	c(k) = \norm{E_-^{-1}(k)}\norm{E_-(k)} = \frac{1}{|D_-(k)|}\Big(1 + \frac{A_-}{|\lambda_- + (k-V/2)|}\Big)^2\,,
\nonumber
\ee
is the condition number of $E_-(k)$. Recall that $\Im \lambda_- \leq 0$ for $k \in \R\cup\C^-\cup\Sigma_{-1}$, and that $c(k) \rightarrow \infty$ as $k \rightarrow p_{-}\,,\conj{p_{-}}$. Thus, given $\epsilon > 0$, we restrict our attention to the domain
\be
	U_\epsilon = \R\cup\C^-\cup\Sigma_{-1}^o\setminus B_\epsilon(\conj{p_{-}})\,,
\nonumber
\ee
where $B_\epsilon(k_o) = \{k \in \C\hspace{1mm}:\hspace{1mm}|k-k_o| < \epsilon\}$. We next prove that, for all $k \in U_\epsilon$ and for all $n \in \Natural$,
\bse
	\[
		\label{w induction}
		\norm{w^{(n)}(x,t,k)} \leq \frac{M^n(x,t)}{n!}\,,
	\]
	where
	\[
		M(x,t) = 2c_\epsilon \int_{-\infty}^x |q(y,t) - q_-(y,t)|dy\,,
	\]
\ese
and $c_\epsilon := \max_{k \in U_\epsilon} c(k)$.
The claim is trivially true for $n=0$. Also note that for all $k \in U_\epsilon$ and for all $y\leq x$ we have $1 + \e^{2\Im \lambda_-(x-y)} \leq 2$. Thus if~\eqref{w induction} holds, then
\be
	\norm{w^{(n+1)}(x,t,k)} \leq \frac{2c_\epsilon}{n!} \int_{-\infty}^x |q(y,t)-q_-(y,t)| M^n(y,t) dy = \frac{1}{n!(n+1)}M^{n+1}(x,t)\,,
\nonumber
\ee
proving the induction step. Thus for all $\epsilon > 0$, if $q(x,t) - q_-(x,t) \in L^{1}(-\infty,a]$ for some $a \in \R$, the Neumann series converges absolutely and uniformly with respect to $k \in U_\epsilon$ for $x \in (-\infty,a)$. This demonstrates that $\mu_{-1}(x,t,k)$ and thus $\phi_{-1}(x,t,k)$ is defined for $k \in \R\cup\C^-\cup\Sigma_{-1}^o\setminus\{\conj{p_{-}}\}$, continuous from the right for $k \in \R\cup\Sigma_{-2}^o$, and analytic for $k\in\C^-\setminus\Sigma_{-2}$.

The arguments for the remaining eigenfunctions are similar.
\endproof
%
\paragraph{Proof of Lemma~\ref{mu asymptotics} (Asymptotics of $\phi_\pm(x,t,k)$ as $k \rightarrow \infty$).}
\label{Asymptotics of phi_pm(x,t,k) as k rightarrow infty appendix}
\bb
We now determine the asymptotic behavior for $\mu_\pm(x,t,k)$ as $k$ goes to infinity under the assumption that $(q-q_\pm)\in L^1_x(\R)$, and find an explicit asymptotic expansion up to $o(1/k)$ in order to reconstruct the potential $q(x,t)$.
\eb
Consider the formal expansion
\bse
\[
	\mu_-(x,t,k) = \sum_{n=0}^\infty \mu^{(n)}(x,t,k)\,,
\]
with
\begin{align}
		\mu^{(0)}(x,t,k) &= E_-(k),
		\\
		\mu^{(n+1)}(x,t,k) &= \int_{-\infty}^x E_-\e^{i\lambda_-(x-y)\sigma_3}E_-^{-1}\Delta \^Q_-(y,t)\mu^{(n)}(y,t,k)\e^{-i\lambda_-(x-y)\sigma_3} dy\,.
\end{align}
\ese
Let $B_d$ and $B_o$ denote the diagonal and off-diagonal parts of a matrix $B$ respectively. The above expression gives an asymptotic expansion for the columns of $\mu_-(x,t,k)$ as $k\rightarrow \infty$ within the appropriate regions of the complex $k$-plane for each column (see Section~\ref{Rigorous definition of the Jost solutions, analyticity and continuous spectrum}).
We now show that if the potential admits a continuous derivative with $(q-q_-)_x \in L_x^1(-\infty,a)$ for some $a\in\R$, then
\be
\label{mu even/odd asymptotics}
\begin{aligned}
		\mu^{(2m)}_d=O\Big(\frac{1}{k^m}\Big), && \mu^{(2m)}_o = O\Big(\frac{1}{k^{m+1}}\Big)\,, && \mu^{(2m+1)}_d = O\Big(\frac{1}{k^{m+1}}\Big), && \mu^{(2m+1)}_o = O\Big(\frac{1}{k^{m+1}}\Big)\,,
\end{aligned}
\ee
within the appropriate region of the complex $k$-plane for each column. Explicitly, the first column is valid for $k \in \C^-$ while the second column is valid for $k \in \C^+$. To aid in the argument, we define 
\[
\label{R definition}
	R^{(n)}(x,t,k) = \int_{-\infty}^x \e^{2i\lambda_-(x-y)\sigma_3}\big(\Delta\^Q_-(y,t) \mu^{(n)}_d(y,t,k) - \frac{iA_-}{\lambda_- + (k-V/2)}\sigma_1\Delta\^Q_-(y,t)\mu^{(n)}_o(y,t,k)\big)dy\,,
\]
and simultaneously show that
\[
\label{R even/odd asymptotics}
	R^{(2m)} = O\Big(\frac{1}{k^{m+1}}\Big), \quad R^{(2m-1)} = O\Big(\frac{1}{k^{m+1}}\Big)\,.
\]
To clarify the logic, in the induction step we will assume that~\eqref{mu even/odd asymptotics} is true for $\mu^{(n-1)}$ and $\mu^{(n)}$, and~\eqref{R even/odd asymptotics} is true for $R^{(n-1)}$. We will then show that~\eqref{R even/odd asymptotics} holds for $R^{(n)}$, which will be used to show that~\eqref{mu even/odd asymptotics} holds for $\mu^{(n+1)}$. 
Defining
$\mu^{(-1)} = 0$ and  $\quad R^{(-1)} = 0$,
and noting that the claim is clearly true for $\mu^{(0)}$ gives the necessary base cases.
Next, integrating by parts we find
\begin{align*}
		R^{(n)} &= \frac{i}{2\lambda_-}\sigma_3\big(\Delta\^Q_-\mu^{(n)}_d - \frac{iA_-}{\lambda_- + (k-V/2)}\sigma_1\Delta\^Q_-\mu^{(n)}_o\big)
		\\
		&\quad - \frac{i}{2\lambda_-}\sigma_3 \int_{-\infty}^x \e^{2i\lambda_-(x-y)\sigma_3}\Big[  (\^Q_- )_x\mu^{(n)}_d - \frac{iA_-}{\lambda_- + (k-V/2)}\sigma_1 ( \^Q_- )_x\mu^{(n)}_o 
		\\
		&\quad + \Delta\^Q_- ( \mu^{(n)}_d )_x - \frac{iA_-}{\lambda_- + (k-V/2)}\sigma_1\Delta\^Q_- ( \mu^{(n)}_o )_x\Big]dy\,,
\end{align*}
so that
\be
\label{R integration by parts}
	R^{(n)} = O\Big(\frac{\mu^{(n)}_d}{k}\Big) + O\Big(\frac{\mu^{(n)}_o}{k^2}\Big) + O\Big(\frac{(\mu^{(n)}_d)_x}{k}\Big) + O\Big(\frac{(\mu^{(n)}_o)_x}{k^2}\Big)\,.
\ee
Now
\begin{align*}
	E_-^{-1}\Delta\^Q_-\mu^{(n)} &= \frac{1}{D_-}\big[\Delta\^Q_-\mu^{(n)}_o - \frac{iA_-}{\lambda_- + (k-V/2)}\sigma_1\Delta\^Q_-\mu^{(n)}_d\big]
		\\
		&\quad+ \frac{1}{D_-}\big[\Delta\^Q_-\mu^{(n)}_d - \frac{iA_-}{\lambda_- + (k-V/2)}\sigma_1\Delta\^Q_-\mu^{(n)}_o\big]\,,
	\end{align*}
where the first term is a diagonal matrix and the second is off-diagonal. Then
\begin{align*}
		\e^{i\lambda_-(x-y)\sigma_3}E_-^{-1}\Delta\^Q_-\mu^{(n)}\e^{-i\lambda(x-y)\sigma_3} 
		&=\frac{1}{D_-}\big[\Delta\^Q_-\mu^{(n)}_o - \frac{iA_-}{\lambda_- + (k-V/2)}\sigma_1\Delta\^Q_-\mu^{(n)}_d\big] 
		\\
		&\quad + \frac{1}{D_-}\e^{2i\lambda_-(x-y)\sigma_3}\big[\Delta\^Q_-\mu^{(n)}_d - \frac{iA_-}{\lambda_- + (k-V/2)}\sigma_1\Delta\^Q_-\mu^{(n)}_o\big]\,,
\end{align*}
and
\begin{align*}
		E_-\e^{i\lambda_-(x-y)\sigma_3}E_-^{-1}\Delta\^Q_-\mu^{(n)}\e^{-i\lambda(x-y)\sigma_3} 
		&=\frac{1}{D_-}\big[\Delta\^Q_-\mu^{(n)}_o - \frac{iA_-}{\lambda_- + (k-V/2)}\sigma_1\Delta\^Q_-\mu^{(n)}_d\big]
		\\
		&\quad + \frac{iA_-}{2\lambda_-}\sigma_1\e^{2i\lambda_-(x-y)\sigma_3}\big[\Delta\^Q_-\mu^{(n)}_d - \frac{iA_-}{\lambda_- + (k-V/2)}\sigma_1 \Delta\^Q_-\mu^{(n)}_o\big]
		\\
		&\quad + \frac{1}{D_-}\e^{2i\lambda_-(x-y)\sigma_3}\big[\Delta\^Q_-\mu^{(n)}_d - \frac{iA_-}{\lambda_- + (k-V/2)}\sigma_1\Delta\^Q_-\mu^{(n)}_o\big]
		\\
		&\quad + \frac{iA_-}{2\lambda_-}\sigma_1\big[\Delta\^Q_-\mu^{(n)}_o - \frac{iA_-}{\lambda_- + (k-V/2)}\sigma_1\Delta\^Q_-\mu^{(n)}_d\big]\,,
\end{align*}
where the first two terms are diagonal and the last two are off-diagonal. Then, we have
\bse
\label{off/diagonal recursions}
\begin{align}
		\begin{split}
			\mu^{(n+1)}_d &= \frac{1}{D_-}\int_{-\infty}^x\big(\Delta\^Q_-\mu^{(n)}_o - \frac{iA_-}{\lambda_- + (k-V/2)}\sigma_1\Delta\^Q_-\mu^{(n)}_d\big)dy +\frac{iA_-}{2\lambda_-}\sigma_1 R^{(n)}\,,
		\end{split}
		\\
		\begin{split}
			\mu^{(n+1)}_o &= \frac{iA_-}{2\lambda_-}\sigma_1\int_{-\infty}^x\big(\Delta\^Q_-\mu^{(n)}_o - \frac{iA_-}{\lambda_- + (k-V/2)}\sigma_1\Delta\^Q_-\mu^{(n)}_d\big)dy +\frac{1}{D_-}R^{(n)}\,.
		\end{split}
\end{align}
\ese
Differentiating and re-indexing, we find
\bse
\label{off/diagonal derivative recursions}
\begin{align}
\begin{split}
		( \mu^{(n)}_d )_x &= \Delta\^Q_-\mu^{(n-1)}_o - A_-\sigma_1\sigma_3R^{(n-1)}\,,
\end{split}
	\\
\begin{split}
		( \mu^{(n)}_o )_x &= \Delta\^Q_-\mu^{(n-1)}_d + i \sigma_3 R^{(n-1)}\,.
\end{split}
\end{align}
\ese
so that with~\eqref{R integration by parts} we have
\be
	R^{(n)} = O\Big(\frac{\mu^{(n)}_d}{k}\Big) + O\Big(\frac{\mu^{(n)}_o}{k^2}\Big) + O\Big(\frac{\mu^{(n-1)}_d}{k^2}\Big) + O\Big(\frac{\mu^{(n-1)}_o}{k}\Big) + O\Big(\frac{R^{(n-1)}}{k}\Big)\,.
\nonumber
\ee
By induction, we see that if $n=2m$, then
$R^{(2m)} = O\Big({1}/{k^{m+1}}\Big)$,
so that with~\eqref{off/diagonal recursions} we have
\begin{align*}
		\mu^{(2m+1)}_d &= O\big(\mu^{(2m)}_o\big) + O\Big(\frac{\mu^{(2m)}_d}{k}\Big) + O\Big(\frac{1}{k^{m+2}}\Big) = O\Big(\frac{1}{k^{m+1}}\Big)\,,
		\\
		\mu^{(2m+1)}_o &= O\Big(\frac{\mu^{(2m)}_o}{k}\Big) + O\Big(\frac{\mu^{(2m)}_d}{k^2}\Big) + O\Big(\frac{1}{k^{m+1}}\Big) = O\Big(\frac{1}{k^{m+1}}\Big)\,.
\end{align*}
while if $n=2m-1$, then
$R^{(2m-1)} = O\Big(\frac{1}{k^{m+1}}\Big)$,
so that with~\eqref{off/diagonal recursions} we have
\begin{align*}
		\mu^{(2m)}_d &= O\big(\mu^{(2m-1)}_o\big) + O\Big(\frac{\mu^{(2m-1)}_d}{k}\Big) + O\Big(\frac{1}{k^{m}}\Big) = O\Big(\frac{1}{k^{m}}\Big)\,,
		\\
		\mu^{(2m)}_o &= O\Big(\frac{\mu^{(2m-1)}_o}{k}\Big) + O\Big(\frac{\mu^{(2m-1)}_d}{k^2}\Big) + O\Big(\frac{1}{k^{m+1}}\Big) = O\Big(\frac{1}{k^{m+1}}\Big)\,.
\end{align*}
which completes the induction. Similar argument gives the corresponding asymptotics for $\mu_+$.
%
\paragraph{Recovery of the potential.}
From the above, we see that
\[
	\mu_- = \mu^{(0)}_d + \mu^{(0)}_o + \mu^{(1)}_d + \mu^{(1)}_o + \mu^{(2)}_d + o(1/k)\,.
\nonumber
\]
Computing these terms explicitly up to order $1/k$, we have
\begin{align*}
		\mu^{(0)}_d &= \I\,,\qquad
		\mu^{(1)}_o = \frac{i}{2k}A_-\sigma_1\,,
		\\
		\mu^{(1)}_d &= \frac{i}{2k}\int_{-\infty}^x\big[\Delta\^Q_-,A_-\sigma_1\big]dy + o(1/k)\,, \qquad
		\mu^{(1)}_o = \frac{i}{2k}\sigma_3\Delta\^Q_- + o(1/k)\,,
		\\
		\mu^{(2)}_d &= \frac{i}{2k}\int_{-\infty}^x \Delta\^Q_-\sigma_3\Delta\^Q_-dy + o(1/k)\,, & &
\end{align*}
where the last two make use of the Riemann-Lebesgue lemma.
Then
\be
	\mu_- = \I + \frac{i}{2k}\sigma_3\e^{2if_-\sigma_3}Q + \frac{i}{2k}\int_{-\infty}^x \big([\e^{2if_-\sigma_3}\Delta Q_-,A_-\sigma_1 ] + \Delta Q_-\sigma_3\Delta Q_- \big)dy + o(1/k)\,.
\nonumber
\ee
The 12-entry of this expression then gives~\eqref{q from mu}.
\endproof

\bb
\paragraph{Proof of Corollary~\ref{c:phi differentiable} (Real differentiability of $\phi_\pm(x,t,k)$).}
\label{c:phi differentiable proof}

We now show that under the assumption $(q-q_\pm)\in L^1_x(\R)$, the Jost solutions $\phi_+(x,t,k)$ and $\phi_-(x,t,k)$ are real differentiable for all $k\in\R$. Consider the first column of the integral equation~\eqref{mu integral equations} for $\mu_{+1}(x,t,k)$,
\[
	\label{e:mu_+1 integral equation}
	\mu_{+1}(x,t,k) = E_{+1}(k) + \int_\infty^x K(x-y,k)\e^{2if_+(y,t)\sigma_3} \Delta Q_+(y,t) \mu_{+1}(y,t,k) dy\,,
\]
where
\[
	K(\xi,k) = E_+(k)\diag\big(1,\e^{-2i\lambda_+\xi}\big)E_+^{-1}(k)\,.
\nonumber
\]
Formally differentiating with respect to $k$, we have 
\begin{multline}
		\frac{\partial\mu_{+1}}{\partial k}(x,t,k) = \frac{\partial E_{+1}}{\partial k}(k) + \int_\infty^x \frac{\partial K}{\partial k}(x-y,k) \e^{2if_+(y,t)\sigma_3}\Delta Q_+(y,t) \mu_{+1}(y,t,k) dy
		\\
		 \quad + \int_\infty^x K(x-y,k)\e^{2if_+(y,t)\sigma_3}\Delta Q_+(y,t) \frac{\partial \mu_{+1}}{\partial k}(y,t,k) dy\,.
\label{e:dmudk integral equation}
\end{multline}

Note that the integral equation~\eqref{e:dmudk integral equation} has exactly the same kernel as~\eqref{e:mu_+1 integral equation}, 
and just the integrated term is different. Analysis of the Neumann iterates for~\eqref{e:dmudk integral equation}, similar to the proof of Theorem~\ref{Jost analyticity}, then shows that if $(q-q_+) \in L^1_x(a,\infty)$ for some $a\in\R$, then $\partial \mu_{+1}/\partial k$ is well-defined and continuous for $k \in \R$, with continuity restricted to $\R$. Similar analysis for the remaining eigenfunctions gives the result.
\endproof
\eb

\paragraph{Proof of Lemma~\ref{l:mu branch point continuity} (Well-defined modified eigenfunctions at the branch points).}
\label{Behavior at the branch points appendix}
\bb
Here we show that under the assumption $(q-q_\pm)\in L^{1,1}_x(\R)$, the modified eigenfunctions $\mu_\pm(x,t,k)$ can be extended to the branch points. Again consider the integral equation~\eqref{e:mu_+1 integral equation} for $\mu_{+1}(x,t,k)$.
\eb
Note that
\[
	\lim_{k \rightarrow -V/2 \pm iA_+} K(\xi,k) = \I + A_+\xi(\sigma_3\e^{i\delta\sigma_3}\sigma_1 \mp \sigma_3)\,.
\nonumber
\]
Analysis of the Neumann iterates similar to the proof of Theorem~\ref{Jost analyticity} shows that if $(q - q_+) \in L_x^{1,1}(a,\infty)$ for some $a\in\R$, then $\mu_{+1}(x,t,k)$ is well-defined and continuous at the branch points $k = p_{+}\,, \conj{p_{+}}$. Note that continuity at $k=\conj{p_{+}}$ is restricted to $k \in\Sigma_{+2}$.
Similar argument for the remaining eigenfunctions at their respective branch points gives:
\begin{itemize}
\advance\itemsep-4pt
	\item 
		If $(q - q_+) \in L_x^{1,1}(a,\infty)$ for some $a\in\R$, then 
		\begin{itemize}
\advance\itemsep-4pt
			\item[$\circ$] 
				$\mu_{+1}(x,t,k)$ is continuous at the branch points $k = p_{+}\,, \conj{p_{+}}$, where continuity at $k = \conj{p_{+}}$ is restricted to $k \in\Sigma_{+2}$,
			\item[$\circ$] 
				$\mu_{+2}(x,t,k)$ is continuous at the branch points $k = p_{+}\,,\conj{p_{+}}$, where continuity at $k = p_{+}$ is restricted to $k \in\Sigma_{+1}$.
		\end{itemize}
	\item  
		If $(q - q_-) \in L_x^{1,1}(-\infty,a)$ for some $a\in\R$, then 
		\begin{itemize}
\advance\itemsep-4pt
			\item[$\circ$]  
				$\mu_{-1}(x,t,k)$ is continuous at the branch points $k = p_{-}\,, \conj{p_{-}}$, where continuity at $k = p_{-}$ is restricted to $k \in \Sigma_{-1}$,
			\item[$\circ$]  
				$\mu_{-2}(x,t,k)$ is continuous at the branch points $k = p_{-}\,, \conj{p_{-}}$, where continuity at $k = \conj{p_{-}}$ is restricted to $k \in \Sigma_{-2}$.
		\end{itemize}
\end{itemize}
\bb
Moreover, the normalizations $\mu_\pm(x,t,k) = E_\pm(k) (1+o(1))$ as $x \to \pm \infty$ imply that the modified eigenfunctions are nonzero at the branch points.
\eb
Lemma~\ref{l:mu branch point continuity} then follows.
\endproof

\paragraph{Proof of Lemma~\ref{l:mu branch point expansion} (Expansion of modified eigenfunctions at the branch points).}
\bb
We now improve the expansions of $\mu_\pm(x,t,k)$ about the branch points from the previous lemma under the more strict assumption that $(q-q_\pm)\in L^{1,2}_x(\R)$.
\eb
We again consider $\mu_{+1}(x,t,k)$. It is convenient to introduce the variable
\[
	z(k) = \lambda_+(k) + (k + V/2)\,,
\nonumber
\]
so that
\[
	k + V/2 = \frac{1}{2}(z - A_+^2/z), \quad \lambda_+ = \frac{1}{2}(z + A_+^2/z)\,.
\nonumber
\]
Note that $|z|\geq A_+$ for all $k \in \C$, with $|z|=A_+$ exactly when $k \in \Sigma_+$. Furthermore, the branch points $k = -V/2 \pm iA_+$ correspond to $z = \pm iA_+$.
\bb
The reason for introducing the variable $z$ is that,
while the derivatives of the eigenfunctions with respect to $k$ are not well-defined at the branch points, 
those with respect to $z$ are,
which will enable us to obtain asymptotic estimates near the branch points.
\eb

With some abuse of notation, we write all $k$-dependence as $z$-dependence,
\bb
 so that the integral equations~\eqref{e:mu_+1 integral equation} and~\eqref{e:dmudk integral equation} become
\be
	\mu_{+1}(x,t,z) = E_{+1}(z) + \int_{\infty}^x K(x-y,z)\e^{2if_+(y,t)\sigma_3}\Delta Q_+(y,t) \mu_{+1}(y,t,z)dy\,,
\label{e:mu+1_intleqn}
\nonumber
\ee
and
\begin{multline}
		\frac{\partial\mu_{+1}}{\partial z}(x,t,z) = \frac{\partial E_{+1}}{\partial z}(z) + \int_\infty^x \frac{\partial K}{\partial z}(x-y,z) \e^{2if_+(y,t)\sigma_3}\Delta Q_+(y,t) \mu_{+1}(y,t,z) dy
		\\
		 \quad + \int_\infty^x K(x-y,z)\e^{2if_+(y,t)\sigma_3}\Delta Q_+(y,t) \frac{\partial \mu_{+1}}{\partial z}(y,t,z) dy\,.
\label{e:dmudz_intleqn}
\end{multline}
respectively.
\eb
Note that
\begin{align*}
		\lim_{z \rightarrow \pm iA_+} K(\xi,z) &= \I + A_+\xi(\sigma_3\e^{i\delta\sigma_3}\sigma_1 \mp \sigma_3)\,,
		\\
		\lim_{z \rightarrow \pm iA_+} \frac{\partial K}{\partial z}(\xi,z) &= -i\xi \,\I - iA_+\xi^2(\sigma_3\e^{i\delta\sigma_3}\sigma_1 \mp \sigma_3)\,.
\end{align*}
Analysis of the Neumann iterates 
\bb
for ~\eqref{e:dmudz_intleqn},
\eb
similar to the proof of Theorem~\ref{Jost analyticity}, then shows that if $(q - q_+) \in L_x^{1,2}(a,\infty)$ for some $a\in\R$, then ${\partial \mu_{+1}}/{\partial z}$ is well-defined and continuous at $z = iA_+$, with continuity restricted to $|z| \geq A_+$. Then
\[
	\frac{\partial\mu_{+1}}{\partial z}(x,t,z) = \frac{\partial\mu_{+1}}{\partial z}(x,t,iA_+) + o(1)\,, \quad z \rightarrow iA_+\,.
\nonumber
\]
Since
\[
	\mu_{+1}(x,t,z) = \mu_{+1}(x,t,iA_+) + \int_{iA_+}^z \frac{\partial\mu_{+1}}{\partial z}(0,0,s)ds\,,
\nonumber
\]
we have
\[
	\mu_{+1}(x,t,z) = \beta^{(0)}_{p_+}(x,t) + \~\beta^{(1)}_{p_+}(x,t)(z-iA_+) + o(z-iA_+), \quad z \rightarrow iA_+\,,
\nonumber
\]
where $\beta^{(0)}_{p_+}(x,t) = \mu_{+1}(x,t,z=iA_+)$ and $\~\beta^{(1)}_{p_+}(x,t) = \frac{\partial \mu_{+1}}{\partial z}(x,t,z=iA_+)$. 
In terms of $k$, we have
\[
	\mu_{+1}(x,t,k) = \beta^{(0)}_{p_+}(x,t) + \~\beta^{(1)}_{p_+}(x,t)(\lambda_+ + k - p_{+}) + o(\lambda_+ + k-p_{+})\,, \quad k \rightarrow p_{+}\,.
\nonumber
\]
Note that
\[
	\lambda_+(k) = ((k+V/2)^2 + A_+^2)^{1/2} = (k - p_{+})^{1/2} (k - \conj{p_{+}})^{1/2}\,,
\nonumber
\]
so that 
\[
	\lambda_+ + k - p_{+} = (2iA_+)^{1/2}(k - p_{+})^{1/2} + o(k-p_{+})^{1/2}\,, \quad k \rightarrow p_{+}\,.
\nonumber
\]
Then
\[
	\mu_{+1}(x,t,k) = \beta^{(0)}_{p_+}(x,t) + \beta^{(1)}_{p_+}(x,t)(k-p_{+})^{1/2} + o(k-p_{+})^{1/2}, \quad k \rightarrow p_{+},
\nonumber
\]
where $\beta^{(1)}_{p_+}(x,t) = (2iA_+)^{1/2}\, \~\beta^{(1)}_{p_+}(x,t)$.

Using similar analysis for the remaining eigenfunctions (instead with $z(k) = \lambda_-(k) + (k-V/2)$ for $\mu_-(x,t,k)$), we see that
Lemma~\ref{l:mu branch point expansion} then follows.
\endproof

\bb
\paragraph{Proof of Lemma~\ref{l:d_pm branch point behavior} (Behavior of $d_\pm(k)$ at the branch points).}

At the branch points $p_\pm$, we have
\begin{align}
	&& && &\lambda_\pm(k) = (k-p_\pm)^{1/2}(k-\conj{p_\pm})^{1/2} = (2iA_\pm)^{1/2}(k-p_\pm)^{1/2} + o(1)\,,&  &k \rightarrow p_\pm\,, && &&
	\nonumber
	\\
	&& && &\lambda_\pm(k) + (k \pm V/2) = iA_\pm + o(1)\,, & &k \rightarrow p_\pm\,, && &&
	\nonumber
\end{align}
so that 
\be
	D_\pm(k) = \Big(\frac{8}{iA_\pm}\Big)^{1/2}(k-p_\pm)^{1/2} + o(1)\,, \quad k \to p_\pm\,.\qquad
	\nonumber
\ee
Similar calculation gives the asymptotic behavior at the branch points $\conj{p_\pm}$. 
\endproof
\eb


\paragraph{Proof of Lemmas~\ref{first symmetry Jost} and~\ref{first symmetry extended} (First symmetry, Jost solutions and scattering coefficients).}
\label{Symmetries appendix}

\bb
Here we show relations between the Jost solutions and their Schwarz conjugates, and the corresponding relations between the scattering coefficients.
\eb

Consider $\psi_\pm(x,t,k)=-\sigma_\ast \conj{\phi_\pm(x,t,\conj{k})} \sigma_\ast$, defined column-wise wherever $\phi_\pm(x,t,\conj{k})$ exists. Then
\be
	(\psi_\pm)_x = -\sigma_\ast \conj{(\phi_\pm)_x(x,t,\conj{k})} \sigma_\ast
											= \big(-\sigma_\ast \conj{X(x,t,\conj{k})}\sigma_\ast \big) \big(-\sigma_\ast\conj{\phi_\pm(x,t,\conj{k})} \sigma_\ast\big)
											= X(x,t,k)\psi_\pm\,.
\nonumber
\ee
Similar calculation shows that $(\psi_\pm)_t = T(x,t,k)\psi_\pm$. Then $-\sigma_\ast \conj{\phi_\pm(x,t,\conj{k})} \sigma_\ast$ satisfies the Lax pair~\eqref{e:LP}. Comparing the asymptotic behavior as $x \rightarrow \pm \infty$, we see that
\begin{align}
\label{first symmetry relation}
	\conj{\phi_\pm(x,t,\conj{k})} = -\sigma_\ast \phi_\pm(x,t,k) \sigma_\ast\,, 
\end{align}
which is to be understood column-wise wherever the appropriate columns are defined. Writing~\eqref{first symmetry relation} in terms of the columns gives~\eqref{phi conjugates} and proves Lemma~\ref{first symmetry Jost}.

In particular,~\eqref{first symmetry relation} holds for all $k \in \R$, which implies that
\be
	\conj{S(\conj{k})} = -\sigma_* S(k) \sigma_* \,, \quad k \in \R\,,
	\nonumber
\ee
so that
\bse
\label{first symmetry}
\begin{align}
		\label{first symmetry a}
		\conj{s_{22}(\conj{k})} &= \phantom{-}s_{11}(k)\,, \quad k \in \R\,,
		\\
		\label{first symmetry b}
		\conj{s_{12}(\conj{k})} &= -s_{21}(k)\,, \quad k \in \R\,.
\end{align}
\ese
Moreover, using~\eqref{phi conjugates}, we see that
\bb
\be
	\conj{s_{22}(\conj{k})} = \Wr[\sigma_*\phi_{+2}(x,t,k),-\sigma_*\phi_{-1}(x,t,k)]
							= \Wr[\phi_{-1}(x,t,k),\phi_{+2}(x,t,k)]
							= s_{11}(k)\,,
\ee
\eb
for $k \in \C^-\setminus\{\conj{p_{+}}, \conj{p_{-}}\}$, so that~\eqref{first symmetry a} can be extended. We can similarly extend~\eqref{first symmetry b} to $k \in \R \cup \Sigma_{+2}^o \cup \Sigma_{-1}^o$.
Recalling~\eqref{R S relations} then completes the proof.
\endproof

\paragraph{Proof of Lemmas~\ref{second symmetry} and~\ref{l:second symmetry scattering coefficients} (Second symmetry, Jost solutions and scattering coefficients).}
\bb
We now determine the discontinuities of the Jost solutions across the appropriate portions of the branch cuts $\Sigma_\pm$, and the corresponding jumps for the scattering coefficients.
\eb

Consider the transformation $\lambda_+ \mapsto -\lambda_+$. Simple algebraic manipulations yield
\bb
\be
	E_\pm(\lambda_\pm \mapsto -\lambda_\pm) = \I + \frac{iA_\pm}{-\lambda_\pm + (k\pm V/2)}\e^{\pm i \delta \sigma_3}\sigma_1 = \Big(\frac{2\lambda_\pm}{iA_\pm D_\pm}\Big)E_\pm\e^{\pm i\delta\sigma_3}\sigma_1\,,
\nonumber
\ee
so that 
with Lemma~\ref{l:d_pm properties} we have
\be
	\left(\frac{1}{d_\pm}\e^{-if_\pm\sigma_3}E_\pm\e^{i\theta_\pm\sigma_3}\right)(\lambda_\pm \mapsto -\lambda_\pm) 
	=-i\frac{1}{d_\pm}\e^{-if_\pm\sigma_3}E_\pm\e^{\pm i\delta\sigma_3}\sigma_1\e^{-i\theta_\pm\sigma_3}
	=-i\frac{1}{d_\pm}\e^{-if_\pm\sigma_3}E_\pm\e^{i\theta_\pm\sigma_3}\e^{\pm i \delta\sigma_3}\sigma_1\,.
\nonumber
\ee
The asymptotics~\eqref{e:phi x asymptotics} then give
\be
	\phi_\pm(\lambda_\pm \mapsto -\lambda_\pm) 
	= -i\phi_\pm\e^{\pm i\delta\sigma_3}\sigma_1(1+o(1))\,, \quad x \rightarrow \pm\infty\,,
\nonumber
\ee
which is to be understood column-wise. 
\eb
This, together with~\eqref{left/right lambda}, proves Lemma~\ref{second symmetry}.


The Wronskian representations~\eqref{e:Wronskians} give
\bb
\be
	s_{22}^+(k) = \Wr\big[\phi_{+1}^+(x,t,k)\,, \phi_{-2}^+(x,t,k)\big]\,, \quad k\in\Sigma_{+1}^o\cup\Sigma_{-1}^o\,.
\nonumber
\ee
\eb
Noting that $\phi_{-1}(x,t,k)$ is analytic on $\Sigma_{+1}$ while $\phi_{+2}(x,t,k)$ is analytic on $\Sigma_{-1}$ for $V\not=0$, we have
\bb
\begin{align*}
	s_{22}^+(k) = -i\e^{-i\delta}\Wr\big[\phi_{+2}(x,t,k)\,,\phi_{-2}(x,t,k)\big] = \phantom{-}i\e^{-i\delta}s_{12}(k)\,, \quad k \in \Sigma_{+1}^o\,,
	\\
	s_{22}^+(k) = -i\e^{-i\delta}\Wr\big[\phi_{+1}(x,t,k)\,,\phi_{-1}(x,t,k)\big] = -ie^{-i\delta}s_{21}(k)\,, \quad k \in \Sigma_{-1}^o\,.
\end{align*}
\eb
\bb
The symmetries~\eqref{e:first symmetry} give the corresponding jumps for $s_{11}(k)$.
%
The jumps for $r_{11}(k)$ and $r_{22}(k)$ are then easily found using~\eqref{R S relations}. \endproof
\eb

\paragraph{Proof of Lemma~\ref{discrete eigenvalues} (Discrete eigenvalues).}
\label{Discrete eigenvalues appendix}
\bb
Here we show the correspondence between zeros of the analytic scattering coefficients and bounded solutions to the Lax pair~\eqref{e:LP} away from the continuous spectrum and branch points.
\eb

Let $s_{22}(k_o) = 0$ for some $k_o \in \C^+\setminus \Sigma$. Then from the Wronskian definition~\eqref{s_22}, we see that $\phi_{-2}(x,t,k_o)$ and $\phi_{+1}(x,t,k_o)$ are linearly dependent so that both decay as $x \rightarrow \pm \infty$, establishing the existence of a bounded solution to the Lax pair~\eqref{e:LP} for $k=k_o$ which decays at both spatial infinities.

Conversely, let $v(x,t)$ be a nontrivial bounded solution to the Lax pair for $k=k_o \in \C^+ \setminus \Sigma$. Suppose $s_{22}(k_o) =: s_o \not= 0$, so that $\Phi(x,t) = \big(\phi_{+1}(x,t,k_o), \phi_{-2}(x,t,k_o)\big)$ is a fundamental matrix solution and
\[
	\label{v(x,t)}
	v(x,t) = \Phi(x,t) c\,,
\nonumber
\]
for some constant vector $c$. Since $k_o \in \C^+\setminus\Sigma$, we have $\Im \lambda_\pm(k_o) > 0$. Correspondingly, the asymptotic behavior~\eqref{e:phi x asymptotics} gives
\be
	\label{phi_{+1},phi_{-2} one-sided decay}
	\lim_{x\rightarrow \infty} \norm{\phi_{+1}(x,t,k_o)} = 0\,,
	\quad \quad
	\lim_{x\rightarrow -\infty} \norm{\phi_{-2}(x,t,k_o)} = 0\,.
\nonumber
\ee 
If $\phi_{+1}(x,t,k_o)$ is bounded for all $x$, then
\[
	s_o = \lim_{x\rightarrow -\infty} \Wr\big[\phi_{+1}(x,t,k_o), \phi_{-2}(x,t,k_o)\big] = 0\,,
\nonumber
\]
which is a contradiction. Arguing similarly for $\phi_{-2}(x,t,k_o)$, we see
\be
	\lim_{x\rightarrow -\infty} \norm{\phi_{+1}(x,t,k_o)} = \infty\,,
	\quad \quad
	\lim_{x\rightarrow \infty} \norm{\phi_{-2}(x,t,k_o)} = \infty\,.
\nonumber
\ee
On the other hand,~\eqref{v(x,t)} and~\eqref{phi_{+1},phi_{-2} one-sided decay} give
\be
	\lim_{x\rightarrow \infty} \norm{v(x,t)} = \lim_{x\rightarrow\infty} \norm{c_2 \phi_{-2}(x,t,k_o)}\,, 
	\quad 
	\lim_{x\rightarrow -\infty} \norm{v(x,t)} = \lim_{x\rightarrow-\infty} \norm{c_1 \phi_{+1}(x,t,k_o)}\,.
\nonumber
\ee
Since $v$ is bounded for all $x$, we must have $c = 0$, which is a contradiction. Thus $s_{22}(k_o) = 0$.

The symmetries~\eqref{phi conjugates} and~\eqref{e:first symmetry} give the corresponding statement for $k \in \C^-\setminus\Sigma$.
\endproof

\paragraph{Proof of Lemma~\ref{nonzero scattering entries on branch cuts} (Non-vanishing scattering coefficients on branch cuts).}
\bb
We now show that the scattering coefficients are non-vanishing on the branch cuts.
\eb
We first show that if $u,v$ are solutions to the scattering problem~\eqref{e:LPX}, then
\vspace*{-1ex}
\[
	\frac{\partial}{\partial x}\big(u^\dagger(x,t,k) v(x,t,k)\big) = 0\,.
\nonumber
\]
%
Indeed, the symmetry $X^\dagger(x,t,k) = -X(x,t,k)$ gives
\vspace*{-1ex}
\be
	\frac{\partial}{\partial x}\big(u^\dagger v\big) = u_x^\dagger v + u^\dagger v_x
	= u^\dagger X^\dagger v + u^\dagger Xv
	= -u^\dagger X v + u^\dagger X v = 0\,.
\nonumber
\ee
For $k \in \Sigma_{+1}^o$, taking $u = v = \phi_{+1}$ or $u=v=\phi_{+2}$ gives
\vspace*{-1ex}
\[
	\frac{\partial}{\partial x}\big(\phi_{+j}^\dagger(x,t,k) \phi_{+j}(x,t,k)\big) = 0\,, \quad j = 1,2\,.
\nonumber
\]
Using the symmetry~\eqref{phi_+1 conjugate} and taking the limit as $x \rightarrow \infty$, we see that
\bb
\[
	\phi^\dagger_{+j}(x,t,k) \phi_{+j}(x,t,k) = 1\,, \quad j=1,2\,.
\nonumber
\]
\eb
If either $s_{12}(k_o)=0$ or $s_{22}(k_o)=0$ for some $k_o \in \Sigma_{+1}^o$, the Wronskians~\eqref{s_12} and~\eqref{s_22} give $\phi_{+2}(x,t,k_o) = c_o\phi_{-2}(x,t,k_o)$ or $\phi_{+1}(x,t,k_o) = c_o\phi_{-2}(x,t,k_o)$ for some $c_o \in \C$. 

With $V\not=0$ and $k \in \Sigma_{+1}$, we have $\phi_{-2}(x,t,k_o) \rightarrow 0$ as $x\rightarrow -\infty$ so that, for the appropriate $j$,
\be
	\phi^\dagger_{+j}(x,t,k)\phi_{+j}(x,t,k) = \lim_{x\rightarrow -\infty} \big(\conj{c_o}\phi^\dagger_{-2}(x,t,k)\big)\big(c_o\phi_{-2}(x,t,k)\big) = 0\,,
\nonumber
\ee
which is a contradiction. Thus $s_{12}(k)\not=0$ and $s_{22}(k)\not=0$ for all $k \in \Sigma_{+1}^o$. The symmetries give $s_{21}(k) \not=0$ and $s_{11}(k) \not=0$ for $k \in \Sigma_{+2}^o$.

Similar argument shows that $s_{21}(k)\not=0$ and $s_{22}(k)\not=0$ for all $k \in \Sigma_{-1}^o$. The symmetries again give $s_{12}(k) \not=0$ and $s_{11}(k) \not=0$ for $k \in \Sigma_{-2}^o$.
\endproof

\paragraph{Proof of Lemma~\ref{l:residues} (Residues of Jost solutions).}
\bb
We now express the residues of $\Phi(x,t,k)$.
From Corollary~\ref{Wronskians}, we see that if $s_{22}(k_o)=0$ for some discrete eigenvalue $k_o \in \C^+\setminus\Sigma$, then
\[
	\Wr[ \phi_{+1}(x,t,k_o),\phi_{-2}(x,t,k_o) ]=0\,, \quad \forall x,t\in \R\,.
	\nonumber
\]
Neither $\phi_{+1}(x,t,k_o)$ nor $\phi_{-2}(x,t,k_o)$ can be identically zero due to the normalizations in~\eqref{e:phi x asymptotics}. Then, %
\[
	\phi_{-2}(x,t,k_o) = C_n \phi_{+1}(x,t,k_o)\,, \quad \forall x,t\in \R\,, \quad C_n \neq 0\,.
	\nonumber
\]
Now $s_{11}(\conj{k_o}) = 0$ by Lemma~\ref{first symmetry extended}. Hence, $\phi_{-1}(x,t,\conj{k_o})$ and $\phi_{+2}(x,t,\conj{k_o})$ are also proportional,  and $\conj{k_o}$ is a discrete eigenvalue as well. In particular, from Lemma~\ref{first symmetry Jost} we have
\be
	-\sigma_\ast \conj{\phi_{-1}(x,t,\conj{k_o})} = C_n \sigma_\ast\conj{\phi_{+2}(x,t,\conj{k_o})}\,, \quad \forall x,t\in\R\,,
	\nonumber
\ee
so that
\[
	\phi_{-1}(x,t,\conj{k_o}) = -\conj{C_n} \phi_{+2}(x,t,\conj{k_o})\,, \quad \forall x,t\in \R\,.
	\nonumber
\]
If $k_o$ is a simple root of $s_{22}(k)$ so that $s_{22}'(k_o) \neq 0$, then
\bb
\bse
\label{e:column residues}
\begin{gather}
		\Res_{k=k_o} \Big[\frac{\phi_{-2}(x,t,k)}{s_{22}(k)}\Big] = \phantom{-}c_n\, \phi_{+1}(x,t,k_o)\,, 
		\\
		\Res_{k=\conj{k_o}} \Big[ \frac{\phi_{-1}(x,t,k)}{s_{11}(k)} \Big] = -\conj{c_n}\, \phi_{+2}(x,t,\conj{k_o})\,, \end{gather}
\ese
\eb
with
\[
	c_n = \frac{C_n}{s_{22}'(k_o)}\,.
	\nonumber
\]
Writing the relations~\eqref{e:column residues} in terms of $\Phi(x,t,k)$ then gives the result.\endproof
\eb

\subsection{Inverse problem}
\label{a:Inverse problem}

\paragraph{Proof of Lemma~\ref{l:jump matrix} (Calculation of the jump matrices).}
\label{Calculation of the jump matrices appendix}
\bb
Here we compute the jump conditions satisfied by $M(x,t,k)$ as defined by~\eqref{M definition}.
\eb
\bb
For simplicity of calculation, we first compute the jump matrices for $\Phi(x,t,k)$ as given by~\eqref{e:Phi}. Noting that
\be
\label{Psi to M}
\Phi(x,t,k) = M(x,t,k)\e^{i\theta_o(x,t,k)\sigma_3}\,, \quad k \in \C \setminus \Sigma\,,
\nonumber
\ee
\eb
the jump condition~\eqref{e:M jump condition generic} for $M(x,t,k)$ is equivalent to the jump condition
\[
\label{e:psi jump}
	\Phi^+(x,t,k) = \Phi^-(x,t,k)J_o(k), \quad k \in \Sigma^o\,,
	\nonumber
\]
for $\Phi(x,t,k)$, where $J(x,t,k)$ and $J_o(k)$ are related by~\eqref{e:JfromJo}.
The symmetries in Lemma~\ref{first symmetry Jost} imply that
\[
\label{J_o symmetry}
	J_o(k) = -\sigma_\ast \conj{J_o(\conj{k})} \sigma_\ast, \quad k \in \Sigma_\pm^o\,. 
\]
We now compute $J_o(k)$ for $V\not= 0$. 
%
\paragraph{Jump for $k \in \R$.}
Rearranging~\eqref{scattering relation}, we have
\bb
\begin{align*}
		\phi_{+1}(x,t,k) &= \frac{\phi_{-1}(x,t,k)}{s_{11}(k)} -\frac{s_{21}(k)}{s_{11}(k)}\phi_{+2}(x,t,k), & &k \in \R\,,
		\\
		\frac{\phi_{-2}(x,t,k)}{s_{22}(k)} &= \frac{s_{12}(k)}{s_{22}(k)} \frac{\phi_{-1}(x,t,k)}{s_{11}(k)} + \Big(1-\frac{s_{21}(k)s_{12}(k)}{s_{11}(k)s_{22}(k)}\Big)\phi_{+2}(x,t,k), & &k \in \R\,.
\end{align*}
\eb
Recalling $\rho(k)$ as defined in~\eqref{rho definition} and the symmetry~\eqref{rho symmetry}, we then have
\be
\label{R jump}
	\Phi^+(x,t,k) = \Phi^-(x,t,k)
\begin{pmatrix}
		1 & \rho(k)\\ 
		\conj{\rho(k)} & 1+\rho(k)\conj{\rho(k)}
\end{pmatrix}, 
	\quad k \in \R\,.
\ee
%
\paragraph{Jumps for $k \in \Sigma_{+}^o$.}
Recall that $\phi_{-2}(x,t,k)$ is analytic for $k \in\Sigma_{+1}^o$. Then
\bb
\begin{align*}
	\Big(\frac{\phi_{-2}(x,t,k)}{s_{22}(k)} \Big)^+ = \frac{s_{22}^-(k)}{s_{22}^+(k)}\Big(\frac{\phi_{-2}(x,t,k)}{s_{22}(k)} \Big)^-, \quad k \in \Sigma_{+1}^o\,.
\end{align*}
Lemma~\ref{l:second symmetry scattering coefficients} then gives
\be
\label{s_22 jump Sigma_{+1}}
	\frac{s_{22}^-(k)}{s_{22}^+(k)} = -\frac{i\e^{i\delta}}{\rho(k)}\,, \quad k\in\Sigma_{+1}^o\,.
	\nonumber
\ee
\eb
Recalling~\eqref{scattering relation 2} which was extended to $\Sigma_{+1}^o$, we have
\be
	\frac{\phi_{-2}(x,t,k)}{s_{22}(k)} = \rho(k)\phi_{+1}(x,t,k) + \phi_{+2}(x,t,k)\,, \quad k \in \Sigma_{+1}^o\,.
\nonumber
\ee
\bb
Solving for $\phi_{+2}(x,t,k)$ and using~\eqref{second symmetry Sigma_{+1}}, we see that
\be
	\phi_{+1}^+(x,t,k) 
        = i\rho(k)\e^{-i\delta}\phi_{+1}(x,t,k) - i\e^{-i\delta}\frac{\phi_{-2}(x,t,k)}{s_{22}(k)}\,,
  \quad k \in \Sigma_{+1}^o\,.
\nonumber
\ee
\eb
Together we have
\be
\label{Sigma_{+1} jump}
	\Phi^+(x,t,k) = \Phi^-(x,t,k) 
\begin{pmatrix}
		i\rho(k)\e^{-i\delta} & 0\\
		-i\e^{-i\delta} & -i\e^{i\delta}/\rho(k)
\end{pmatrix}, 
	\quad k \in \Sigma_{+1}^o\,.
\ee
The symmetry~\eqref{J_o symmetry} gives
\be
\label{Sigma_{+2} jump}
	\Phi^+(x,t,k) = \Phi^-(x,t,k)
\begin{pmatrix}
		i\e^{-i\delta}/\conj{\rho(\conj{k})} & -i\e^{i\delta}\\
		0 & -i\conj{\rho(\conj{k})}\e^{i\delta}
\end{pmatrix}, 
	\quad k \in \Sigma_{+2}^o\,.
\ee
%
\paragraph{Jumps for $k \in \Sigma_{-}^o$.}
Recall that $\phi_{+1}(x,t,k)$ is analytic for $k\in\Sigma_{-1}^o$. Then
\bb
\be
	\phi_{+1}^+(x,t,k) = \phi_{+1}^-(x,t,k)\,, \quad k \in \Sigma_{-1}^o\,.
\nonumber
\ee
\eb
Solving~\eqref{linear combination 2} for $\phi_{-1}(x,t,k)$ we have
\be
	\phi_{-1}(x,t,k) = \frac{1}{r_{11}(k)}\phi_{+1}(x,t,k) - \frac{r_{21}(k)}{r_{11}(k)}\phi_{-2}(x,t,k)\,, \quad k \in \Sigma_{-1}^o\,.
\nonumber
\ee
\bb
Lemma~\ref{l:second symmetry scattering coefficients} then gives
\be
	\Big(\frac{\phi_{-2}(x,t,k)}{s_{22}(k)}\Big)^+ = \frac{1}{s_{21}(k)r_{11}(k)} \phi_{+1}(x,t,k) - \frac{r_{21}(k)s_{22}(k)}{s_{21}(k)r_{11}(k)}\Big(\frac{\phi_{-2}(x,t,k)}{s_{22}(k)}\Big), \quad k \in \Sigma_{-1}^o\,.
\nonumber
\ee
\eb
From~\eqref{R S relations} we see 
\be
	s_{21}(k)r_{11}(k) = -r_{21}(k)s_{22}(k) = -1/r(k), \quad k \in \R\cup\Sigma_{-1}^o,
\nonumber
\ee
with $r(k)$ as defined in~\eqref{r definition}.
Together we have
\be
\label{Sigma_{-1} jump}
	\Phi^+(x,t,k) = \Phi^-(x,t,k) 
\begin{pmatrix}
		1 & -r(k)\\
		0 & 1
\end{pmatrix}, 
	\quad k \in \Sigma_{-1}^o\,.
\ee
The symmetry~\eqref{J_o symmetry} gives
\be
\label{Sigma_{-2} jump}
	\Phi^+(x,t,k) = \Phi^-(x,t,k) 
\begin{pmatrix}
		1 & 0\\
		\conj{r(\conj{k})} & 1
\end{pmatrix}, 
	\quad k \in \Sigma_{-2}^o\,.
	\endproof
\ee

\bb
\paragraph{Proof of Lemmas~\ref{nonzero velocity growth conditions generic case} and~\ref{nonzero velocity growth conditions exceptional case} (Growth conditions).}

The growth conditions follow immediately from the definition of $M(x,t,k)$ and Corollaries~\ref{phi branch point continuity} and~\ref{wr branch point continuity} in the generic case, or Corollaries~\ref{phi branch point expansion} and~\ref{wr branch point expansion} in the considered exceptional case. To see that the matrices $B_{p_\pm}^{(0)}$ and $B_{\conj{p_\pm}}^{(0)}$ are invertible in the generic case, taking determinants of both sides of~\eqref{e:growth condition generic case} yields
\be
	\det M(x,t,k) = \begin{cases}
		\det\big[B_{p_\pm}^{(0)}(x,t) + o(1)\big] \,, &k\rightarrow p_{\pm}\,,
		\\
		\det\big[B_{\conj{p_\pm}}^{(0)}(x,t) + o(1)\big] \,, &k\rightarrow \conj{p_{\pm}}\,.
	\end{cases}
	\nonumber
\ee
Since $M(x,t,k)$ has unit determinant for all $k\in\C\setminus\Sigma$, the invertibility of $B_{p_\pm}^{(0)}$ and $B_{\conj{p_\pm}}^{(0)}$ follows.

On the other hand, for the considered exceptional case, taking determinants of both sides of~\eqref{e:growth condition exceptional case} yields
\be
	\det M(x,t,k) = \begin{cases}
	 \det\big[B_{p_{\pm}}^{(0)}(x,t) + O(k-p_\pm)^{1/2}\big](k-p_+)^{-1/2}\,, &k\rightarrow p_{\pm}\,,
		\\
		\det\big[B_{\conj{p_\pm}}^{(0)}(x,t) + O(k-p_\pm)^{1/2}\big](k-\conj{p_\pm})^{-1/2}\,, &k\rightarrow \conj{p_{+}}\,,
	\end{cases}
	\nonumber
\ee
from which we see that we must have $\det B_{p_\pm}^{(0)}(x,t) = \det B_{\conj{p_\pm}}^{(0)}(x,t) = 0$. \endproof
\eb

\paragraph{Proof of Theorem~\ref{t:integral equation for M} (Linear algebraic-integral equations).}

\bb
Here we convert the RHP~\ref{nonzero velocity RHP} to a set of linear algebraic-integral equations.
\eb

\begin{figure}[t!]
	\centering
\begin{subfigure}{.5\textwidth}
		\centering
		\includegraphics[width=0.8\textwidth]{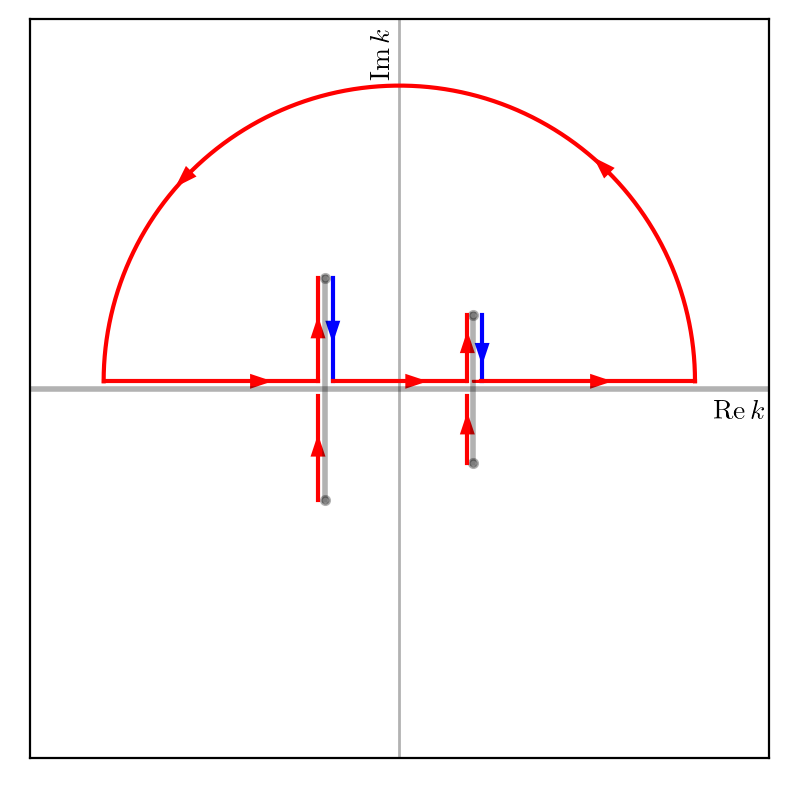}
\end{subfigure}%
\begin{subfigure}{.5\textwidth}
		\centering
		\includegraphics[width=0.8\textwidth]{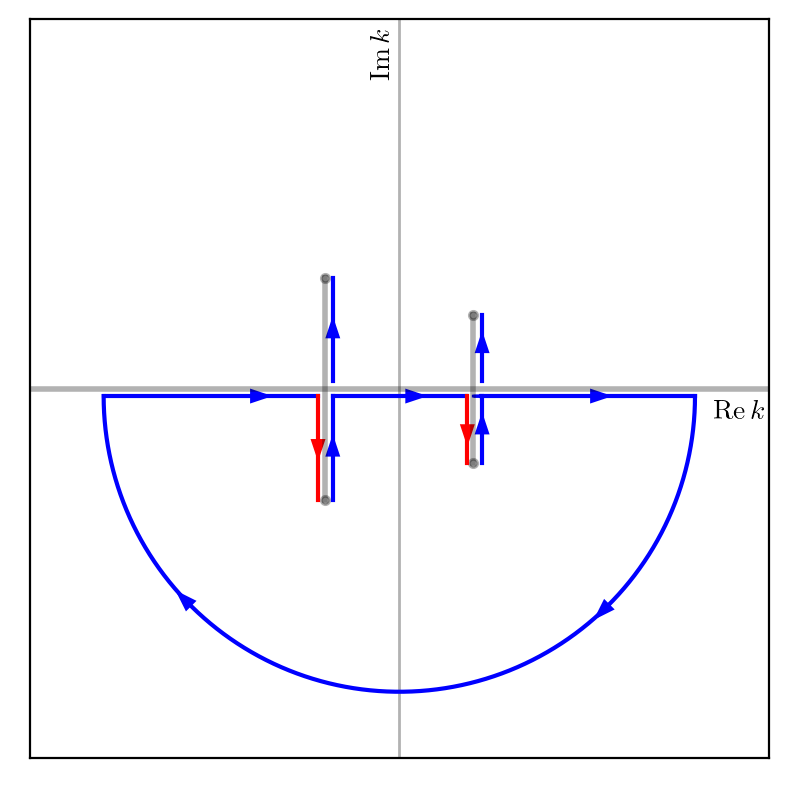}
\end{subfigure}
	\caption{
	\bb
	Closing the contours of integration for the Cauchy projector $P$ applied to $\^M^+$ (left) and $\^M^-$ (right). Red and blue contours involve integrals with $\^M^+$ and $\^M^-$ respectively. The integrals over the blue (red) contours on the left (right) are added and subtracted in order to close the contour in the upper (lower) half-plane.
	\eb
	}
\label{f:cauchy projector contours}
\end{figure}

Letting
\be
	\^M(k) = M(k) - \I - \sum_{n=1}^N \frac{\Res_{\xi=k_n} [M(\xi)]}{k-k_n} - \sum_{n=1}^N \frac{\Res_{\xi=\conj{k_n}} [M(\xi)]}{k-\conj{k_n}}\,,
\nonumber
\ee
we see that $\^M(k)$ satisfies a modified RHP, similar to the RHP~\ref{nonzero velocity RHP}, but where the jump condition~\eqref{e:M jump condition generic} is replaced by
\be
\^M^+(k) = \^M^-(k) + M^-(k)(J(k) - \I)\,, \quad k \in \Sigma^o\,,
\nonumber
\ee
and $\^M(k) = O(1/k)$ as $k\rightarrow \infty$.
Introducing the Cauchy projector,
\be
	P[f](k) = \frac{1}{2\pi i} \int_\Sigma \frac{f(\xi)}{\xi - k} d\xi, \quad k \in \C\setminus\Sigma\,,
\nonumber
\ee
for $f:\Sigma \rightarrow \C$ and applying $P$ to $\^M^\pm(k)$, we see that
\be
	P[\^M^\pm](k) = \frac{1}{2\pi i} \Bigg[\int_{\Sigma_{+2}\cup\Sigma_{-2}} \frac{\^M^+(\xi)}{\xi - k}d\xi + \int_{\Sigma_{+1}\cup\Sigma_{-1}} \frac{\^M^-(\xi)}{\xi - k}d\xi \Bigg]
	+ \begin{cases}
		\pm \^M(k), & k \in \C^\pm \setminus \Sigma\,,\\
		0,					 & k \in \C^\mp \setminus \Sigma\,.
		\end{cases}
\nonumber
\ee
The above expression is obtained using the analyticity of $\^M(k)$ in $\C\setminus\Sigma$ to close the contour in the appropriate half-plane. To do so, one must add and subtract the integral along the ``opposite" side of $\Sigma_+$ and $\Sigma_-$ in that half-plane.
\bb
Specifically, consider $P[\^M^+](k)$. Writing
\begin{align}
	P[\^M^+](k) &= \frac{1}{2\pi i}\Bigg[\int_{\Sigma_{+2}\cup\Sigma_{-2}} \frac{\^M^+(\xi)}{\xi-k}d\xi + \int_{\Sigma_{+1}\cup\Sigma_{-1}} \frac{\^M^-(\xi)}{\xi-k}d\xi \Bigg]
	\nonumber
	\\
	&\quad+ \frac{1}{2\pi i}\Bigg[\int_{\R \cup \Sigma_{+1}\cup\Sigma_{-1}} \frac{\^M^+(\xi)}{\xi-k}\d\xi + \int_{-(\Sigma_{+1}\cup\Sigma_{-1})} \frac{\^M^-(\xi)}{\xi-k}d\xi\Bigg]\,,
	\nonumber
\end{align}
we see that the contour for the last two terms can be closed in the upper half-plane (see Fig.~\ref{f:cauchy projector contours}). When $k$ is in the upper half-plane, these last two terms then reduce to $\^M(k)$, while they reduce to $0$ when $k$ is in the lower half-plane. Similar argument gives the corresponding result for $P[\^M^-](k)$.

\eb
Applying $P$ to the jump condition for $\^M(k)$ then gives
\be
	\~M(k) = \frac{1}{2\pi i} \int_\Sigma \frac{M^-(\xi)(J(\xi)-I)}{\xi-k}d\xi, \quad k \in \C\setminus\Sigma\,.
\nonumber
\ee
This then gives the algebraic-integral equation
\be
\label{integral equation for M}
	M(k) =\I+ \sum_{n=1}^N \frac{\Res_{\xi=k_n} [M(\xi)]}{k-k_n} 
	+ \sum_{n=1}^N \frac{\Res_{\xi=\conj{k_n}} [M(\xi)]}{k-\conj{k_n}} 
	+ \frac{1}{2\pi i} \int_\Sigma \frac{M^-(\xi)(J(\xi)-I)}{\xi-k}d\xi\,, \quad k \in \C\setminus\Sigma\,.
\ee
%
%
Evaluating the first and second columns of~\eqref{integral equation for M} at $k=k_n$ and $k=\conj{k_n}$ respectively 
\bb
and using the residue conditions~\eqref{residues}
\eb
then gives
\eqref{e:M algebraic 1} and~\eqref{e:M algebraic 2},
which closes the system.

Recall that $M^-(k)$ has singularities at the branch points. As such, the convergence of the above improper integrals over $\Sigma$ must be considered. The growth conditions~\eqref{e:growth condition generic case} show that $M(k)$ has at worst fourth-root singularity at the branch points. 
%
\bb
On the other hand, Corollary~\ref{c:reflection coefficients branch point behavior generic case} shows that $J(k)$ is continuous at the branch points.
\eb
Thus we see that the integrands in each improper integral have at worst fourth-root singularity at the branch points, so that the integrals converge.
\endproof

\paragraph{Proof of Lemma~\ref{l:M Lax pair} ($M(x,t,k)$ satisfies the modified Lax pair).}
\bb
We now show that if $M(x,t,k)$ solves the RHP~\ref{nonzero velocity RHP} then it also satisfies the modified Lax pair~\eqref{e:LP} with $Q(x,t)$ given by~\eqref{e:Q(x,t) inverse problem definition}, under the assumption of an appropriate vanishing lemma. Specifically, we assume that a modified RHP~\ref{nonzero velocity RHP} with the normalization instead given as $M(x,t,k) = O(1/k)$, $k \to \infty$ has only the trivial solution. 
\eb

We first define $\~X(x,t,k)$ and $\~T(x,t,k)$ by
\begin{align*}
		\~X(x,t,k)M(x,t,k) &= M_x(x,t,k) - ik\big[\sigma_3\,,M(x,t,k)\big] - Q(x,t)M(x,t,k)\,,
		\\
		\begin{split}
			\~T(x,t,k)M(x,t,k) &= M_t(x,t,k) + 2ik^2\big[\sigma_3\,, M(x,t,k)\big]
			\\
			&\quad - i\sigma_3\big(Q_x(x,t) - Q^2(x,t)\big)M(x,t,k) + 2kQ(x,t)M(x,t,k)\,,
		\end{split}
	\end{align*}
where $Q(x,t)$ is given by~\eqref{e:Q(x,t) inverse problem definition}. In particular,
$	Q(x,t) = -i\big[\sigma_3\,, M^{(1)}(x,t)\big]$,
where $M^{(1)}(x,t)$ is defined by the asymptotic expansion
\[
\label{e:M expansion}
	M(x,t,k) = \I + \frac{M^{(1)}(x,t)}{k} + \frac{M^{(2)}(x,t)}{k^2} + O(1/k^3)\,.
\]
We will show that $\~X(x,t,k)M(x,t,k)$ and $\~T(x,t,k)M(x,t,k)$ satisfy the same jump condition as $M(x,t,k)$ but are $O(1/k)$ as $k\to\infty$. 
\bb
Assumption the above stated vanishing lemma,
\eb
we then conclude that 
\[
\label{e:vanishing lemma}
	\~X(x,t,k)M(x,t,k) = 0\,, \quad \~T(x,t,k)M(x,t,k) = 0\,.
\]
Indeed, using the asymptotic expansion~\eqref{e:M expansion}, simple algebra yields $\~X(x,t,k)M(x,t,k) = O(1/k)$ as $k \rightarrow \infty$. Moreover,
\begin{align*}
	(\~X(x,t,k)M(x,t,k))^+ &= \~X(x,t,k)(M^-(x,t,k)J(x,t,k)) 
												 \\
												 &= M_x^-(x,t,k)J(x,t,k) + M^-(x,t,k) J_x(x,t,k) 
												 \\
												 &\quad-ik\big[\sigma_3, M^-(x,t,k)J(x,t,k)] - Q(x,t)M^-(x,t,k)J(x,t,k)\,,
\end{align*}
where $J(x,t,k)$ is the jump matrix~\eqref{nonzero velocity jump matrix}.
Noting that $J_x(x,t,k) = ik\big[\sigma_3\,, J(x,t,k)\big]$ and more algebra gives
\[
	(\~X(x,t,k)M(x,t,k))^+ = (\~X(x,t,k)M(x,t,k))^-J(x,t,k)\,, \quad k\in\Sigma\,,
\nonumber
\]
and so we conclude that 
\[
  \label{e:XM=0}
	\~X(x,t,k)M(x,t,k) = 0.
\]
To see that $\~T(x,t,k)M(x,t,k)$ is $O(1/k)$, note that
\begin{align*}
	Q^2(x,t) &= i\sigma_3 Q(x,t)M^{(1)}(x,t) + iQ(x,t)M^{(1)}(x,t)\sigma_3\,,
	\\
	Q_x(x,t) &= 2\big[M^{(2)}\sigma_3\,, \sigma_3\big] - i\big[\sigma_3\,, Q(x,t)M^{(1)}(x,t)\big]\,.
\end{align*}
In the above, we have used the facts that $\sigma_3$ and $Q(x,t)$ anti-commute and that~\eqref{e:XM=0} implies
\[
	M^{(1)}_x(x,t) = i\big[\sigma_3\,, M^{(2)}(x,t)\big] + Q(x,t)M^{(1)}(x,t)\,.
\nonumber
\]
From there, straightforward algebra shows that $\~T(x,t,k)M(x,t,k) = O(1/k)$ as $k \to \infty$. Arguing as with the jump for $\~X(x,t,k)M(x,t,k)$, instead noting that $J_t(x,t,k) = -2ik^2\big[\sigma_3\,, J(x,t,k)\big]$, shows that 
\[
	(\~T(x,t,k)M(x,t,k))^+ = (\~T(x,t,k)M(x,t,k))^-J(x,t,k)\,,\quad k\in\Sigma\,,
\nonumber
\]
from which we conclude~\eqref{e:vanishing lemma}. Equivalently, $M(x,t,k)$ satisfies the modified Lax pair~\eqref{e:modified Lax pair} with $Q(x,t)$ defined by~\eqref{e:Q(x,t) inverse problem definition}.
\endproof

\paragraph{Proof of Corollary~\ref{c:reconstruction formula} (Reconstruction formula).}
\bb
We now find the solution $q(x,t)$ corresponding to a solution $M(x,t,k)$ to the RHP~\ref{nonzero velocity RHP}, assuming the same vanishing lemma used in Lemma~\ref{l:M Lax pair}.
\eb

Lemma~\ref{l:M Lax pair} shows that $M(x,t,k)\e^{i\theta_o(x,t,k)\sigma_3}$ satisfies the Lax pair~\eqref{e:LP} with $Q(x,t)$ given by~\eqref{e:Q(x,t) inverse problem definition}.
Then defining
\be
\label{q from M}
	q(x,t) := Q_{12}(x,t) = -2i\lim_{k\to\infty} [k M_{12}(x,t,k)]\,,
\ee
we see that $q(x,t)$ satisfies the NLS equation~\eqref{e:fNLS} as the compatibility condition of the Lax pair~\eqref{e:LP}.
Moreover, from~\eqref{integral equation for M} we see
\be
\label{M asymptotics}
	M(k) = \I + \frac{1}{k}\Big[\sum_{n=1}^N \Res_{\xi=k_n} [M(\xi)] + \sum_{n=1}^N \Res_{\xi=\conj{k_n}} [M(\xi)] + \frac{1}{2\pi i} \int_\Sigma M^-(\xi)(J(\xi)-I)d\xi \Big] + o(1/k)\,, \quad k\rightarrow \infty\,.
\ee
Combining~\eqref{residue conditions},~\eqref{q from M} and~\eqref{M asymptotics} then gives~\eqref{e:reconstruction formula}. 
\endproof

\bb
\paragraph{Proof of Theorem~\ref{c:alternative RHP existence and uniqueness} (Existence and uniqueness of solutions for the modified RHP~\ref{d:alternative RHP}).}

Here we establish the existence of a unique solution for the modified RHP~\ref{d:alternative RHP}. We do so by showing that the jump matrix $\~J(x,t,k)$ satisfies the conditions of Lemma~\ref{l:Zhou's lemma}.

Condition (a) is satisfied by the assumptions that $\rho(k) \in C^1(\R\setminus(-R,R))$ and $C(k)\in C^1(\partial B_R \cap \C^\pm)$. Condition (b) is satisfied by the assumption $C(k)C^\dagger(k) = \I$. It is straightforward to verify that condition (c) is met due to the structure of $\~J(x,t,k)$. Indeed, for $k \in \~\Sigma\cap\R$ we have $\~J^\dagger(x,t,k) = \~J(x,t,k)$, so that $\~J(x,t,k)$ is an invertible Hermitian matrix and is thus positive definite. We then conclude that $\Re\~J(x,t,k)$ is positive definite.
\endproof

\paragraph{Proof of Theorem~\ref{t:RHP uniqueness} (Uniqueness of solutions for the original RHP~\ref{nonzero velocity RHP}).}
The proof relies on the existence of a solution of the original RHP at $(x,t)=(0,0)$ to construct the map $F_o$. 
We need to show that $\~M(x,t,k) = F_o(M)(x,t,k)$ solves the modified RHP~\ref{d:alternative RHP} (after filling all removable singularities) for any solution $M(x,t,k)$ of the original RHP~\ref{nonzero velocity RHP}.
The key here is that, even though $C_o(k)$ is fixed by this given solution, 
when applied to \textit{any other} solution $M(x,t,k)$ of the original RHP, the map $F_o(M)$ still produces a solution of the modified RHP.
Essentially, this is because $\Phi(x,t,k) = M(x,t,k)\,\e^{i\theta_o(x,t,k)\sigma_3}$
[cf.\ \eqref{M definition}] is a fundamental matrix solution of the Lax pair.
However, given any fundamental matrix solution $\phi(x,t,k)$ of both parts of the Lax pair, $\phi(x,t,k)c(k)$ is also a fundamental matrix solution of both parts of the Lax pair for any invertible matrix $c(k)$.

It is immediately clear that the normalization at infinity is satisfied, and that the jump condition is satisfied on $(-\infty,R)\cup(R,\infty)$. It is straightforward to check that $\~M(x,t,k)$ satisfies the jump condition on $\partial B_R$. Indeed, we have
\be
	\~M^+(x,t,k) = M(x,t,k) = \~M^-(x,t,k)\e^{i\theta_o(x,t,k)\sigma_3}C_o(k)\e^{-i\theta_o(x,t,k)\sigma_3}\,, \quad k \in \partial B_R \cap \C^+\,,
	\nonumber
\ee
which verifies the jump on $\partial B_R \cap \C^+$, with similar calculation for the jump on $\partial B_R \cap \C^-$. Continuing, it is easy to show that the jump on $\Sigma\cap B_R$ is removed. Indeed, for $k\in\Sigma\setminus B_R$ we have
\be
	\begin{aligned}
	\~M^+(x,t,k) &= M^+(x,t,k)\e^{i\theta_o(x,t,k)\sigma_3}(C_o^+(k))^{-1}\e^{-i\theta_o(x,t,k)\sigma_3}
	\\
					     &= M^-(x,t,k)J(x,t,k)\e^{i\theta_o(x,t,k)\sigma_3}(C_o^-(k)J_o(k))^{-1}\e^{-i\theta_o(x,t,k)\sigma_3}\\
							 &= M^-(x,t,k)J(x,t,k)\e^{i\theta_o(x,t,k)\sigma_3}J_o^{-1}(k)\e^{-i\theta_o(x,t,k)\sigma_3}\e^{i\theta_o(x,t,k)\sigma_3}(C_o^-(k))^{-1}\e^{-i\theta_o(x,t,k)\sigma_3}
							\\
							 &= \~M^-(x,t,k)\,,
	\end{aligned}
	\nonumber
\ee
since $J(x,t,k)\e^{i\theta_o(x,t,k)\sigma_3}J_o^{-1}(k)\e^{-i\theta_o(x,t,k)\sigma_3} = \I$.

It remains to be shown that $\~M(x,t,k)$ is analytic at the discrete eigenvalues and branch points. Straightforward algebra using the residue conditions for $M(x,t,k)$ and $C_o(k)$ shows that $\~M(x,t,k)$ has no residues at the discrete eigenvalues. Finally, to see that $\~M(x,t,k)$ is analytic at the branch points, the growth conditions~\eqref{e:growth condition generic case} for $M(x,t,k)$ and $C_o(k)$ in the generic case show that $\~M(x,t,k)$ has well-defined limit and is indeed analytic at the branch points (note that applying determinants to the growth conditions shows that the limiting matrix values must be invertible).
We then see that $\~M(x,t,k)$ satisfies the RHP~\ref{d:alternative RHP} with $C_o(k)$ in place of $C(k)$. As in Theorem~\ref{c:alternative RHP existence and uniqueness}, the assumptions that $\rho(k) \in C^1(\R\setminus(-R,R))$ and $C_o(k) \in C^1(\partial B_R \cap \C^\pm)$ together with $C_o(k)C_o^\dagger(k)=\I$ show that this modified RHP admits a unique solution.

We now show that solutions of the original RHP are unique.
Suppose that $M(x,t,k)$ and $M'(x,t,k)$ are two distinct solutions of the original RHP,
and let $\~M(x,t,k) = F_o(M)(x,t,k)$ 
and $\~M'(x,t,k) = F_o(M')(x,t,k)$.
Since we have already proved that solutions of the modified RHP are unique, 
we have that $\~M(x,t,k) = \~M'(x,t,k)$.
But since the map~\eqref{e:M to tilde M} is invertible, 
the equality of $\~M(x,t,k)$ and $\~M'(x,t,k)$ immediately implies the equality of 
$M(x,t,k)$ and $M'(x,t,k)$.
\endproof
\eb

\bb
\subsection{Reductions}
\eb

\paragraph{Proof of Lemma~\ref{l:second symmetry scattering coefficients zero velocity} (Second symmetry, scattering coefficients with $V=0$).}

\bb
Here we recompute the jumps of the scattering coefficients across the branch cuts in the case that $V=0$.
\eb

The Wronskian representations~\eqref{e:Wronskians zero velocity} give
\bb
\be
	s_{22}^+(k) = \Wr\big[\phi_{+1}^+(x,t,k)\,, \phi_{-2}^+(x,t,k)\big]\,, \quad k\in\Sigma_{+1}^o\,.
\nonumber
\ee
\eb
The calculation for $k \in \Sigma_{+1}^o\setminus\Sigma_{-1}$ follows exactly as in the proof of Lemma~\ref{l:second symmetry scattering coefficients} for $k \in \Sigma_{+1}^o$. We proceed with the calculation for $k\in\Sigma_{+1}^o\cap\Sigma_{-1}^o$. From Lemma~\ref{second symmetry} 
\bb
and the Wronskian representation~\eqref{s_11}
\eb
we see
\bb
\be
	s_{22}^+(k) = \e^{-2i\delta}s_{11}(k)\,, \quad k \in \Sigma_{+1}^o\cap\Sigma_{-1}^o\,,
\nonumber
\ee
\eb
The symmetries~\eqref{e:first symmetry} give the corresponding jump for $s_{11}(k)$.
\bb
The jumps for $r_{11}(k)$ and $r_{22}(k)$ are then given by~\eqref{R S relations}.
\eb
\endproof

\paragraph{Proof of Lemma~\ref{l:jump matrix zero velocity} (Calculation of the jump matrices for $V=0$).} 
\label{Zero velocity appendix}

We now compute the jumps in the special case $V = 0$.
We need only compute the jumps for $k \in \Sigma_{+1}^o \cap \Sigma_{-1}^o$ and $k \in \Sigma_{+2}^o \cap \Sigma_{-2}^o$, as remaining jumps on $\R$, $\Sigma_{+1}^o\setminus \Sigma_{-1}$ and $\Sigma_{+2}^o\setminus \Sigma_{-2}$ will match the corresponding jumps on $\R$, $\Sigma_{+1}^o$ and $\Sigma_{+2}^o$ with $V\not=0$. Note that~\eqref{scattering matrix relation} and~\eqref{R scattering matrix relation} can now be extended to $(\Sigma_{+1}^o \cap \Sigma_{-1}^o) \cup (\Sigma_{+2}^o \cap \Sigma_{-2}^o)$.

We first consider the jump for $k \in \Sigma_{+1}^o \cap \Sigma_{-1}^o$. Unlike in the case of nonzero velocity, both $\phi_{+1}(x,t,k)$ and $\phi_{-2}(x,t,k)$ have jumps across the contour. However, we again find
\bb
\[
	\phi_{+1}^+(x,t,k) = i\rho(k)\e^{-i\delta}\phi_{+1}(x,t,k) - i\e^{-i\delta}\frac{\phi_{-2}(x,t,k)}{s_{22}(k)}\,, \quad k \in \Sigma_{+1}^o \cap \Sigma_{-1}^o\,.
\nonumber
\]
\eb
From Lemma~\ref{c:extendedscattering zero velocity}, we have
\[
	\frac{\phi_{-1}(x,t,k)}{s_{11}(k)} = (1 + \rho(k)\conj{\rho(\conj{k})})\phi_{+1}(x,t,k) - \conj{\rho(\conj{k})}\frac{\phi_{-2}(x,t,k)}{s_{22}(k)}\,, \quad k \in \Sigma_{+1}^o \cap \Sigma_{-1}^o\,.
\nonumber
\]
Combining with~\eqref{second symmetry Sigma_{-1}} and Lemma~\ref{e:second symmetry scattering coefficients zero velocity} gives
\bb
\be
	\Big(\frac{\phi_{-2}(x,t,k)}{s_{22}(k)}\Big)^+ 
			 = -i(1+\rho(k)\conj{\rho(\conj{k})})\e^{i\delta}\phi_{+1}(x,t,k) + i\conj{\rho(\conj{k})}\e^{i\delta}\frac{\phi_{-2}(x,t,k)}{s_{22}(k)}\,,
			\quad k\in\Sigma_{+1}^o \cap \Sigma_{-1}^o\,.
\nonumber
\ee
\eb
Together, we have
\[
\Phi^+(x,t,k) = \Phi^-(x,t,k)
\begin{pmatrix} 
		i\rho(k)\e^{-i\delta} & -i(1+\rho(k)\conj{\rho(\conj{k})})\e^{i\delta} \\
		-i\e^{-i\delta} & i\conj{\rho(\conj{k})}\e^{i\delta}
\end{pmatrix},
	\quad k \in \Sigma_{+1}^o \cap \Sigma_{-1}^o\,.
\]
The symmetry~\eqref{J_o symmetry} gives
\[
	\Phi^+(x,t,k) = \Phi^-(x,t,k) 
\begin{pmatrix} 
		-i\rho(k)\e^{-i\delta} & -i\e^{i\delta} \\
		-i(1+\rho(k)\conj{\rho(\conj{k})})\e^{-i\delta} & -i\conj{\rho(\conj{k})}\e^{i\delta}
\end{pmatrix},
	\quad k \in \Sigma_{+2}^o \cap \Sigma_{-2}^o\,.
	\endproof
\]

\subsection{Riemann problems}
\label{a:Riemann problems}
%
\paragraph{Proof of Lemma~\ref{Pure two-sided step IC, Vnot=0 zeros} (Pure two-sided step with counterpropagating flows).}
\label{Pure two-sided step IC, Vnot=0 appendix}
\bb
Here we consider the possible zeros and poles of $s_{22}(k)$, $\rho(k)$ and $r(k)$ for $V\not=0$ and $0<A_- \leq A_+$.
\eb

Suppose $s_{22}(k) = 0$. Then 
\[
	(\lambda_- + (k-V/2))A_+\e^{i\delta} + (\lambda_+ - (k +V/2))A_-\e^{-i\delta} = 0
\nonumber
\]
and so
$	|\lambda_- + (k-V/2)|A_+ = |\lambda_+ - (k+V/2)|A_-
$.
Note that 
\begin{align*}
		|\lambda_- + (k-V/2)| &\geq A_-\,, \hspace{4mm} k \in \C\,,
		\\
		|\lambda_- + (k-V/2)| &= A_- \Leftrightarrow k\in \Sigma_-\,,
		\\
		|\lambda_+ - (k+V/2)| &\leq A_+\,, \hspace{4mm}k \in \C\,,
		\\
		|\lambda_+ - (k+V/2)| &= A_+ \Leftrightarrow k\in \Sigma_+\,.
\end{align*}
Since $\Sigma_+$ and $\Sigma_-$ are disjoint, at least one of the above inequalities is strict. Then the above equality implies
$A_-A_+ < A_+A_-$.
Thus, $s_{22}(k) \not= 0$ for all $k \in \C$, meaning that there are no discrete eigenvalues.

Additionally, with $\delta=0$, if $\rho(k)=0$ then
\[
	(\lambda_+ + (k+V/2))A_- = (\lambda_- + (k-V/2))A_+\,.
\nonumber
\]
Squaring, expressing $A_\pm^2$ in terms of $\lambda_\pm, k$, and canceling common factors gives
\[
	(\lambda_+ + (k+V/2))(\lambda_- - (k-V/2)) = (\lambda_- + (k-V/2))(\lambda_+ - (k+V/2))\,.
\nonumber
\]
Expanding and simplifying gives
$	(k-V/2)\lambda_+ = (k+V/2)\lambda_-\,,
$
which after squaring again and simplifying gives
\[
\label{rho root quadratic}
	(A_+^2 - A_-^2)k^2 - V(A_+^2 + A_-^2) k + \frac{V^2}{4}(A_+^2 - A_-^2) = 0\,.
\]
If $A_+ = A_-$ then $\rho(k)$ has a possible zero at $k=0$. Plugging back into $\rho(k)$ shows that this is not an actual zero.

On the other hand,
if $A_+ \not= A_-$, we get two possible zeros, at
\[
	k_\pm = \frac{V}{2}\Big(\frac{A_+ \pm A_-}{A_+ \mp A_-}\Big)\,.
\nonumber
\]
Plugging back into $\rho(k)$, we see that $k_+$ is indeed
a zero of $\rho(k)$ for $A_+ \not= A_-$, while $k_-$ is not.

If $A_+ = A_-$, then $r(k)$ has no singularities. On the other hand,
if $A_+ \not= A_-$, then $r(k)$ has singularity at $k=k_+$, finishing the argument.
\endproof 

\paragraph{Behavior at the branch points for the pure two-sided step with counterpropagating flows.}
\bb
Here we consider the linear dependence of the modified eigenfunctions $\mu_{+1}(x,t,k)$ and $\mu_{-2}(x,t,k)$ at the branch points for $V \not=0$ and $0<A_-\leq A_+$.
\eb

From~\eqref{Pure two-sided step IC, Vnot=0 Jost solutions}, we see that
\bb
\[
	\Wr\big[\mu_{+1}(x,t,k)\,, \mu_{-2}(x,t,k)\big] = 1 + \frac{A_+A_-}{(\lambda_+ + (k+V/2))(\lambda_- + (k-V/2))}\e^{-2i\delta}\,.
\nonumber
\]
\eb
Suppose that 
\bb
$\Wr\big[\mu_{+1}(x,t,p_{+})\,, \mu_{-2}(x,t,p_{+})\big] = 0$.
\eb 
Straightforward algebra shows that the requiring the right-hand side of the above expression to
vanish implies
$A_-\cos(2\delta) - A_+ = iV$\,.
Since $V\not=0$, this cannot be true. Similar calculation shows that 
\bb
$\Wr[\mu_{+1}(x,t,p_{-}), \mu_{-2}(x,t,p_{-})] \not=0$.
\eb
%
\paragraph{Proof of Lemma~\ref{Pure two-sided step IC, V=0 zeros} (Pure two-sided step without counterpropagating flows).}
\label{Pure two-sided step IC, V=0 appendix}
\bb
Here we consider the possible zeros and poles of $s_{22}(k)$ and $\rho(k)$ for $V=0$ and $0<A_-\leq A_+$.
\eb

Suppose $s_{22}(k) = 0$ with $\delta=0$. Following the same argument as for $V\not=0$, we now must have
$A_+(\lambda_+ + k) = -A_-(\lambda_+ - k)$,
and $k \in \Sigma_+ \cap \Sigma_- = \Sigma_-$.
Letting $k = is$ with $s \in [-A_-,A_-]$, we have
\[
	A_+\big(\sqrt{A_-^2-s^2} + is\big) = -A_-\big(\sqrt{A_+^2-s^2} - is\big)\,.
\nonumber
\]
If $A_+ = A_- = A$, then $s_{22}(k)=0$ at $k = \pm iA$. In such case, $\rho(k) =0$ identically and the jumps across $\Sigma_{+1}\setminus\Sigma_{-1}$ and $\Sigma_{+2}\setminus\Sigma_{-2}$ disappear. On the other hand,
if $A_+ \not=A_-$, then $s_{22}(k) \not=0$ for any $k \in \C$, and there are no discrete eigenvalues.

Note that the calculation done to arrive at~\eqref{rho root quadratic} can again be used here, now with $V=0$. Then if $\rho(k) = 0$, we have
$\left(A_+^2 - A_-^2\right) k^2 = 0\,$.
If $A_+ = A_-$, then as stated before $\rho(k)$ is identically zero. On the other hand, if $A_+ \not= A_- >0$, then $\rho(k)$ has a possible zero at $k=0$. Plugging back into $\rho(k)$ verifies that this is indeed a zero.
\endproof
%
\paragraph{Behavior at the branch points for the pure two-sided step without counterpropagating flows.}
\bb
Here we consider the linear dependence of the modified eigenfunctions $\mu_{+1}(x,t,k)$ and $\mu_{-2}(x,t,k)$ at the branch points for $V=0$ and $0<A_-\leq A_+$.
\eb
From~\eqref{e:jost rp-3}, we see that
\vspace*{-1ex}
\bb
\[
	\Wr\big[\mu_{+1}(x,t,k), \mu_{-2}(x,t,k)\big] = 1 + \frac{A_+A_-}{(\lambda_+ + k)(\lambda_- + k)}\e^{-2i\delta}\,.
\nonumber
\]
\eb
Straightforward algebra shows that requiring the right-hand side of the above expression to vanish implies
$A_-\cos(2\delta) = A_+\,$.
Since $A_-\leq A_+$, this can only happen when $A_-=A_+=A$ and $\delta = n\pi$ (i.e. no phase difference).
\bb
In such case, we have $s_{22}(k) \equiv 1$.
\eb

\section{Conclusions}
\label{s:conclusions}

In summary, we presented the formulation of the inverse scattering transform for the focusing nonlinear Schr\"odinger equation with a general class of nonzero boundary conditions at infinity consisting of counterpropagating waves.
The spectrum of the scattering problem is characterized by the presence of four distinct branch points.
Thus, even if one takes into account the multivaluedness of the asymptotic eigenvalues by introducing a suitable two-sheeted Riemann surface, the resulting curve has genus one and, therefore, it is not possible to introduce a uniformization variable to map it back to a single copy of the complex plane.
Accordingly, we developed the formalism by explicitly taking into account the non-analyticity of the asymptotic eigenvalues and by making a suitable choice of branch cuts.
We also explicitly studied the limiting behavior of the Jost eigenfunctions and scattering coefficients at the branch points.
We formulated the inverse problem as a matrix Riemann-Hilbert problem with jumps along the real axis and the branch cuts, converted the problem to a set of linear algebraic-integral equations,
and obtained a reconstruction formula for the potential.
We discussed several exact reductions as special cases. 
One of them is the case when no counterpropagating flows are present, namely $V=0$, which had been studied in \cite{DPMV2014}.
Even in that case, however, our formalism is slightly different from that of \cite{DPMV2014} in a few respects
(such as proof of analyticity, different sectionally meromorphic matrix,  etc.).
Finally, we considered various Riemann problems as specific examples.

The availability of the inverse scattering transform makes it possible to calculate the long-time asymptotic behavior of solutions with the given class of initial conditions.  
Similar problems were recently considered in \cite{B2018} using the genus-one Whitham modulation equations,
and it was shown that, in many cases,
Whitham theory provides an effective asymptotic description for the behavior of solutions.
However, it was also shown there that there are many cases in which the genus-one Whitham equations are 
not sufficient to fully characterize the behavior of solutions.
To fully describe those cases, either higher-genus theory or the full power of the IST are needed.
Moreover, even when effective, Whitham theory is only a formal perturbation theory, and does not provide rigorous estimates.
The long-time asymptotics using the inverse scattering transform was computed in \cite{BM2016,BM2017} in the special case $A_+ = A_-$ and $V=0$.
Until recently, the case $V\ne0$ was only studied in a special case (a Riemann problem with equal amplitudes and $V>0$) in \cite{BV2007}.
Moreover, even in that case, the analysis only applies to the case of large $V$.

While in the process of finalizing the present manuscript, we learned that a similar problem was also independently considered in a recent preprint\cite{MLS2020}, 
where the inverse scattering transform was concisely formulated and 
various scenarios for the long-time asymptotics were presented and discussed.
The main differences between the formalism of the inverse scattering transform in \cite{MLS2020} and the one in the present work are that
a different normalization was used for the Jost eigenfunctions, that
%
no discrete spectrum for the scattering problem, and consequently no poles in the Riemann-Hilbert problem, were allowed in \cite{MLS2020},
\bb
and that the issue of existence and uniqueness of solutions of the Riemann-Hilbert problem was not addressed in \cite{MLS2020}.
\eb
As is well known, each discrete eigenvalue  contributes a soliton to the solution
of the NLS equation.
On one hand, as  shown in \cite{BLM2020}, discrete eigenvalues greatly complicate the long-time asymptotics;
on the other hand, as shown in \cite{BLM2018}, the presence of discrete eigenvalues leads to very interesting interaction phenomena 
between solitons and radiation, including transmission, trapping, and the emergence of
soliton-generated wakes.

As usual, the inverse scattering transform was developed under the assumption of existence of solution.
However, one could  use the results of the present work to prove well-posedness in appropriate function spaces.
At the same time, 
\bb
as was discussed at length,
\eb
the issue of existence and uniqueness of solutions of the Riemann-Hilbert problem is nontrivial.
This is because of the fact that the associated jumps occur along an open contour.
\bb
Here, we addressed this problem by introducing a modified Riemann-Hilbert problem using a similar approach as in \cite{BilmanMiller2019}. 
\eb
However, even in the case $V=0$, it is not entirely clear what conditions must be included in the 
\bb 
original 
\eb
Riemann-Hilbert problem in order to ensure the existence of solutions. 
This question is left as a topic for future work.
Another complication that 
\bb
is addressed by introduction of the modified Riemann-Hilbert problem
\eb
is the possible presence of zeros of the 
analytic scattering coefficients on the continuous spectrum $\Sigma$ (cf.~\eqref{e:Sigmadef}).  
Such zeros lead to so-called spectral singularities in the Riemann-Hilbert problem \cite{Z1989-2}.
We have shown (cf. Lemma~\ref{nonzero scattering entries on branch cuts}) that, when $V\ne0$, no such singularities are possible on $\Sigma_+^o$ and $\Sigma_-^o$.
However, one could have singularities for $k\in\Real$.
Moreover, when $V=0$ the scattering coefficients could vanish on the portion of
$\Sigma$ where all four Jost eigenfunctions are defined (cf. Lemma~\ref{l:nonvanishing scattering coefficients zero velocity}).
The presence of spectral singularities not only complicates the analysis, but also leads to
different long-time asymptotic behavior for the solutions
(e.g., see \cite{AS1981,Kamvissis} for the case of zero boundary conditions).

\bb
\subsection*{Acknowledgments}

We thank the anonymous referees for their constructive criticism, which helped to improve the manuscript.  
This work was partially supported by the National Science Foundation under grant number DMS-2009487.
\eb

\def\doibase{http://dx.doi.org/}

\end{document}